\documentclass{jfm}
\usepackage{amsmath}
\usepackage{color}
\usepackage{graphicx}
\usepackage{epstopdf, epsfig}
\usepackage{overpic}
\usepackage[colorlinks=true]{hyperref}
\usepackage[normalem]{ulem}

\newcommand{\ps}   [2]{\ensuremath{\left(\left. {#1} \, \right| \, {#2} \right) }}
\newcommand\be{\begin{equation}}
\newcommand\ee{\end{equation}}

\newcommand{\tcr}[1]{\textcolor{red}{#1}}

\newcommand{\tcgreen}[1]{\textcolor[rgb]{0,0.6,0}{#1}}

\def\00{\mathbf{0}}
\def\AAA{\mathbf{A}}

\def\FF{\mathbf{F}}
\def\II{\mathbf{I}}
\def\KK{\mathbf{K}}
\def\LL{\mathbf{L}}
\def\MM{\mathbf{M}}
\def\NN{\mathbf{N}}
\def\RR{\mathbf{R}}
\def\SS{\mathbf{S}}
\def\TT{\mathbf{T}}
 	
\def\UUa{\mathbf{U}^\dag}		
\def\UU{\mathbf{U}}

\def\aa{\mathbf{a}}
\def\bb{\mathbf{b}}

\def\ff{\mathbf{f}}

\def\nn{\mathbf{n}}
\def\uu{\mathbf{u}}

\def\xx{\mathbf{x}}
 
\def\nnu{\Rey^{-1}}

\DeclareMathOperator*{\argminA}{arg\,min}

\shorttitle{Second-order adjoint-based sensitivity analysis}
\shortauthor{E. Boujo}

\title{Second-order adjoint-based sensitivity
\\ for hydrodynamic stability and control}

\author{Edouard Boujo}

\affiliation{Laboratory of Fluid Mechanics and Instabilities, \'Ecole Polytechnique F\'ed\'erale de Lausanne, CH-1015 Lausanne, Switzerland}

\begin{document}

\maketitle

\begin{abstract}
Adjoint-based sensitivity analysis is routinely used today to assess efficiently the effect of  open-loop control on the linear stability properties of unstable flows. Sensitivity maps identify regions where small-amplitude control is the most effective, i.e. yields the largest first-order (linear) eigenvalue variation. In this study an adjoint method is proposed for computing a second-order (quadratic) sensitivity operator, and applied to the flow past a circular cylinder, controlled with a steady body force or a passive device model. Maps of second-order eigenvalue variations are obtained, without computing controlled base flows and eigenmodes. For finite control amplitudes, the second-order analysis improves the accuracy of the first-order prediction, and informs about its range of validity, and  whether it underestimates or overestimates the actual eigenvalue variation. Regions are identified where control has little or no first-order effect but a second-order effect. In the cylinder wake, the effect of a control cylinder tends to be underestimated by the first-order sensitivity, and including second-order effects yields larger regions of flow restabilisation. Second-order effects can be decomposed into two mechanisms: second-order base flow modification, and interaction between first-order modifications of the base flow and eigenmode. Both contribute equally in general in sensitive regions of the cylinder wake. Exploiting the second-order sensitivity operator, the optimal control maximising the total second-order stabilisation is computed via a quadratic eigenvalue problem. The approach is applicable to other types of control (e.g. wall blowing/suction and shape deformation) and  other eigenvalue problems (e.g.  amplification of time-harmonic perturbations, or resolvent gain, in stable flows).
\end{abstract}


\section{Introduction}

Over the past decades, adjoint-based sensitivity analysis has become a standard tool for estimating the effect of flow control.
The key underlying idea is to  compute the gradient 
of a quantity of interest with respect to control by solving so-called adjoint equations, only once. 
This approach contrasts with the brute-force method, where the gradient is obtained by solving the direct equations (e.g. Navier--Stokes equations) once for each control degree of freedom. 
When the control has many degrees of freedom, for instance when it depends on space or time, the adjoint method dramatically reduces the computational cost.
This efficient  calculation is crucial in iterative gradient-based methods for optimal control, where the gradient is repeatedly evaluated. 
This is true in general for systems governed by partial differential equations  \citep{Lions71}, and in particular for a wide range of problems in fluid mechanics: 
shape optimisation for aerodynamics
 or mixing \citep{Jameson98, Pironneau01};
optimal wall actuation for turbulent drag reduction \citep{Bewley01}  
or mixing \citep{Foures14};
optimal kinematics for thin-film coating \citep{BoujoSellier2019};
and optimal perturbations (initial perturbations undergoing the largest possible transient growth), especially for time-varying base flows or nonlinear amplification \citep{Schmid2007}, the latter being relevant to transition to turbulence \citep{Pringle2010, Monokrousos11}.

Adjoint equations also appear naturally in fluid mechanics when investigating how linear stability properties (growth rate and frequency, characterised by a linearised eigenvalue problem) are affected by flow control \citep{Luchini14}. 
Sensitivity maps are obtained that allow one to identify the most sensitive regions at a glance and thus to design effective controls easily.
This approach is very efficient: unlike trial-and-error techniques, it never actually solves for controlled flows, and only requires one adjoint calculation. 
The method has been applied extensively in the flow past a circular cylinder, a prototypical globally unstable open flow: the sensitivity of the leading eigenvalue has been computed with respect to  passive control  (namely, a model of a  small secondary cylinder acting on both the base flow and the perturbations) \citep{Hill92AIAA}, to a localised feedback force proportional to the perturbation flow velocity \citep{Giannetti07}, and to flow modification and steady forcing in the bulk \citep{Marquet08cyl}.  
To some extent, these  studies correctly identified  restabilising regions where vortex shedding is suppressed by a small secondary cylinder, first identified by the systematic experiment of \cite{Strykowski1990}.
Other studies include sensitivity to base flow modification in the parallel Couette flow \citep{Bottaro03},  
a compressible axisymmetric body wake \citep{Meliga10} 
controlled with steady forcing in the bulk
(with sources of mass, momentum or energy) and steady  wall control (with blowing/suction or
heating),
the wake past a spheroidal bubble \citep{Tchoufag2013},
a three-dimensional T-junction \citep{Fani2013},
and a thermoacoustic system \citep{Magri13}.

\begin{figure} 
\centerline{
   \hspace{-0.3cm}
   \begin{overpic}[height=5.5cm, trim=0mm 70mm 100mm 55mm, clip=true]{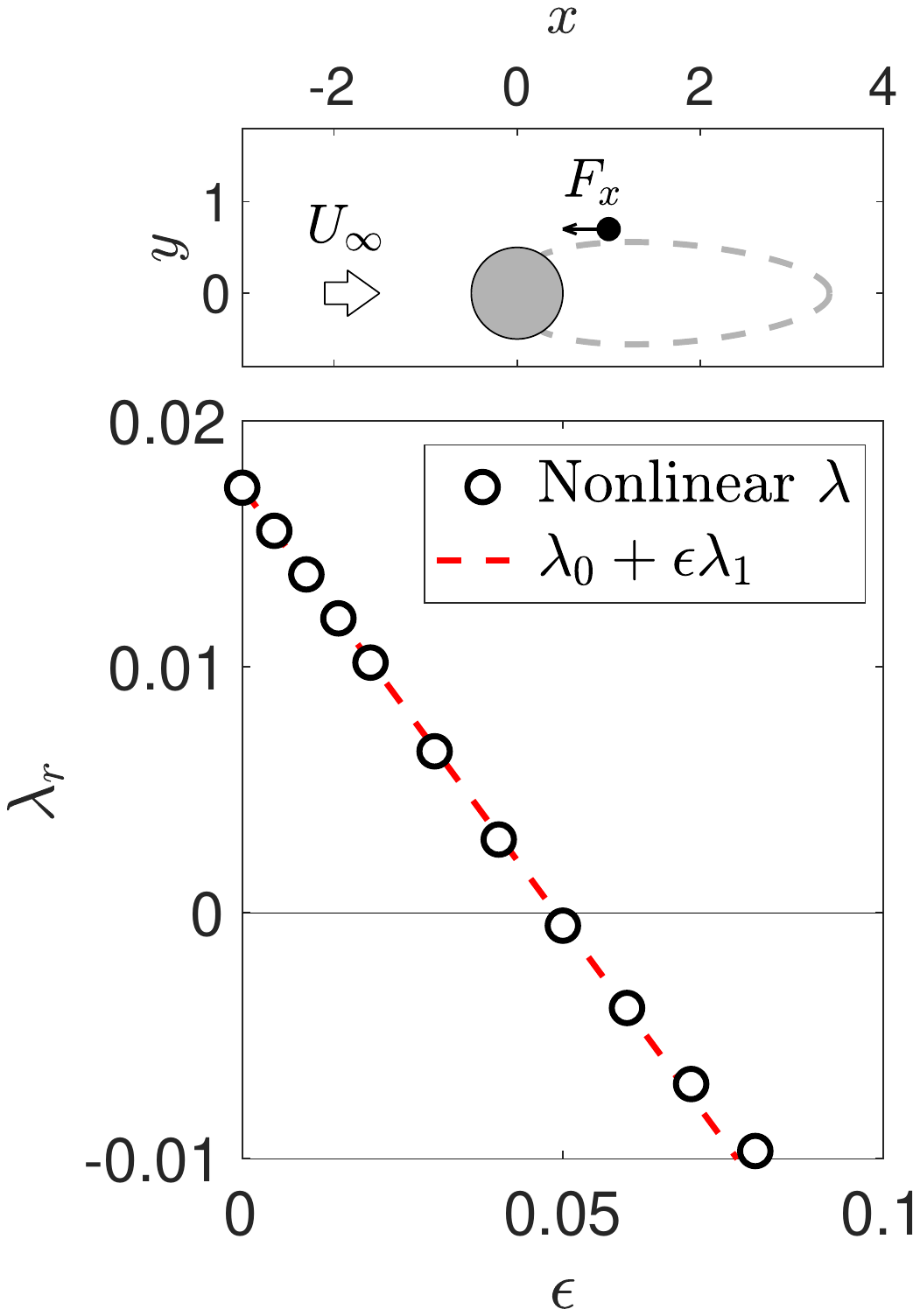}
      \put(6,85){$(a)$}
   \end{overpic}  
   \hspace{0.1cm}
   \begin{overpic}[height=5.5cm, trim=15mm 70mm 100mm 55mm, clip=true]{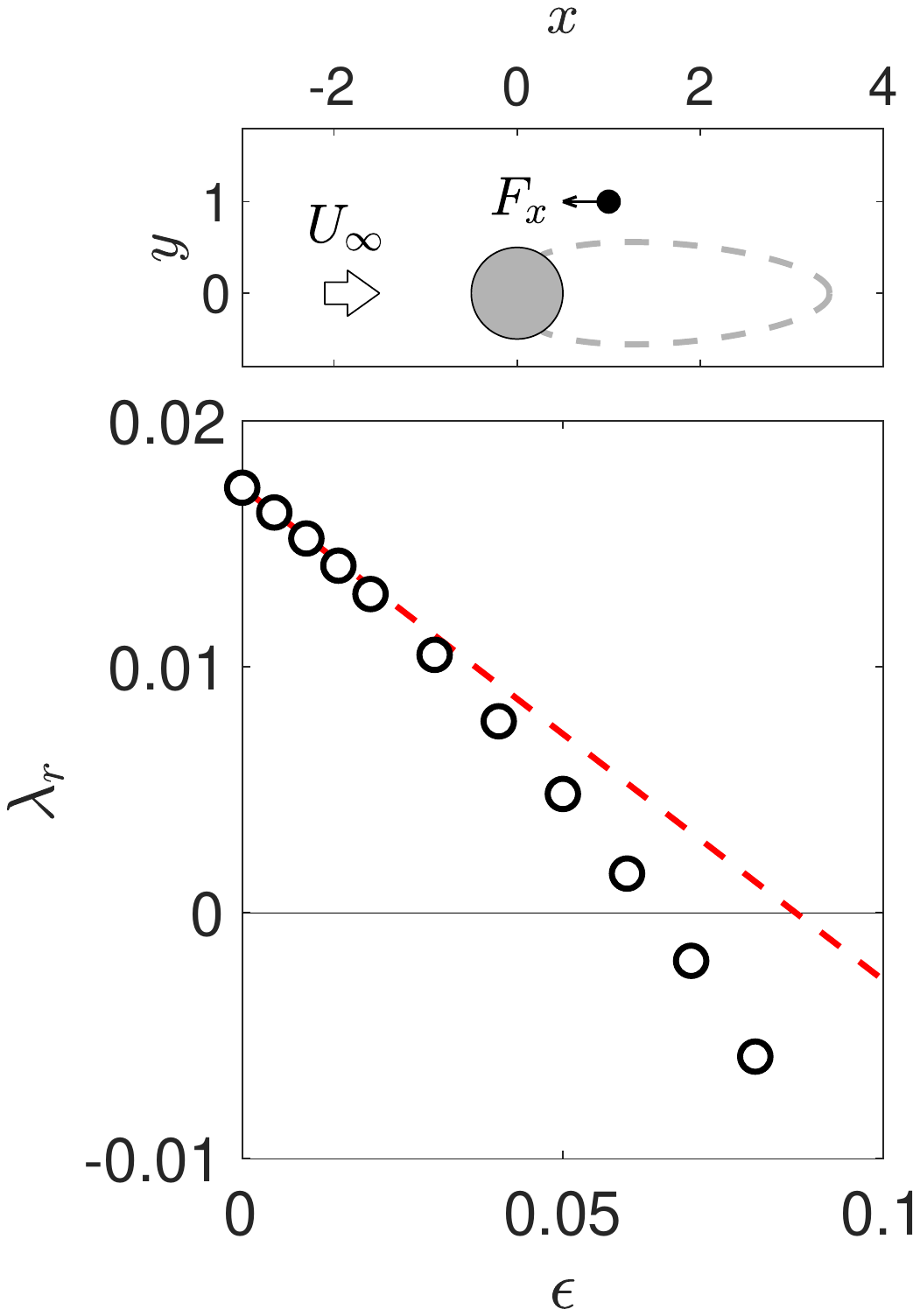}
      \put(-2,85){$(b)$}
   \end{overpic}  
   \hspace{0.1cm}
   \begin{overpic}[height=5.5cm, trim=15mm 70mm 100mm 55mm, clip=true]{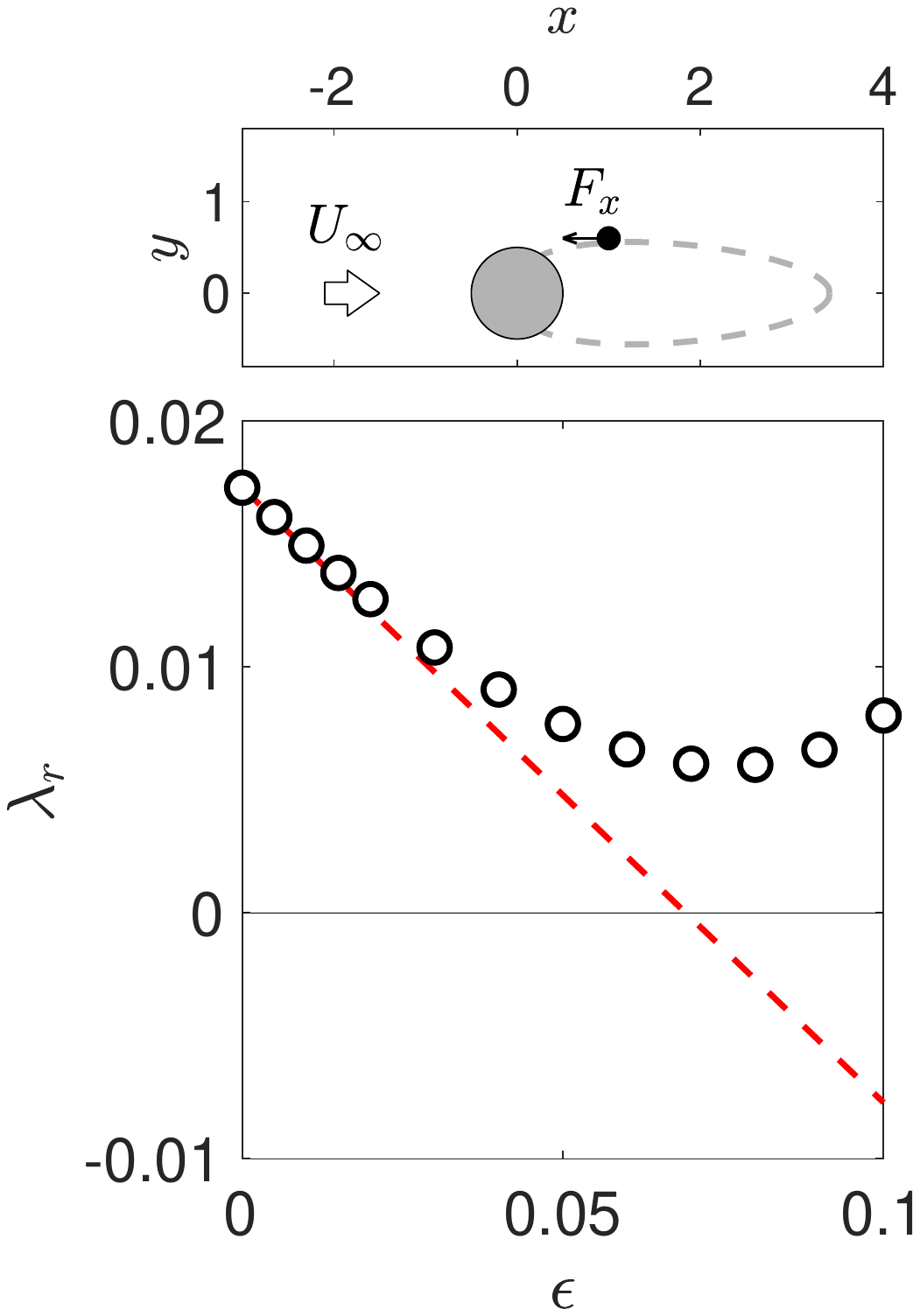}
      \put(-2,85){$(c)$}
   \end{overpic}  
   \hspace{0.1cm}
   \begin{overpic}[height=5.5cm, trim=15mm 70mm 100mm 55mm, clip=true]{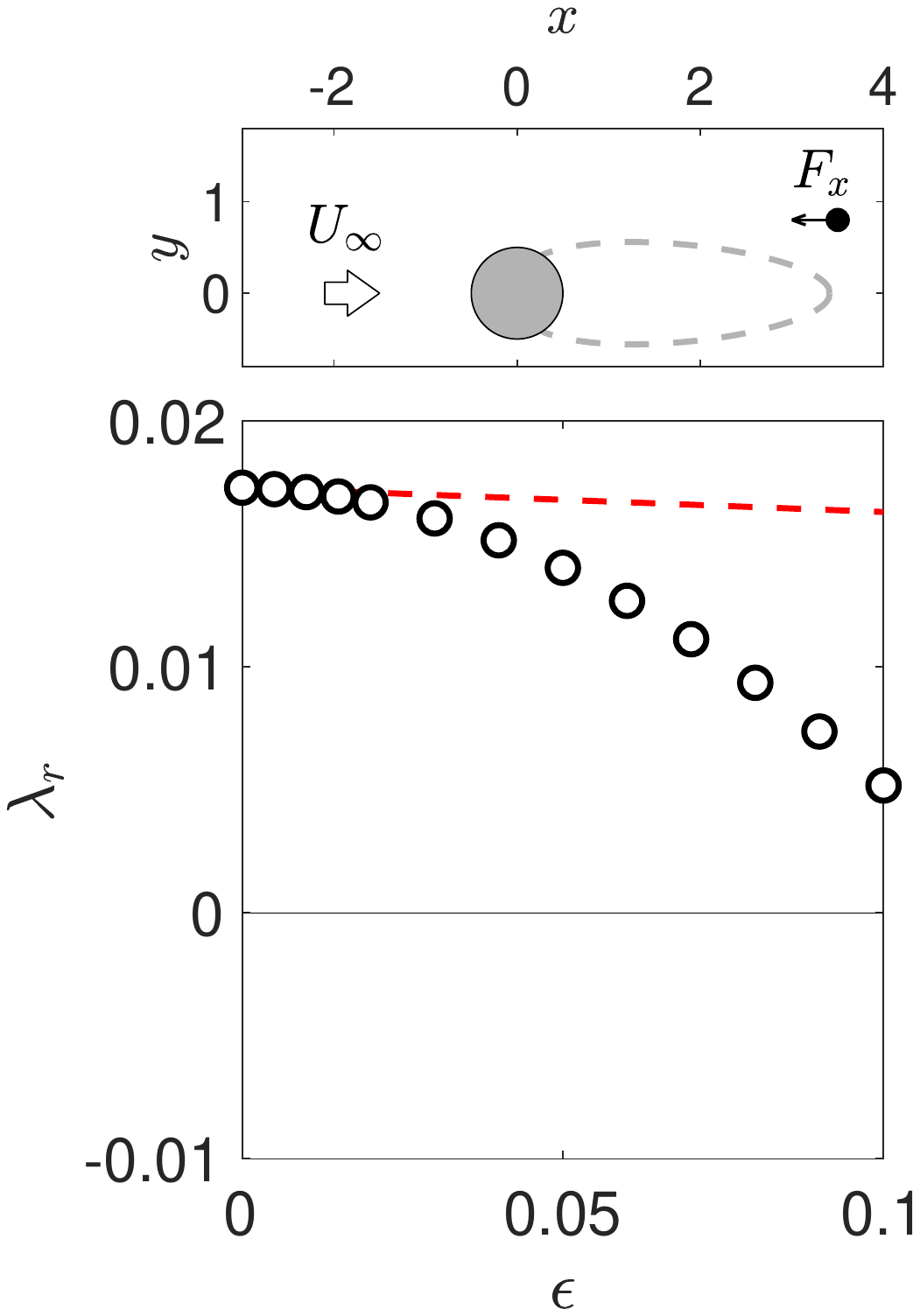}
      \put(-2,85){$(d)$}
   \end{overpic}  
}
\caption{
Variation of the leading eigenmode's growth rate for the flow past a circular cylinder at $\Rey=50$, induced by a  body force oriented along $-x$, of amplitude $\epsilon$, and localised in
$(a)$~$\xx_c=(1,0.7)$,
$(b)$~$\xx_c=(1,1)$,
$(c)$~$\xx_c=(1,0.6)$,
and
$(d)$~$\xx_c=(3.5,0.8)$.
Symbols: nonlinear calculations;
dashed line: first-order sensitivity.
}
\label{fig:valid_1}
\end{figure}

Because standard sensitivity analysis computes a gradient,
it is linear by nature and  expected to provide meaningful results in the limit of infinitesimal  flow control only. 
For finite-amplitude control, nonlinear effects come into play, and the actual variation of the quantity of interest inevitably departs from the sensitivity prediction. 
This is illustrated in figure~\ref{fig:valid_1}, which shows the effect of a localised body force on the leading growth rate $\lambda_r$ of the cylinder flow. At $\Rey=50$ the uncontrolled flow is slightly unstable, $\lambda_r(\epsilon=0)>0$.
In all four control locations considered, the body force  has a stabilising effect: the growth rate computed about the nonlinearly controlled base flow  (symbols) initially decreases.   
Sensitivity analysis (dashed lines) perfectly captures the slope of the growth rate reduction at zero amplitude, $\mathrm{d}\lambda_{r}/\mathrm{d}\epsilon|_{\epsilon=0}$.
It does not, however, provide any information about finite amplitudes $\epsilon>0$:
depending on the control location, sensitivity analysis is accurate up to smaller or larger amplitudes, and may or may not predict well the critical stabilising amplitude;
it may also  underestimate or overestimate the actual growth rate variation. This information cannot be obtained except with nonlinear calculations of the controlled flow.

Given this limitation, 
it is tempting to investigate whether adding one or more higher-order terms in the sensitivity analysis can improve the prediction accuracy for small but finite amplitudes.
In some scientific fields, second-order sensitivity is sometimes calculated as a means to speed up the convergence of iterative gradient-based optimisation, where the modified state and the sensitivity need to be  repeatedly recomputed. 
In hydrodynamic stability, iterative optimisation is seldom performed, and only first-order sensitivity is routinely calculated.
One notable exception concerns  the three-dimensional control of nominally two-dimensional (or axisymmetric) flows: when the control is periodic in the spanwise (or azimuthal) direction, the standard first-order sensitivity is exactly zero, and at leading order the effect of the control is quadratic
\citep{Hwang2013, DelGuercio2014globalwake, DelGuercio2014globalcyl, DelGuercio2014parallel}.
In other words, expressing the eigenvalue variation with the control amplitude $\epsilon$ as $\lambda = \lambda_0 + \epsilon \lambda_1 + \epsilon^2 \lambda_2 + \ldots$,
the aforementioned periodic configuration is such that $\lambda_1=0$, and one needs to compute $\lambda_2$.
This has triggered a number of studies that either evaluated the second-order variation induced by a given control
\citep{Cossu14secondorder, Tammisola2014}, or computed optimal spanwise-periodic flow modification or  control \citep{Tammisola2017, Boujo15secondorderJFM, Boujo19sensit}. 
To the best of the author's knowledge, 
the second-order sensitivity of  eigenvalues has never been computed in non-parallel flows subject to external control in the general case where  $\lambda_1 \neq 0$,
although the steps of the derivation are similar.
Very recently, a related approach was proposed by \cite{Mensah2020JSV} to compute second- and higher-order eigenvalue variations $\lambda_n$ induced by some scalar
parameter modification. That approach, which  explicitly computes eigenvector modifications, was applied to the parallel Poiseuille flow for variations of the Reynolds number,
and to a two-dimensional time-delayed thermoacoustic system for variations of the time delay.

The first aim of the present study is to propose a method for computing efficiently the second-order sensitivity of an eigenvalue with respect to control, in the context of hydrodynamic instability.
Some emphasis is put on exploiting adjoint operators to derive a sensitivity that is valid for any control, instead of simply evaluating the second-order variation $\lambda_2$ for a specific control.
Specifically, and postponing rigorous definitions to \S~\ref{sec:Theoretical},
 it might help to recall that the first-order coefficient of the eigenvalue variation
 can be expressed as  
 $\lambda_1 = \ps{\SS_1}{\FF}$, the inner product of a control $\FF$ with a first-order sensitivity $\SS_1$ that depends only on the uncontrolled base flow; therefore, as $\SS_1$ is independent of the control,  it can be computed once and for all, without computing controlled base flows and eigenmodes. 
Similarly, the present study will express the second-order variation as $\lambda_2 = \ps{\FF}{\SS_2 \FF}$,
where the second-order sensitivity $\SS_2$  depends only on the uncontrolled base flow.
The method will be illustrated with the global instability of the two-dimensional cylinder flow, controlled by a steady localised force or by a small control cylinder.
The second aim of this study is to leverage second-order sensitivity to find the optimal control for stabilisation, i.e. the control yielding the largest growth rate reduction up to second order, $\epsilon \lambda_{1r} + \epsilon^2 \lambda_{2r}$.

The paper is organised as follows.
Section~\ref{sec:Theoretical} introduces the theoretical framework for the first- and second-order sensitivities of eigenvalues with respect to control (\S\S~\ref{sec:base}-\ref{sec:sensit_ops}).
It also discusses  the generalisation to higher orders (\S~\ref{sec:higher_order}) and the computational cost of the method (\S~\ref{sec:cost}).
Section~\ref{sec:config_num} presents the flow configuration and numerical methods.
Results for the growth rate of the leading eigenmode of the cylinder flow at $\Rey=50$ are given in \S~\ref{sec:results}:
sensitivity to a steady force (\S~\ref{sec:results_F}),
sensitivity to a small control cylinder
(\S~\ref{sec:results_pass_device}),
and an analysis of the stabilisation induced by the small control cylinder when located nearly optimally
(\S~\ref{sec:analysis}).
Finally, section~\ref{sec:opt} deals with  optimal controls that maximise the growth rate reduction at first or second order separately, and first and second orders simultaneously.
In addition, Appendix~\ref{app:freq}
briefly presents results for the sensitivity  of the leading mode's frequency;
Appendix~\ref{app:operators}
details the derivation of the sensitivity operators;
and Appendix~\ref{app:gain}
outlines an extension of the method to the sensitivity of another quantity  defined as an eigenvalue problem: the linear amplification of time-harmonic forcing (resolvent gain).

\section{Theoretical framework}
\label{sec:Theoretical}

\subsection{Base flow and stability analysis}
\label{sec:base}

Consider a steady fluid
flow satisfying the stationary incompressible Navier-Stokes (NS) equations
\begin{eqnarray} 
\UU \bcdot \bnabla\UU + \bnabla  P - \nnu \bnabla^2  \UU &=& \00,
\label{eq:NS_momentum}
\\
\bnabla \bcdot \UU  &=& 0,
\label{eq:NS_continuity}
\end{eqnarray}
where  $P(\xx)$  is the pressure field  and  $\UU(\xx)=(U,V)^T$ or $(U,V,W)^T$ is the velocity vector in two or three dimensions.
Equations are made dimensionless with a characteristic velocity $U_\infty$, a characteristic length scale $D$  and the fluid kinematic viscosity $\nu$, thus defining the Reynolds number $\Rey=U_\infty D /\nu$. 
In the following, all velocity fields are  incompressible and the continuity equation is omitted.

The linear stability of the base flow 
is determined by the temporal evolution of small perturbations 
$\uu'$.
Considering, in particular, the normal mode ansatz
 $\uu'(\xx,t) = \uu(\xx) e^{\lambda t} + c.c.$,
the (complex) eigenmodes $\uu(\xx)$ are solutions of the linearised NS equations
\begin{eqnarray} 
\lambda\uu +  \UU \bcdot \bnabla\uu  + \uu \bcdot \bnabla\UU    +   \bnabla  p - \nnu \bnabla^2  \uu = \00.
\label{eq:evp_uncontrolled_momentum}
\end{eqnarray}
The real and imaginary parts of an eigenvalue $\lambda$  represent the linear growth rate $\lambda_r$  and linear frequency $\lambda_i$ of the associated eigenmode.
The base flow is linearly unstable if at least one mode has a positive growth rate.
In compact form, equations (\ref{eq:NS_momentum})-(\ref{eq:NS_continuity}) for the steady base flow and (\ref{eq:evp_uncontrolled_momentum}) for the eigenmodes can be expressed as
\begin{eqnarray} 
\NN(\UU) &=& \00,
\label{eq:NS_compact}
\\
(\lambda \II +\AAA) \uu &=& \00.
\label{eq:evp_compact}
\end{eqnarray}
Here,
$\NN$ and $\AAA$ are  the nonlinear and linearised Navier-Stokes operators, 
and $\II$ is the identity operator:
\begin{eqnarray}
\NN(\UU) &=& 
\UU \bcdot \bnabla\UU + \bnabla  P - \nnu \bnabla^2  \UU,
\\
 \AAA(\UU) \uu &=& 
\UU \bcdot\bnabla\uu + \uu\bcdot \bnabla\UU    +\bnabla p - \nnu \bnabla^2\uu  .
\end{eqnarray}

\subsection{Eigenvalue sensitivity to  small-amplitude steady control}
\label{sec:steady}

Assume now that a small-amplitude control is applied via a body force  acting on the steady base flow:
\begin{eqnarray} 
\NN(\UU) &= \epsilon \FF,
\label{eq:NS_compact_F}
\end{eqnarray}
where $||\FF||=1$ and $0< \epsilon \ll 1$.
This control modifies the base flow, eigenmodes and eigenvalues, which can be expressed as power series expansions \citep{Hinch1991}:
\begin{eqnarray} 
\UU &=& \UU_0+\epsilon\UU_1+\epsilon^2\UU_2+\ldots,
\label{eq:exp_Q}
\\
\uu &=& \uu_0+\epsilon\uu_1+\epsilon^2\uu_2+\ldots,
\label{eq:exp_q}
\\
\lambda &=& \lambda_0 +\epsilon \lambda_1 +\epsilon^2 \lambda_2 +\ldots.
\label{eq:exp_ev}
\end{eqnarray} 
Injecting the expansion (\ref{eq:exp_Q}) into the base flow equation (\ref{eq:NS_compact_F}) yields the following at orders $\epsilon^0$, $\epsilon^1$ and $\epsilon^2$:
\begin{eqnarray} 
\NN(\UU_0) &=& \00,
\label{eq:Q0}
\\
\AAA_0 \UU_1 &=& \FF,
\label{eq:Q1}
\\
\AAA_0 \UU_2 &=& - \UU_1 \bcdot \bnabla \UU_1,
\label{eq:Q2}
\end{eqnarray}
where $\AAA_0=\AAA(\UU_0)$ is the NS operator linearised about the uncontrolled base flow $\UU_0$,
i.e. $\AAA_0 \UU_n = 
\UU_0 \bcdot\bnabla\UU_n + \UU_n\bcdot \bnabla\UU_0 +\bnabla P_n - \nnu \bnabla^2 \UU_n$ for $n=1,2$.
Although the focus of this study is on first and second orders, note that the steady force $\FF$ modifies the base flow at higher orders too, due to the nonlinear  term of the NS operator. 

Similarly, injecting the expansions  (\ref{eq:exp_Q})-(\ref{eq:exp_ev}) into the eigenvalue problem (\ref{eq:evp_compact}) yields the following at orders $\epsilon^0$, $\epsilon^1$ and $\epsilon^2$:
\begin{eqnarray} 
(\lambda_0 \II+\AAA_0) \uu_0 &=& \00,
\label{eq:evp0}
\\
(\lambda_0 \II+\AAA_0) \uu_1 &=& -(\lambda_1\II+\AAA_1) \uu_0,
\label{eq:evp1}
\\
 (\lambda_0\II+\AAA_0) \uu_2 &=&
-(\lambda_1\II+\AAA_1) \uu_1 
-(\lambda_2\II+\AAA_2) \uu_0,
\label{eq:evp2}
\end{eqnarray}
where the  operators $\AAA_1$ and $\AAA_2$ are linear in $\UU_1$ and $\UU_2$, respectively, and do not depend on any other field,
\begin{eqnarray}
\AAA_1 =
   \UU_1 \bcdot\bnabla(*)+ (*) \bcdot \bnabla \UU_1,
\quad
\AAA_2 = 
\UU_2 \bcdot\bnabla(*)   + (*)\bcdot \bnabla \UU_2.
\end{eqnarray}

Before moving on to determining the first- and second-order eigenvalue variations $\lambda_1$ and $\lambda_2$,
note that the operator $\lambda_0\II + \AAA_0$ is singular, since (\ref{eq:evp0}) holds. 
Therefore, according to the Fredholm alternative, commonly used in the context of weakly nonlinear expansions (see e.g. \cite{Sipp2007}), 
 (\ref{eq:evp1})-(\ref{eq:evp2}) can be solved for $\uu_1$ and $\uu_2$ if and only if their right-hand sides have no component in the direction of the eigenmode $\uu_0$,
 i.e. no projection on the adjoint mode $\uu_0^\dag$. 
Recall that the adjoint mode is a solution of
\begin{eqnarray}
\left(  \overline\lambda_0\II + \AAA_0^\dag \right) \uu_0^\dag  = \00,
\label{eq:adjoint}
\end{eqnarray}
where the overbar stands for complex conjugation,
and $\AAA_0^\dag$ is the adjoint NS operator for the $L^2$ inner product 
$\ps{\aa}{\bb} 
= 
\iint 
\overline\aa^T 
\bb \,
\mathrm{d}\xx$ 
for any $\aa$, $\bb$,
\begin{eqnarray}
\AAA_0^\dag \uu_0^\dag= 
-\UU_0\bcdot\bnabla \uu_0^\dag + \uu_0^\dag \bcdot\bnabla\UU_0^T  -\bnabla p_0^\dag - \nnu \bnabla^2 \uu_0^\dag   ,
\end{eqnarray}
such that $\ps{ \aa}{ \AAA_0 \bb} = \ps{ \AAA_0^\dag \aa}{\bb}$ for any $\aa$, $\bb$.
In particular, projecting the left-hand side of  (\ref{eq:evp1})-(\ref{eq:evp2}) on $\uu_0^\dag$ 
necessarily yields zero:
\begin{eqnarray}
\ps{ \uu_0^\dag}{ ( \lambda_0\II + \AAA_0) \uu_n}
=
\ps{ \left(  \overline\lambda_0\II + \AAA_0^\dag \right) \uu_0^\dag}{\uu_n} = 0,
\quad n=1,2.
\end{eqnarray}

Choosing the normalisation $\ps{ \uu_0^\dag } { \uu_0}=1$,
the eigenvalue variations are  obtained by projecting 
(\ref{eq:evp1})-(\ref{eq:evp2}) on $\uu_0^\dag$ \citep{Hinch1991, Chomaz05, Giannetti07}:
\begin{eqnarray}
\lambda_1  &=& -\ps{ \uu_0^\dag } {{\AAA}_1 \uu_0},
\label{eq:ev1}
\\
\lambda_2 &=& -\ps{ \uu_0^\dag }{
\underbrace{\AAA_2\uu_0}_{\text{I}} 
+
\underbrace{(\lambda_1\II+\AAA_1)\uu_1}_{\text{II}}
}.
\label{eq:ev2}
\end{eqnarray}
Any arbitrary component along $\uu_0$ can be added to  $\uu_1$ and  (\ref{eq:evp1}) will still hold, because of (\ref{eq:evp0}). 
This arbitrary component does not influence $\lambda_2$, because (\ref{eq:evp1}) also implies that $\ps{\uu_0^\dag} {(\lambda_1 \II + \AAA_1)\uu_0}=0$.

The two terms in (\ref{eq:ev2}) correspond to different mechanisms: 
term I is the effect of the second-order base flow modification $\UU_2$ (via the first-order flow modification and the nonlinear term of the NS operator);
term II is the effect of the interaction between the first-order flow modification $\UU_1$ and the first-order eigenmode modification $\uu_1$.
As will be discussed in \S~\ref{sec:results}, these two terms can either compete or collaborate.

Given a steady force $\FF$, one can
compute the base flow modifications $\UU_1$ and $\UU_2$ from (\ref{eq:Q1})-(\ref{eq:Q2}), 
build $\AAA_1$ and $\AAA_2$,
and use expressions (\ref{eq:ev1})-(\ref{eq:ev2}) to estimate the eigenvalue variation up to first order,
$\lambda = \lambda_0 + \epsilon \lambda_1 + O(\epsilon^2)$, 
and up to second order, 
$\lambda = \lambda_0 + \epsilon \lambda_1 + \epsilon^2 \lambda_2 + O(\epsilon^3)$. 
This  involves solving linear systems only, which avoids computing the nonlinear controlled flow $\UU$ and solving the eigenvalue problem for the controlled mode $\uu$, thus reducing the computational cost. 
For instance, the dashed lines  in figure~\ref{fig:valid_1} may be obtained by computing $\lambda_{1r}$ this way.
However, the procedure must be repeated  every time a different force $\FF$ is considered, which may become prohibitively expensive.
More useful expressions for $\lambda_1$ and $\lambda_2$ can be obtained that do not depend explicitly on $\UU_1$ and $\UU_2$, as explained in the next section.

\subsection{Sensitivity operators}
\label{sec:sensit_ops}

Because the operator $\AAA_1$  is linear in $\UU_1$, which itself depends linearly on $\FF$, the first-order eigenvalue variation (\ref{eq:ev1}) can be recast as 
\begin{eqnarray}
\lambda_1 &= \ps{\SS_1}{\FF},
\label{eq:sensit1}
\end{eqnarray}
where the vector field $\SS_1$ is the usual sensitivity to a steady force \citep{Marquet08cyl, Meliga10}, and depends only on the uncontrolled base flow $\UU_0$ and the uncontrolled direct and adjoint modes $\uu_0$ and $\uu_0^\dag$ (see Appendix~\ref{app:operators}).
This formulation offers a significant advantage: $\SS_1$ can be calculated once and for all, and then used to predict the first-order effect of any steady force. 
Since no base flow modification $\UU_1$ is ever calculated, evaluating $\lambda_1$ for a large number of steady forces becomes straightforward. 
For instance, figure~\ref{fig:F}$(a)$ shows the real part of the streamwise component of $\SS_1$. 
The value displayed at each location $\xx_c$ is the first-order sensitivity of the growth rate to  a steady force $\FF=(\delta(\xx-\xx_c),0)^T$ localised at that point and oriented along the streamwise direction.

In a similar way, because (\ref{eq:ev2}) is quadratic in $\UU_1$ and thus in $\FF$, the second-order eigenvalue variation can be recast as 
\begin{eqnarray}
\lambda_2 &= \ps{\FF}{\SS_2 \FF},
\label{eq:sensit2}
\end{eqnarray}
where $\SS_2$ is a linear operator
that, again, depends only on the uncontrolled fields $\UU_0$, $\uu_0$ and $\uu_0^\dag$.
The derivation steps from (\ref{eq:ev2}) to (\ref{eq:sensit2}) introduce suitable adjoint operators (see Appendix~\ref{app:operators}), following the same steps as \cite{Boujo19sensit} for spanwise-periodic controls in nominally spanwise-invariant flows (where $\lambda_1=0$ and the expression for $\SS_2$ is slightly simpler).
Again, this formulation suppresses the need to calculate the base flow modifications $\UU_1$ and $\UU_2$. 
Once $\SS_2$ is available, $\lambda_2$ can be readily evaluated for any steady force.
The dashed lines in figure~\ref{fig:valid_1} can now be obtained simply by probing $\SS_1(\xx_c)$ at each control location $\xx_c$ of interest. 

\begin{figure} 
\centerline{
   \hspace{-0.3cm}
   \begin{overpic}[height=5.5cm, trim=0mm 70mm 99mm 55mm, clip=true]{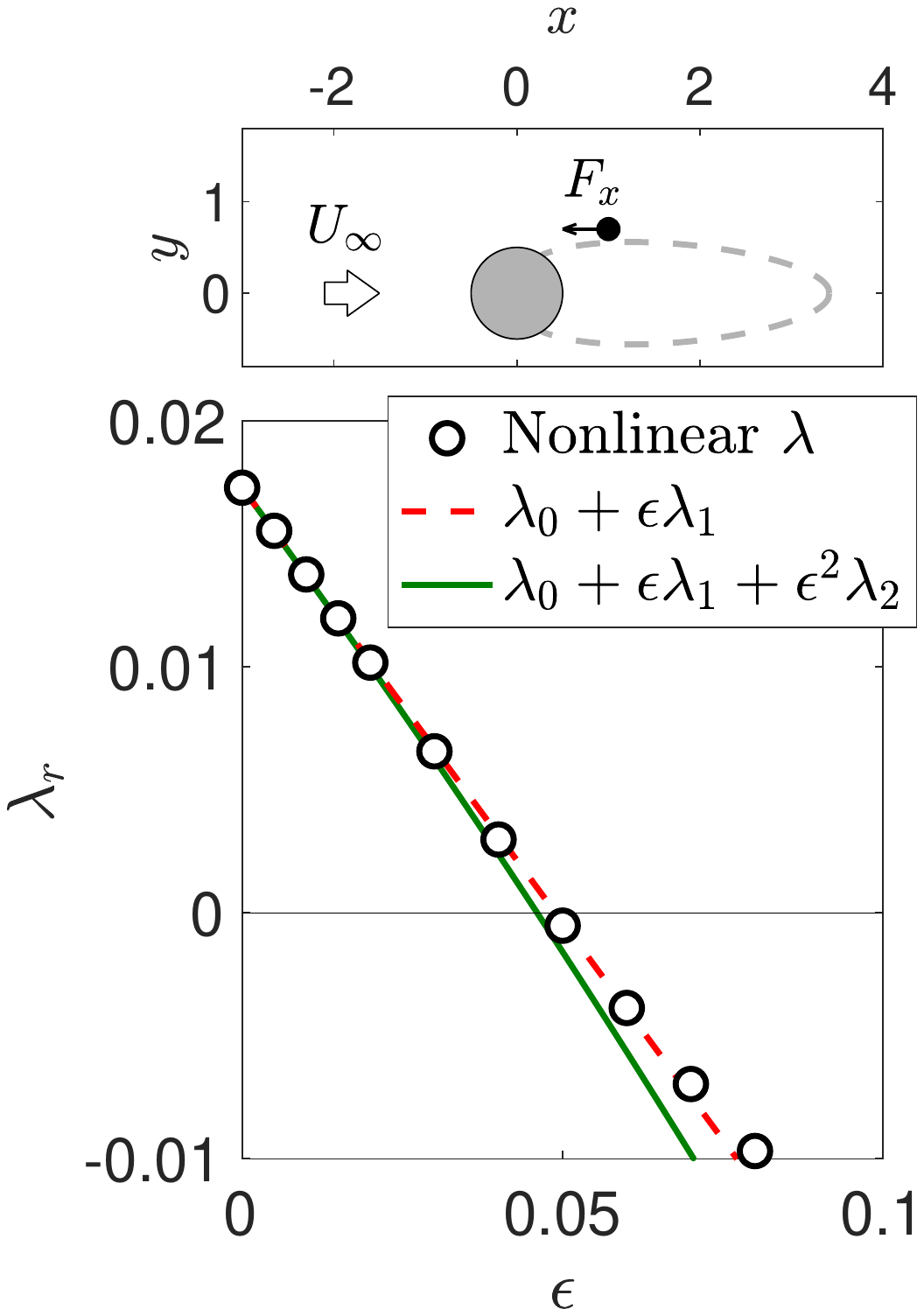}
      \put(6,85){$(a)$}
   \end{overpic}  
   \hspace{0.1cm} 
   \begin{overpic}[height=5.5cm, trim=15mm 70mm 100mm 55mm, clip=true]{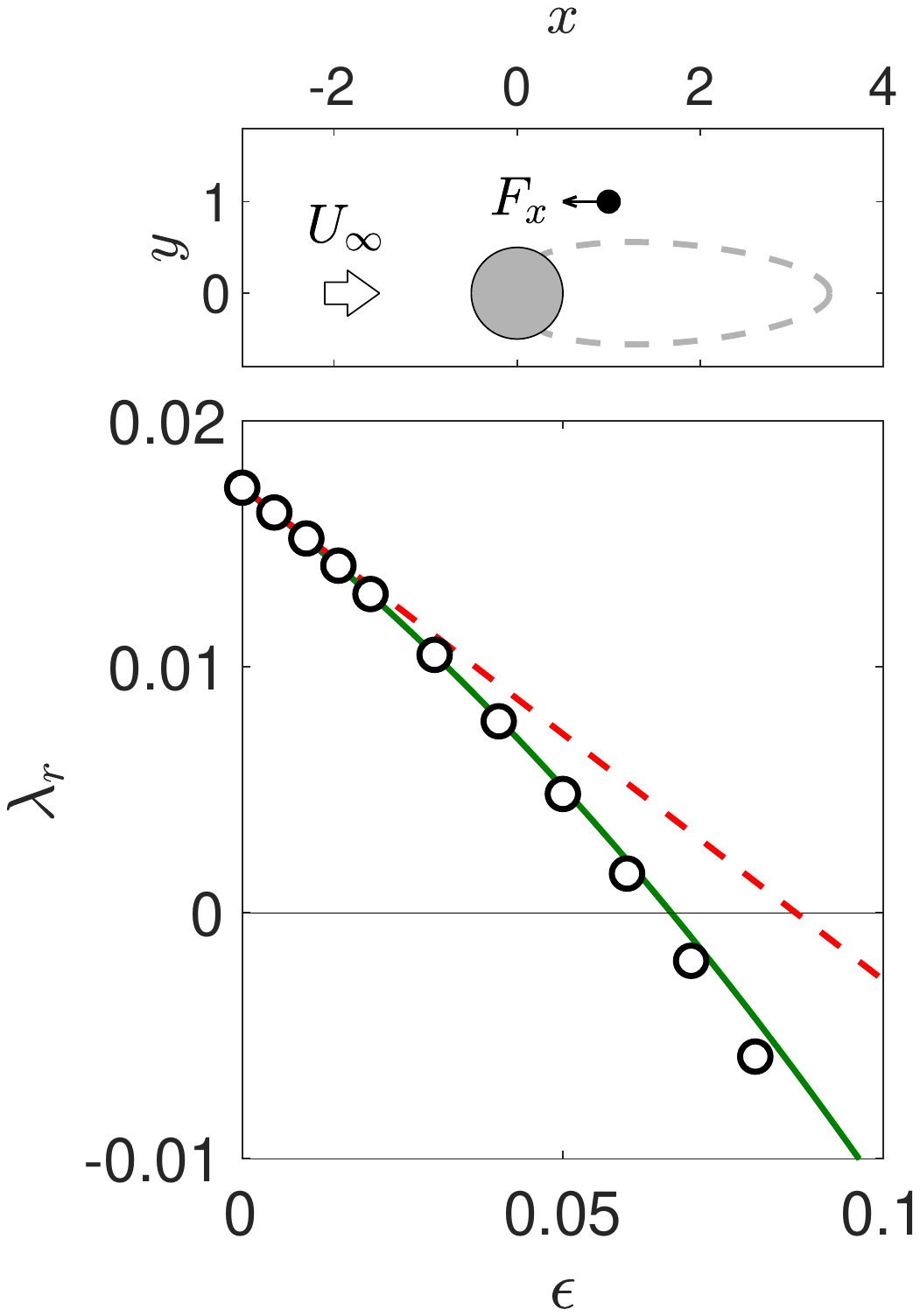}
      \put(-2,85){$(b)$}
   \end{overpic}  
   \hspace{0.1cm}   
   \begin{overpic}[height=5.5cm, trim=15mm 70mm 100mm 55mm, clip=true]{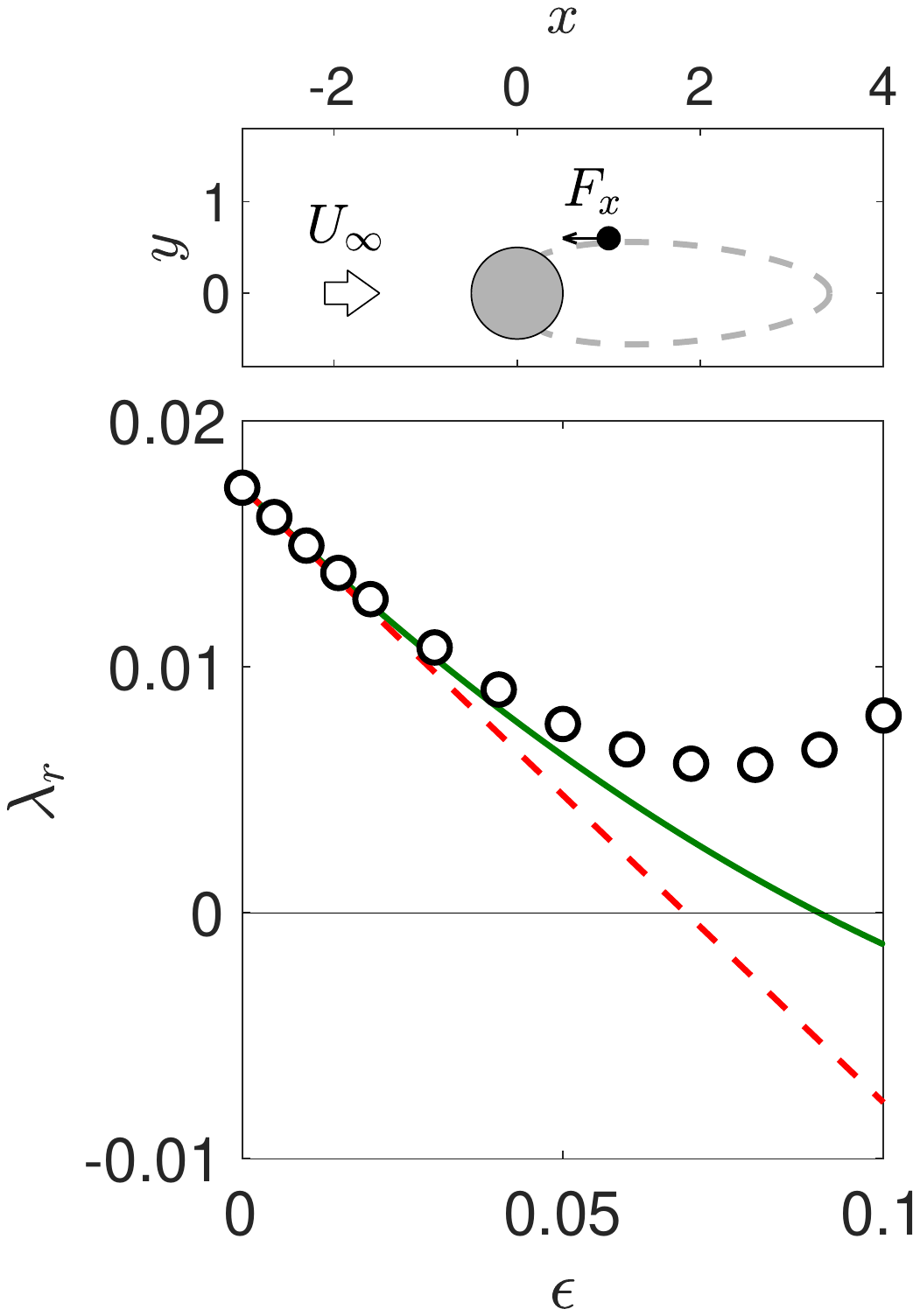}
      \put(-2,85){$(c)$}
   \end{overpic}  
   \hspace{0.1cm} 
   \begin{overpic}[height=5.5cm, trim=15mm 70mm 100mm 55mm, clip=true]{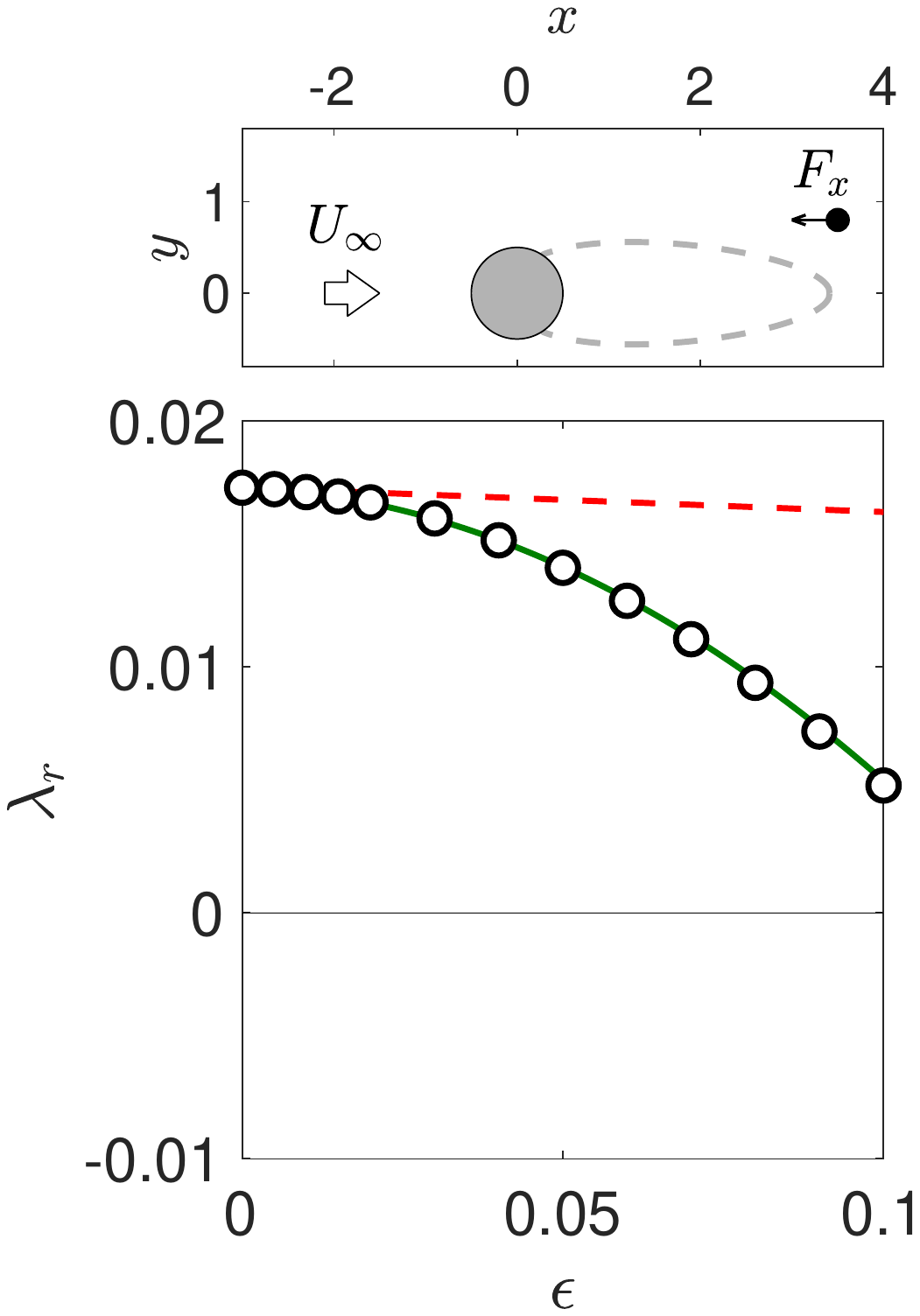}
      \put(-2,85){$(d)$}
   \end{overpic}  
}
%
%
\centerline{  
   \hspace{0.1cm}
   \begin{overpic}[width=3.11cm, trim=13mm 70mm 100mm 110mm, clip=true]{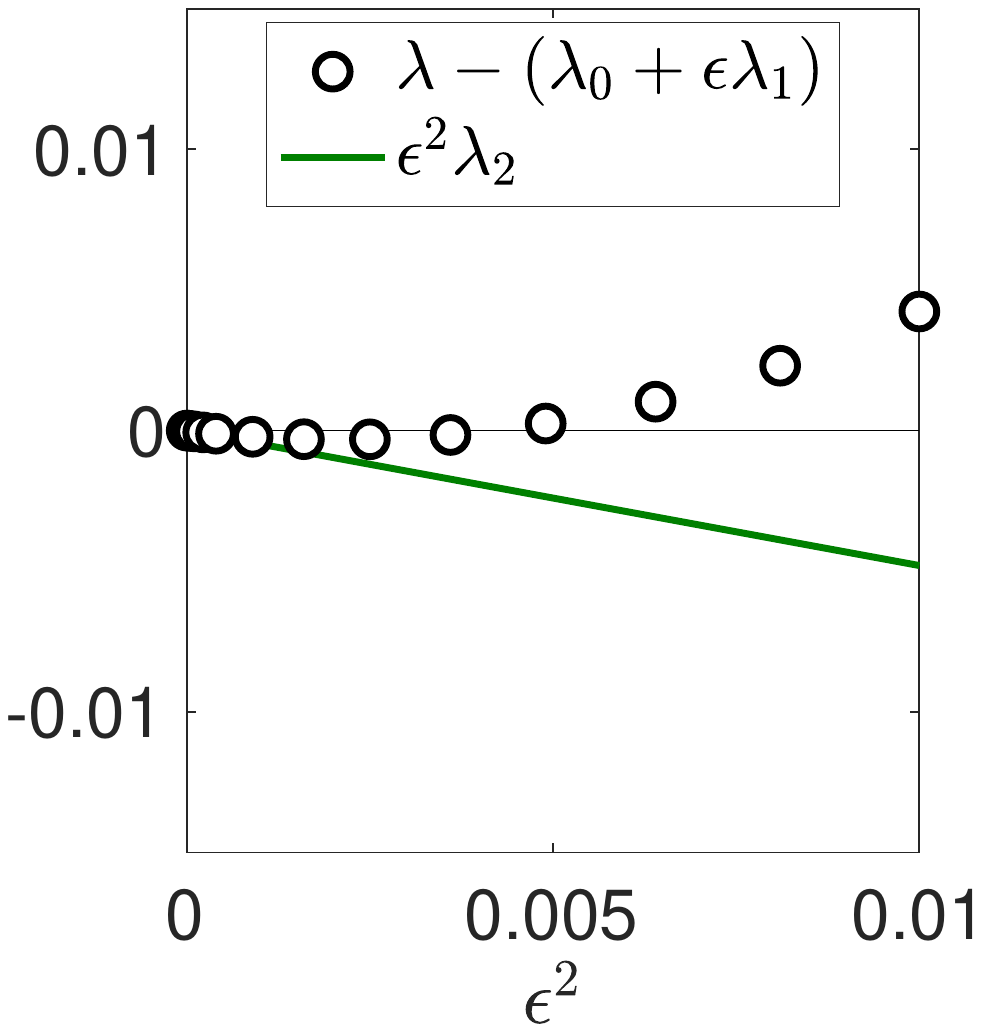}
      \put(0,90){$(e)$}
   \end{overpic}   
   \hspace{0.045cm}
   \begin{overpic}[width=3.11cm, trim=13mm 70mm 100mm 110mm, clip=true]{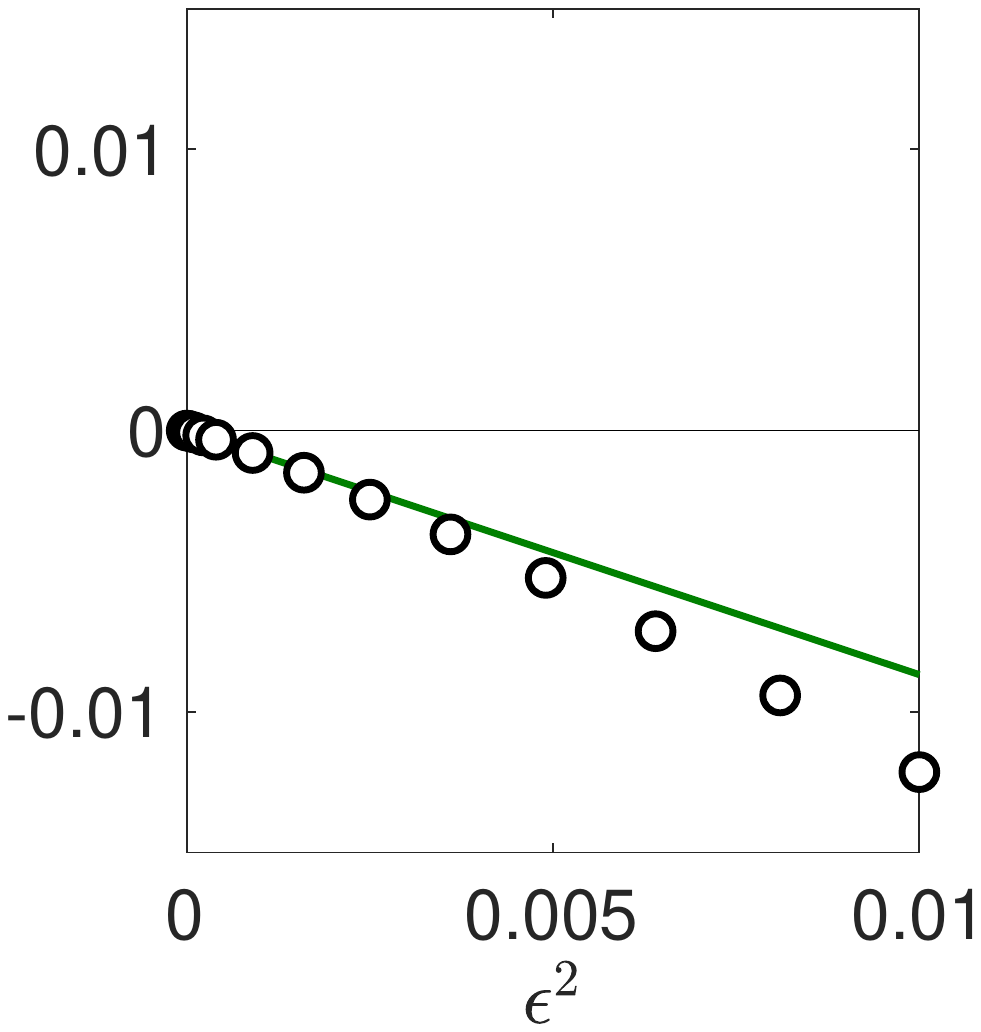}
      \put(0,90){$(f)$}
   \end{overpic}  
   \hspace{0.045cm}    
   \begin{overpic}[width=3.11cm, trim=13mm 70mm 100mm 110mm, clip=true]{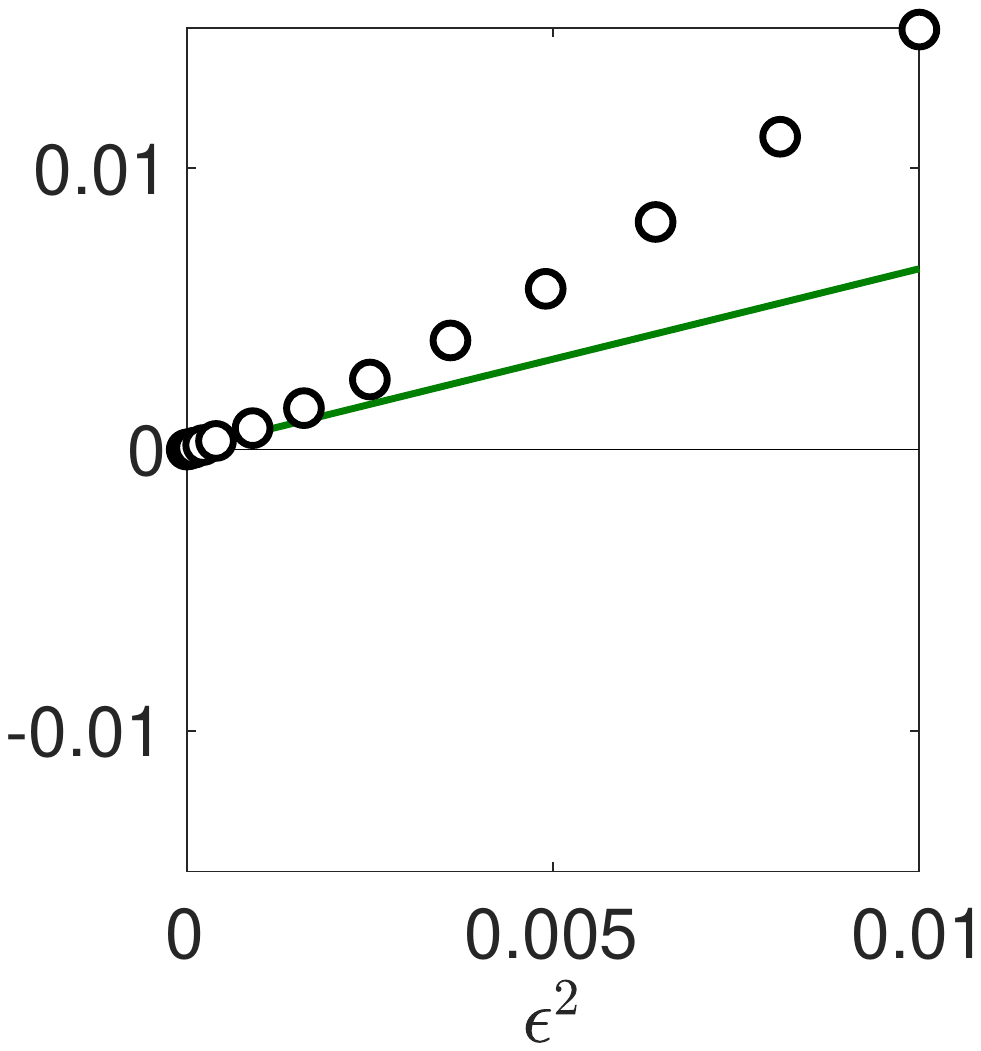}
      \put(0,90){$(g)$}
   \end{overpic}   
   \hspace{0.045cm}
   \begin{overpic}[width=3.11cm, trim=13mm 70mm 100mm 110mm, clip=true]{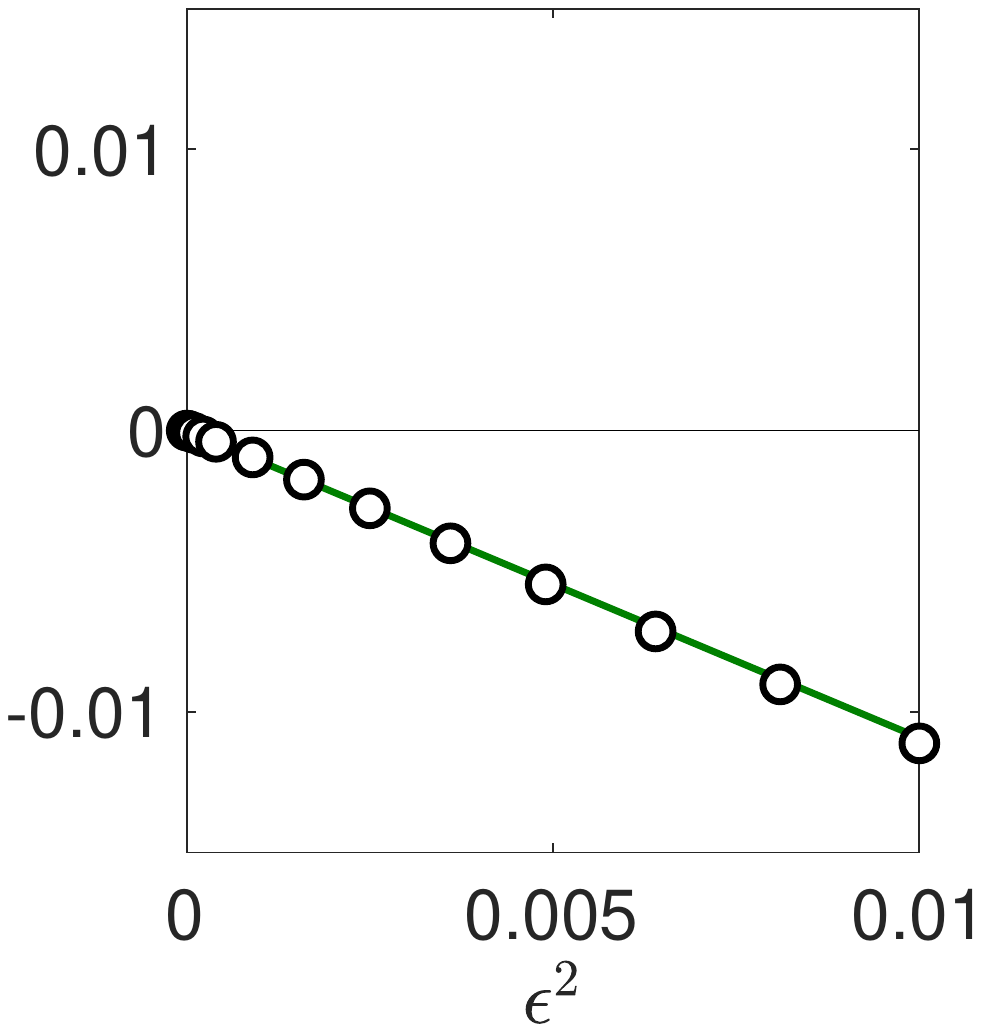}
      \put(0,90){$(h)$}
   \end{overpic}  
}
\caption{
Second-order sensitivity improves the prediction of growth rate variation.
$(a)$-$(d)$~Same data as figure~\ref{fig:valid_1}, together with second-order prediction (solid line).
$(e)$-$(h)$~Higher-order variation: nonlinear data, i.e. all terms of order $n \geq 2$ (symbols) and second-order sensitivity (solid line).
}
\label{fig:valid_2}
\end{figure}

Second-order variations are obtained just as easily, and results for the few control locations considered earlier are shown as solid lines in figure~\ref{fig:valid_2}. The predicted growth rate variation is generally improved. 
In figure~\ref{fig:valid_2}$(b,d)$  for instance, the second-order prediction  follows closely the actual growth rate variation up to much larger amplitudes than the first-order prediction. 
In other locations, however, like in figure~\ref{fig:valid_2}$(a,c)$, the improvement is less significant, owing to higher-order variations. 
Figure~\ref{fig:valid_2}$(e)$-$(h)$ highlights these higher-order variations, confirming their importance (figure~\ref{fig:valid_2}$e,g$) or lack thereof (figure~\ref{fig:valid_2}$f,h$).

As will be discussed in \S~\ref{sec:results}, sensitivity maps 
can be produced that allow one to identify at a glance regions where a steady force alters the eigenvalue most effectively. 
Further, the relative signs and magnitudes of the first- and second-order eigenvalue variations will  characterise the usefulness of the first-order sensitivity.

Before moving on to the next sections,
it is worth mentioning that the method can be applied to the sensitivity of other quantities, as soon as they are defined as eigenvalue problems. 
To illustrate this point, Appendix~\ref{app:gain} derives the first- and second-order sensitivities of the linear amplification of harmonic forcing (resolvent gain).

%

\subsection{Higher-order sensitivity}
\label{sec:higher_order}

Higher-order terms $\UU_n$ for the base flow modification are governed by 
\begin{eqnarray} 
\AAA_0 \UU_3 &=& - \UU_1 \bcdot \bnabla \UU_2 - \UU_2 \bcdot \bnabla \UU_1,
\label{eq:Q3}
\\
& \ldots
\\
\AAA_0 \UU_n &=& \sum_{1 \leq m \leq n-1} - \UU_m \bcdot \bnabla \UU_{n-m}.
\label{eq:Qn}
\end{eqnarray}
Similarly, higher-order terms $\uu_n$ for the eigenmode modification are governed by
\begin{eqnarray} 
 (\lambda_0 \II+\AAA_0) \uu_3 &=&
-(\lambda_1 \II+\AAA_1) \uu_2 
-(\lambda_2 \II+\AAA_2) \uu_1 
-(\lambda_3 \II+\AAA_3) \uu_0,
\label{eq:evp3}
\\
& \ldots
\\
(\lambda_0\II+\AAA_0) \uu_n &=&
\sum_{1\leq m \leq n} -(\lambda_m \II+\AAA_m) \uu_{n-m},
\end{eqnarray}
which, upon  projection onto $\uu_0^\dag$, yields the eigenvalue variations \citep{Hinch1991, Mensah2020JSV}:
\begin{eqnarray} 
\lambda_3 &=&
-\ps{ \uu_0^\dag }{ \AAA_3 \uu_0 }
-\ps{ \uu_0^\dag }{ (\lambda_1 \II+\AAA_1) \uu_{2} + (\lambda_2 \II+\AAA_2) \uu_{1} },
\\
& \ldots
\\
\lambda_n &=&
-\ps{ \uu_0^\dag }{ \AAA_n \uu_0 }
-\sum_{1\leq m \leq n-1} \ps{ \uu_0^\dag }{ (\lambda_m \II+\AAA_m) \uu_{n-m} }.
\end{eqnarray}
Just like $\lambda_1$ and $\lambda_2$ are linear and quadratic in $\UU_1$, respectively,
each of the above expressions is exactly proportional to $\UU_1^n$, and thus to $\FF^n$.
In principle, one can therefore generalise expressions
(\ref{eq:sensit1})-(\ref{eq:sensit2}), which involve the vector $\SS_1$ (tensor of order one) and the matrix $\SS_2$ (tensor of order two), and introduce tensors $\SS_n$ of order $n$ such that
\begin{eqnarray}
\lambda &=& \lambda_0 + \epsilon \lambda_1 +  \epsilon^2 \lambda_2 + \epsilon^3 \lambda_3 + \ldots + \epsilon^n \lambda_n + \ldots 
\nonumber \\
&=& \lambda_0 
+ \epsilon   \iint \left( \SS_1 \right)_i     \FF_i    \,\mathrm{d}\xx 
+ \epsilon^2 \iint \left( \SS_2 \right)_{ij}  \FF_i \FF_j  \,\mathrm{d}\xx
+ \epsilon^3 \iint \left( \SS_3 \right)_{ijk} \FF_j \FF_j \FF_k \,\mathrm{d}\xx
+ \ldots 
\nonumber \\
&& + \epsilon^n \iint \left( \SS_n \right)_{i_1 i_2 \ldots i_n} \FF_{i_1} \FF_{i_2} \ldots \FF_{i_n} \,\mathrm{d}\xx + \ldots,
\end{eqnarray}
with Einstein notation for repeated indices.
Conceptually, the method for obtaining the higher-order sensitivity operators $\SS_n$ is similar to that described in Appendix~\ref{app:operators}, and involves a combination of the following steps:
(i)~redefine linear forms like  $\AAA_n \uu_0$, $(\lambda_m \II+\AAA_m) \uu_{n-m}$, etc., so as to make explicit the dependence on the first-order flow modification $\UU_1$, and eventually on the force $\FF = \AAA_0 \UU_1$;
(ii)~introduce adjoint operators so as to isolate $\FF$, and identify the remaining control-independent operator as the sensitivity  $\SS_n$.

It should be noted that adding more terms to the power series (\ref{eq:exp_ev}) does not necessarily improve its accuracy, and  it certainly does not for amplitudes larger than the radius of convergence $r$ of the expansion. In general, $r$ depends on both the type and location of the control. 
In order to rigorously assess the validity of a second-order or higher-order sensitivity prediction, one must therefore compute the eigenvalue $\lambda$ of the actual nonlinear controlled flow, similar to validation calculations for first-order sensitivity prediction.

\subsection{Computational cost}
\label{sec:cost}

The cost of computing the effect of a steady force on the eigenvalue is estimated in table~\ref{tab:cc}.
Different methods are compared: 
computing the fully nonlinear controlled flow and associated eigenvalue $\lambda$;
and computing the first- and second-order eigenvalue variations $\lambda_1$ and $\lambda_2$ using sensitivity operators.
In what follows, $N$ is the total number of degrees of freedom after numerical discretisation, and $M$ is the number of independent forcing locations.
The uncontrolled base flow $\UU_0$ and leading eigenmode $\uu_0$ are computed prior to considering any control.

For the sake of simplicity, it is reasonable to assume that $1 \ll M \ll N$ when estimating the leading-order computational cost. 
That is, $M$ must be rather large so as to obtain sufficiently fine-grained sensitivity maps, and $N$ must be large enough to compute the eigenvalue and its variation accurately.
To fix ideas, 10 different values for both $x_c$ and $y_c$ already yield $M=100$ control locations to be evaluated.
Further, with a finite-element method, a minimum of $N=10^3$ to $10^4$ degrees of freedom seem necessary.
In this study, $M \simeq 10^4$ and $N \simeq 6 \times 10^5$.

The computational cost of the different methods is as follows.
\begin{itemize}
\item
Recomputing the fully nonlinear controlled flow and the corresponding eigenvalue $\lambda$ for each forcing location (second column of table~\ref{tab:cc}) involves two steps:
(i)~computing $M$ nonlinear base flows $\UU$, for instance with a Newton method requiring $k$ linear system resolutions (typically five to ten iterations) of complexity $O(N^3)$;
(ii)~solving $M$ eigenvalue problems for $\uu$, for instance with an implicitly restarted Arnoldi method, of complexity proportional to $O(N^2)$.
Omitting constant factors for simplicity, the total cost  scales like $M \times O(N^3)$.

\item
Estimating the first- and second-order eigenvalue variations with (\ref{eq:sensit1})-(\ref{eq:sensit2}), i.e. with  sensitivity operators (third and fourth columns of  table~\ref{tab:cc}), 
involves the following steps (see details in Appendix~\ref{app:operators}):
(i)~computing once and for all the (uncontrolled) adjoint mode $\uu_0^\dag$, with a cost proportional to $O(N^2)$;
(ii)~computing once and for all the lower--upper (LU) decompositions of complexity $O(N^3)$ of $\AAA_0^\dag$  and $(\lambda_0\II + \AAA_0)$ for  $\lambda_1$ and $\lambda_2$, respectively;
(iii)~evaluating a few matrix--vector products, with a cost $O(N^2)$ for each forcing location.
The total cost therefore scales like $O(N^3)$, for both
$\lambda_1$ and $\lambda_2$.
\end{itemize}

\begin{table}
  \begin{center}
\def~{\hphantom{0}}
  \begin{tabular}{l c c c}
      & Fully nonlinear     & First-order & Second-order \\
      & controlled eigenvalue     & eigenvalue variation  & eigenvalue variation \\
& $\lambda(\FF)$ & $\lambda_1 = \ps{\SS_1}{\FF}$ & $\lambda_2 = \ps{\FF}{\SS_2 \FF}$ \\[6pt]
      Nonlinear base flow   & $ M \times O(k N^3)$ & -          & - \\
      Eigenvalue problem    & $M \times O(p N^2)$ & $O(p N^2)$ & - \\
      LU decomposition      & -            & $O(N^3)$   & $O(N^3)$ \\
      Matrix-vector product & -            & $M \times O(N^2)$ & $M \times O(N^2)$ \\[6pt]
      \textbf{Dominant contribution} & $M \times O(N^3)$   & $O(N^3)$  & $O(N^3)$ 
  \end{tabular}
  \caption{
  Computational cost for the eigenvalue variation induced by a steady force, in a system discretised with $N$ degrees of freedom and forced at $M$ locations. 
  The dominant contribution is derived assuming $1 \ll M \ll N$.
Recomputing the controlled base flow and the  corresponding eigenvalue for each forcing location is substantially more expensive than evaluating the sensitivities.
}
  \label{tab:cc}
  \end{center}
\end{table}

In conclusion, computing $\lambda_2$ involves an additional cost similar to that of computing $\lambda_1$.
It is  much smaller than that of recomputing the nonlinear eigenvalue $\lambda$ for each forcing location. 
The advantage of adjoint methods therefore applies to both first and second orders. 
Of course, this is true only when a large number $M$ of control locations are considered, e.g. when constructing sensitivity maps. Conversely, when only a few control locations are of interest, calculating the actual eigenvalue $\lambda$ is more accurate and not significantly more computationally expensive.

In the above analysis, memory requirements  have not been considered. 
Storage is not an issue for two-dimensional configurations,  and for spatial discretisation methods that yield sparse matrices (e.g. finite-element method), but may become prohibitive for three-dimensional configurations or methods that yield dense matrices (e.g. spectral methods). This is a practical limitation of the proposed approach. 
For standard eigenvalue calculations and first-order sensitivity analysis, one can use  matrix-free time-stepping techniques as an  alternative to matrix-based techniques (\cite{Tuckerman2000timestepping}). Whether such an approach is possible for second-order sensitivity analysis remains to be determined.

%
\section{Flow configuration and numerical method}
\label{sec:config_num}

The two-dimensional, incompressible flow past a circular cylinder of diameter $D$ with free-stream velocity $(U_\infty,0)^T$ is considered. 
In the remainder of this study, the Reynolds number is set to $\Rey=50$ unless otherwise stated.

%
\subsection{Base flow}

A two-dimensional triangulation of the domain
\begin{eqnarray}
\Omega=\{ (x,y) | -10 \leq x \leq 50, |y| \leq 10, \sqrt{x^2+y^2}\geq 0.5  \}
\end{eqnarray}
 is generated with the finite-element software FreeFem++ \citep{Hecht2012}, resulting in approximately 136'000 elements. Velocity and pressure fields are discretised with P2 and P1 Taylor--Hood elements, respectively, yielding a total of $N \simeq 615'000$ degrees of freedom. 
All discrete operators  are built from their continuous  expressions (see details in Appendix~\ref{app:operators}) in variational form. In particular,  this means that the ``differentiate then discretise” approach is used for adjoint operators, as opposed to the ``discretise  then differentiate” approach.

\begin{figure} 
\centerline{  
   \begin{overpic}[width=6.5cm, trim=8mm 15mm 10mm 150mm, clip=true]{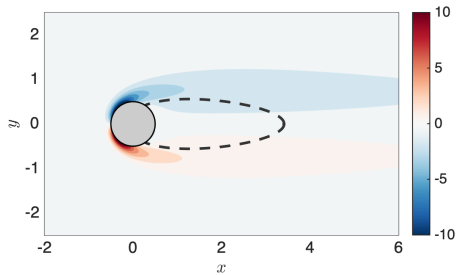}
   \end{overpic}  
}
\caption{
Vorticity of the base flow at $\Rey=50$. Dashed line: recirculation region.
}
\label{fig:base_flow}
\end{figure}

The uncontrolled steady base flow $\UU=\UU_0$  is obtained by solving (\ref{eq:NS_compact})
with a Newton method,  iterated until residuals are smaller than $10^{-12}$. 
Boundary conditions are imposed as follows:
uniform free-stream velocity at the inlet, 
no-slip boundary condition on the cylinder wall, 
 outflow boundary condition  $-P\nn + \Rey^{-1}\bnabla\UU\bcdot\nn =\00$ (with $\nn$ the normal vector) at the outlet, 
and symmetry condition on the lateral sides of the domain. 
Figure~\ref{fig:base_flow} shows the vorticity $\omega=\omega_0=\partial_x V_0 - \partial_y U_0$ of the base flow at $\Rey=50$. Shear layers of opposite vorticity are created on both sides of the cylinder. The recirculation region (dashed line) extends over three diameters downstream.

Controlled base flows $\UU$ are computed for validation purposes, solving (\ref{eq:NS_compact_F}) with the same method. For steady forces $\FF$ that are localised in space, Dirac delta functions are smoothed out numerically into Gaussians of variance $0.0025$.

%
\subsection{Stability analysis}

\begin{figure} 
\centerline{
\hspace{0.8cm}
   \begin{overpic}[height=5.5cm, trim=10mm 70mm 95mm 67mm, clip=true]{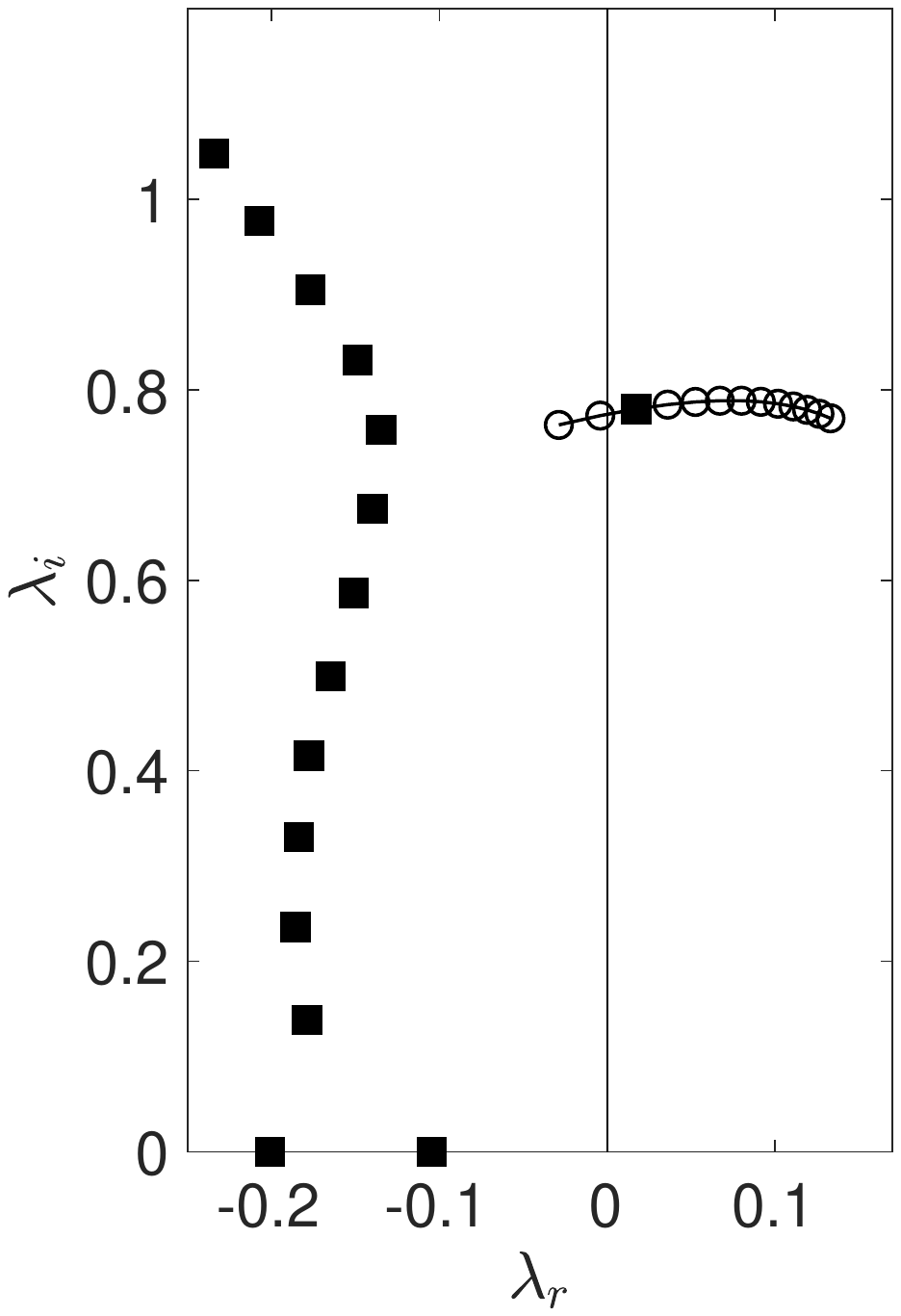}   
      \put(-2,87){$(a)$}  
      \put(21,73){\tiny $\Rey=50$}
      \put(35,57){\tiny $40$}
      \put(42,57.7){\tiny $50$}
      \put(53,57){\tiny $100$}
   \end{overpic}  
\hspace{0.3cm}
   \begin{overpic}[width=9cm, trim=10mm 22mm 10mm 150mm, clip=true]{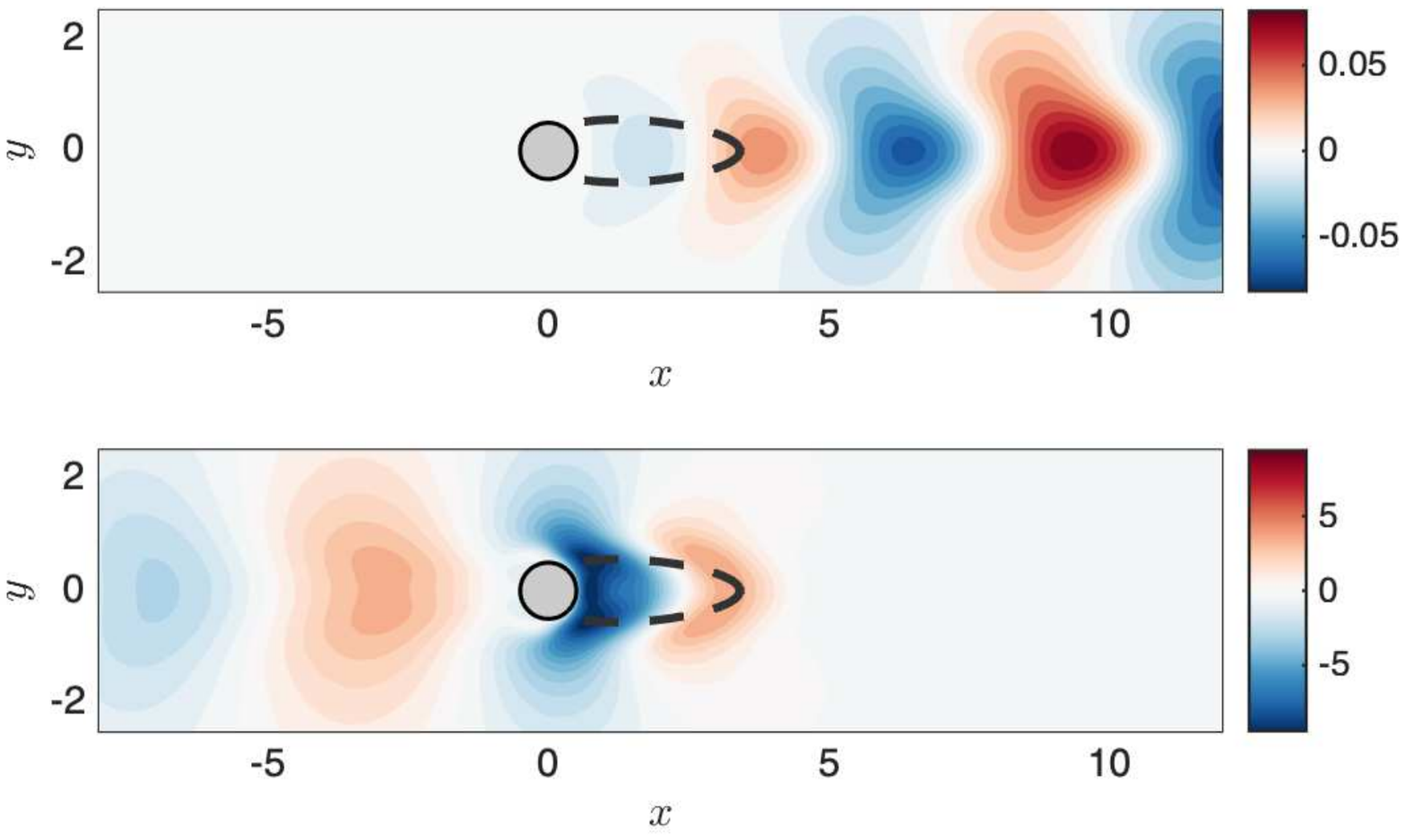}     
      \put(0,53){$(b)$}  
      \put(0,25){$(c)$}
   \end{overpic}  
}
\caption{
$(a)$~Eigenvalues of the cylinder flow at $\Rey=50$ (filled squares), and leading eigenvalue at $\Rey=40$,  $45$, $\ldots$, 100 (empty circles). 
The full spectrum is symmetric with respect to $\lambda_i=0$.
$(b)$~Leading eigenmode and $(c)$~leading adjoint mode (real part, cross-stream velocity) at $\Rey=50$, normalised such that $\ps{\uu_0^\dag}{\uu_0}=1$ and $||\uu_0||=1$. 
}
\label{fig:modes}
\end{figure}

The eigenvalue problem (\ref{eq:evp_compact}) is solved with Matlab using an implicitly restarted Arnoldi method with shift-and-invert preconditioning.
This study focuses on the leading eigenmode $\uu=\uu_0$, which becomes unstable at $\Rey \simeq 47$ via a Hopf bifurcation, as a pair of complex conjugate eigenvalues cross the imaginary axis, as illustrated in the half-plane $\lambda_i>0$ in figure~\ref{fig:modes}$(a)$ (the other half is symmetric with respect to $\lambda_i=0$).
The leading eigenmode at $\Rey=50$,
 associated with the eigenvalue $\lambda \simeq 0.0173+0.7797i$ is shown in figure~\ref{fig:modes}$(b)$.
It is largest a few diameters downstream of the recirculation region, as perturbations are advected by the base flow.
With its wave packet structure and its complex eigenvalue, this mode breaks both the spatial and temporal symmetries, leading to periodic vortex shedding and to the B\'enard--von K\'arm\'an street in the cylinder wake.

The adjoint problem (\ref{eq:adjoint}) is solved with the same method. 
The leading adjoint mode $\uu_0^\dag$ shown in figure~\ref{fig:modes}$(c)$ is largest in the recirculation region, and adjoint perturbations travel upstream, a consequence of upstream advection in the adjoint NS operator.

\subsection{Sensitivity}

First- and second-order sensitivity maps are computed for localised control forces $\FF$.
The control is moved within the subdomain 
$x \in [-2,6]$, $y \in [0,3]$,
with a step size $\Delta x = \Delta y=0.05$, leading to approximately 
$M \simeq 10'000$ control locations.

The second-order sensitivity operator $\SS_2$ defined by (\ref{eq:sensit2}) contains inverse operators (see detailed expression in Appendix~\ref{app:operators}) and is therefore not formed explicitly.
Instead, the LU decomposition of each operator to be inverted is precomputed once and for all, such that each subsequent matrix inversion  is replaced with two simple matrix--vector products.
(Note that $\SS_2$ is a second-order tensor; by contrast, the first-order sensitivity $\SS_1$  defined by (\ref{eq:sensit1}) is a vector  that can be formed explicitly and plotted without further difficulty. 
In this study,  sensitivity maps for  $\lambda_1$ and $\lambda_2$ are evaluated location by location.)

%

\section{Second-order sensitivity of the growth rate}
\label{sec:results}

This section investigates the effect of control
on the  first- and second-order  variations of the leading growth rate $\lambda_r$. 
(For the effect on the linear frequency $\lambda_i$, see Appendix~\ref{app:freq}.)

\subsection{Sensitivity to a steady body force}
\label{sec:results_F}

\begin{figure} 
\centerline{
   \hspace{0.3cm}
   \begin{overpic}[width=6.2cm, trim=10mm 40mm 10mm 170mm, clip=true]{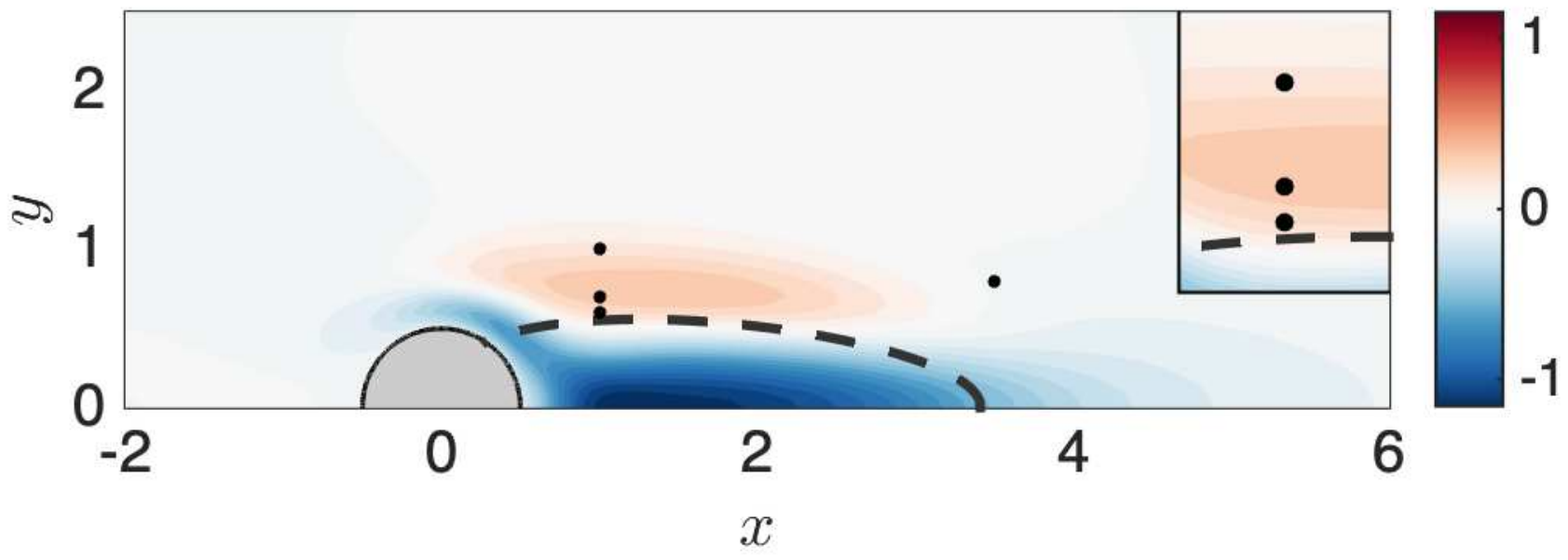}
      \put(-2,30){$(a)$}
   \end{overpic}  
   \begin{overpic}[width=6.2cm, trim=10mm 40mm 10mm 170mm, clip=true]{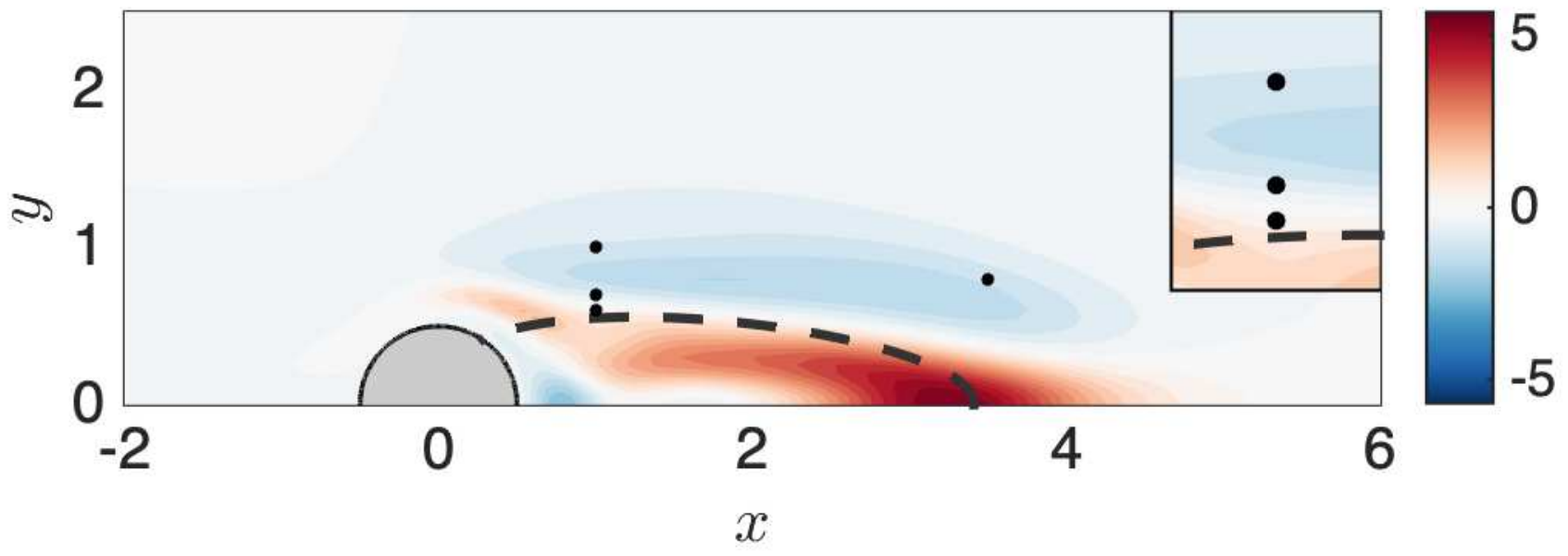}
      \put(-2,30){$(b)$}
   \end{overpic}  
}
\centerline{
   \hspace{0.3cm}
   \begin{overpic}[width=6.2cm, trim=10mm 40mm 10mm 170mm, clip=true]{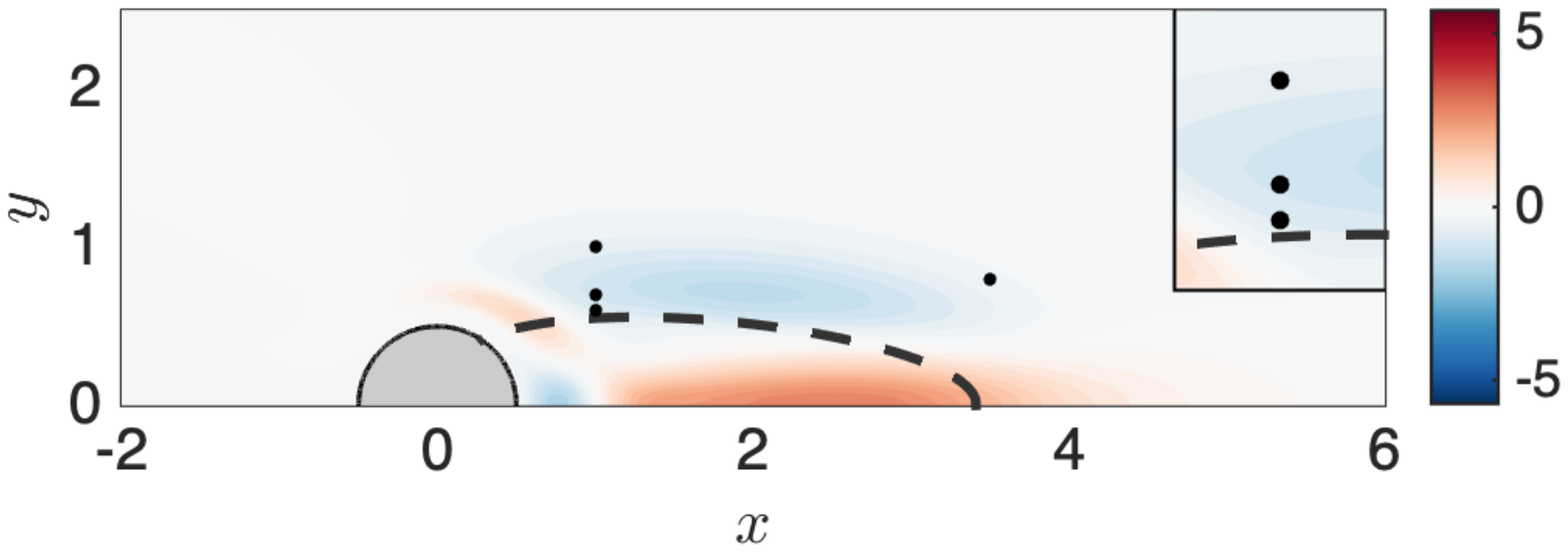}
      \put(-2,30){$(c)$}
   \end{overpic}  
   \begin{overpic}[width=6.2cm, trim=10mm 40mm 10mm 170mm, clip=true]{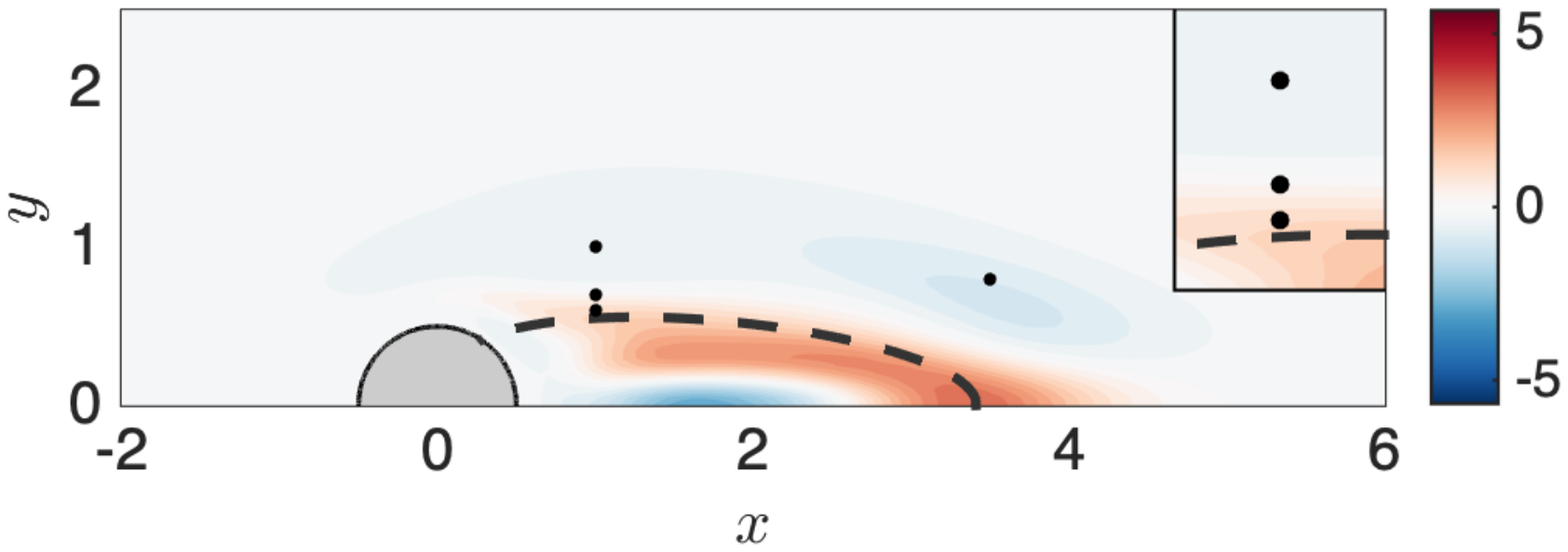}
      \put(-2,30){$(d)$}
   \end{overpic}  
}
\centerline{
   \hspace{0.3cm}
   \begin{overpic}[width=6.2cm, trim=10mm 40mm 10mm 170mm, clip=true]{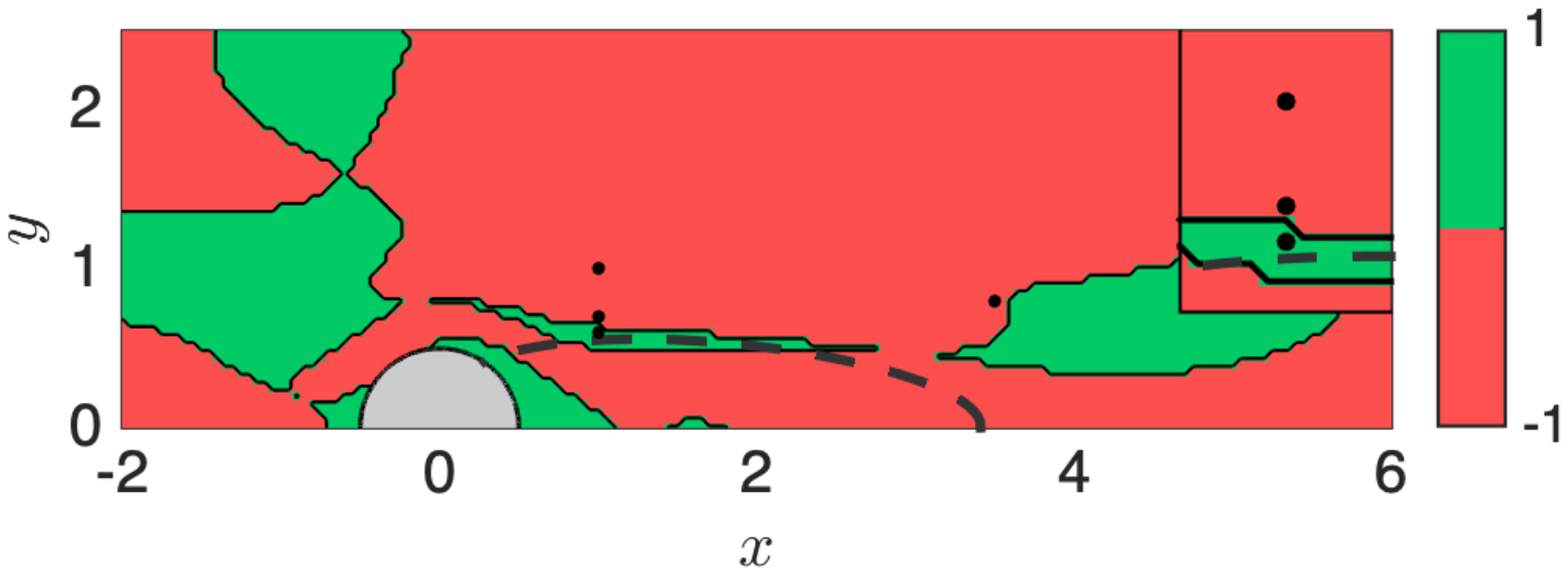}
      \put(-2,30){$(e)$}
   \end{overpic}  
   \begin{overpic}[width=6.2cm, trim=10mm 40mm 10mm 170mm, clip=true]{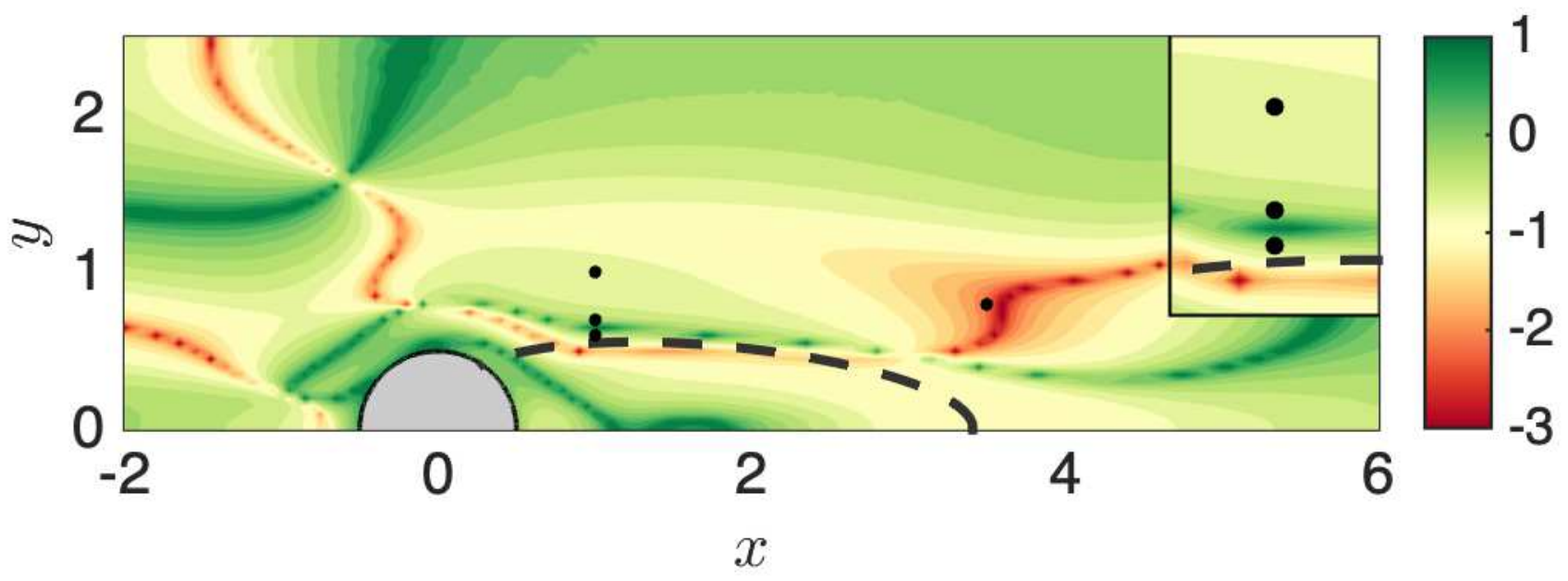}
      \put(-2,30){$(f)$}
   \end{overpic}  
}
\caption{
Sensitivity of the leading mode's growth rate to a localised steady force oriented along the $x$ direction, at $\Rey=50$.
All fields are symmetric with respect to $y=0$.
Black dots show the control locations considered in figures~\ref{fig:valid_1}-\ref{fig:valid_2}.
$(a)$~First-order variation $\lambda_{1r}$.
$(b)$~Second-order variation $\lambda_{2r}$.
$(c)$~Term I and $(d)$~term II in the decomposition (\ref{eq:ev2}) of the second-order variation.
$(e)$~Sign of the product $\lambda_{1r} \lambda_{2r}$.
$(f)$~Relative importance of first- and second-order variations, quantified by the threshold amplitude (\ref{eq:eps_thresh}), shown here as $\log_{10}(\epsilon_t)$.
Insets: close-up views of the region $0.7 \leq x \leq 1.3$, $0.4 \leq y \leq 1.2$.
}
\label{fig:F}
\end{figure}

Let us consider first a  generic steady body force.
Figure~\ref{fig:F}$(a)$ shows the real part of the $x$ component of the first-order sensitivity $\SS_1(\xx)$ to such a steady force. 
As shown by (\ref{eq:sensit1}), the value at each location $\xx=\xx_c$ is also the value of the first-order variation $\lambda_{1r}$ when choosing a localised force along the $x$ direction, $\FF=(\delta(\xx-\xx_c),0)^T$.
The sensitivity is large and negative on the sides of the cylinder and inside the recirculation region, and positive on the sides of recirculation region, in agreement with \cite{Marquet08cyl} (figure~9$(a)$ therein). 
Note that changing the sign of $F_x$ changes the sign of $\lambda_1$, such that stabilising regions ($\lambda_{1r}<0$, blue) become destabilising ($\lambda_{1r}>0$, red) and \textit{vice versa}.

The second-order sensitivity $\SS_2(\xx)$ is visualised in figure~\ref{fig:F}$(b)$, which shows the second-order growth rate variation $\lambda_{2r}$ evaluated according to (\ref{eq:sensit2}) for the same localised force $\FF=(\delta(\xx-\xx_c),0)^T$. 
Overall, and in absolute value, sensitive regions are similar at  first and second orders, namely the domain approximately delimited by $0\leq x \leq 4, |y|\leq 1$, and containing the sides of the cylinder, the recirculation region and the shear layers.
Note that, unlike $\SS_1$, the sign of $\SS_2$ does not change with the sign of $F_x$. 

With these two maps available, it is now possible to explain the results of figure~\ref{fig:valid_2}. 
The three control locations $\xx_c=(1,0.7)$,  $\xx_c=(1,1)$ and  $\xx_c=(1,0.6)$ lie in a region of similar first-order sensitivity (figure~\ref{fig:F}$a$), and therefore induce similar first-order reductions $\lambda_{1r}$ (figure~\ref{fig:valid_2}$a$-$c$).
The second-order variations, however, differ substantially between these three locations (figure~\ref{fig:F}$b$): small in $\xx_c=(1,0.7)$, negative in $\xx_c=(1,1)$ and positive in $\xx_c=(1,0.6)$. As a result, the second-order prediction is not much different from the first-order one in figure~\ref{fig:valid_2}$(a)$, and yields a larger growth rate reduction in figure~\ref{fig:valid_2}$(b)$ and a smaller one in figure~\ref{fig:valid_2}$(c)$. The second-order prediction generally follows more closely the nonlinear results than the first-order one.
In the last control location, $\xx_c=(3.5,0.8)$, the first-order sensitivity is small (figure~\ref{fig:F}$a$), yielding a weak first-order variation in figure~\ref{fig:valid_2}$(d)$. 
The second-order variation, however, is clearly negative (figure~\ref{fig:F}$b$), and the actual growth rate reduction  is well captured by the second-order prediction (figure~\ref{fig:valid_2}$d$).

Considering the large differences observed between different control locations, and the potential impact on flow restabilisation, 
it would be useful to find a simple way to address the following questions:
(i)  What is the range of control amplitude where  the first-order sensitivity yields an accurate prediction? 
(ii) Outside this range, does it underestimate or overestimate the actual variation?
One step towards answering the first question is possible with the ratio of first- to second-order variations. Recalling the expansion
\begin{eqnarray}
\lambda = \lambda_0 + \epsilon \lambda_1 + \epsilon^2 \lambda_2 + O(\epsilon^3),
\end{eqnarray}
it appears that the second-order correction $\epsilon^2 |\lambda_2|$ is of the same order of magnitude as the first-order variation $|\lambda - \lambda_0| = \epsilon |\lambda_1|$ for the threshold amplitude
\begin{eqnarray}
\epsilon_t = \left| \dfrac{\lambda_1}{\lambda_2} \right|.
\label{eq:eps_thresh}
\end{eqnarray}
For small enough amplitudes $\epsilon \ll \epsilon_t$,  the first-order variation predicts  the actual variation accurately,
as the second-order correction is negligible.
Conversely, for large  amplitudes $\epsilon \gg \epsilon_t$, the second-order variation dominates the first-order one. 
In between, the second-order variation becomes important and cannot be ignored when the control amplitude reaches some fraction of the threshold amplitude, say $\epsilon_t/10$.

Obviously, the analysis needs to be refined when $\lambda_2=0$. Taking into account $\epsilon^3 \lambda_3$ or the next non-zero higher-order correction $\epsilon^n \lambda_n$, the threshold amplitude becomes
$\epsilon_t=|\lambda_1/\lambda_n|^{1/(n-1)}.$
Note that the threshold amplitude decreases as the relative importance of $\epsilon^2 \lambda_2$ grows; this latter term becomes the leading-order term in the limiting case $\lambda_1=0$ (e.g. for the spanwise-periodic control of spanwise-invariant flows), and the threshold amplitude then becomes 
$\epsilon_t=|\lambda_2/\lambda_n|^{1/(n-2)}.$

Figure~\ref{fig:F}$(f)$ shows the threshold amplitude (\ref{eq:eps_thresh}), i.e. the ratio of the maps in 
panels $(a)$ and $(b)$, in logarithmic scale. 
Focusing on regions where $\lambda_{1r}$ and $\lambda_{2r}$ are not both small, it appears that the first-order prediction is especially  accurate up to large amplitudes ($\log_{10}(\epsilon_t)>0$, green) near the cylinder, 
downstream of the cylinder on the symmetry axis up to $x=2$, and in a thin strip running along and outside the recirculation region. 
Conversely, the second-order prediction must be taken into account 
($\log_{10}(\epsilon_t)<-1$, yellow and red) in other regions both inside and outside the recirculation region, particularly in a thin strip running along and inside it. 
The proximity of those two strips warns about locating a steady force near the separatrix, or in any region where $\epsilon_t$ has a strong gradient: slight, unintentional shifts can dramatically increase the amplitude of the second-order variation and ruin the accuracy of the first-order prediction.

Figure~\ref{fig:F}$(f)$ confirms observations from figure~\ref{fig:valid_2}: $\epsilon_t$ is large and the first-order prediction is accurate over a  wide range of control amplitudes in $\xx_c=(1,1)$ and $\xx_c=(1,0.7)$, 
while $\epsilon_t$ is small and the second-order variation quickly becomes important in $\xx_c=(1,0.6)$ and $\xx_c=(3.5,0.8)$.

The second question above is answered by considering the signs of $\lambda_{1r}$ and $\lambda_{2r}$.
If both signs are identical, the second-order correction strengthens the effect of the first-order variation: 
when $\lambda_{1r}$, $\lambda_{2r}<0$ the flow is stabilised even more than predicted by  $\lambda_{1r}$ alone (and destabilised even more when $\lambda_{1r}$, $\lambda_{2r}>0$), such that a smaller control amplitude is actually sufficient to obtain the desired effect.
Conversely, if the signs are opposite, the effect is weakened: for example, when $\lambda_{1r}<0$ and $\lambda_{2r}>0$, the flow is not stabilised as efficiently as predicted by $\lambda_{1r}$ alone, such that a larger control amplitude is actually required to obtain the desired effect.
As a way to distinguish between those two situations, figure~\ref{fig:F}$(e)$ shows the sign of the product $\lambda_{1r} \lambda_{2r}$.
Focusing again on regions where $\lambda_{1r}$ and $\lambda_{2r}$ are not both small,
this map indicates that `safe' regions where  $\lambda_{1r} \lambda_{2r}>0$ (green) are rather few and apart (mainly near the cylinder and along the separatrix), the rest being `dangerous' regions where  $\lambda_{1r} \lambda_{2r}<0$ (red).

Consider again the four control configurations of figure~\ref{fig:valid_2}, where $F_x<0$   (recall that the sign of $\lambda_{1r}$ changes when the sign of $F_x$ is changed, which swaps the `safe' and `dangerous' regions).
Figure~\ref{fig:F}$(e)$ confirms that the first-order prediction underestimates the growth rate reduction (compared to first- and second-order predictions together) in $\xx_c=(1,1)$, $\xx_c=(1,0.7)$ and $\xx_c=(3.5,0.8)$,
and overestimates it in $\xx_c=(1,0.6)$.

Let us come back to figure~\ref{fig:F}$(c,d)$, which shows the two terms I  and II in  the second-order sensitivity equation (\ref{eq:ev2}), i.e. the effects of $\UU_2$ and of the $\UU_1$--$\uu_1$ interaction, respectively. 
The map in figure~\ref{fig:F}$(b)$ is the sum of those two maps, and all three colour scales are identical. 
Overall, terms I and II are of the same order of magnitude. Both terms display regions of positive and negative sensitivity. 
They collaborate to yield positive sensitivity near the downstream end of the recirculation region, and negative sensitivity on the side of the recirculation region. 
Conversely, they compete on part of the symmetry axis inside the recirculation region, and on part of the separatrix, resulting in a weak total sensitivity.
Although the map of term I bears an overall qualitative similarity to the map of total sensitivity, term II makes a significant contribution everywhere; 
in other words, the steady control force modifies the growth rate at second order  by changing not only the base flow but also the eigenmode that develops on that base flow.

\subsection{Sensitivity to a small control cylinder}
\label{sec:results_pass_device}

The sensitivity analysis is now applied to a practical flow control strategy, namely inserting a small passive device in order to reduce the growth rate of the leading mode.
Following \cite{Hill92AIAA}, and later \cite{Marquet08cyl} and \cite{Meliga10}, the effect of a small circular cylinder of diameter $d$ located in $\xx_c$ is modelled as a steady force acting on the base flow, equal and opposite to the drag force that would be felt by that cylinder in a uniform flow with the local velocity
\begin{align}
\epsilon \FF(\xx) = - \frac{1}{2} d C_d(\xx) ||\UU_0(\xx)|| \UU_0(\xx) \delta(\xx-\xx_c).
\label{eq:small_cyl_model}
\end{align}
The  drag coefficient $C_d$ of the control cylinder depends on the local Reynolds number $\Rey_d = ||\UU_0(\xx)|| d/\nu $ and is modelled here with the power law $C_d(\Rey_d) = 0.8558 + 10.05 \Rey_d^{-0.7004}$ \citep{Boujo14lengthJFM, Meliga14} meant to approximate data from the literature \citep{Finn1953, Tritton1959} and in-house numerical simulations in the range of interest $1 \lesssim \Rey_d \lesssim 15$.
In the following, results are illustrated with $d=0.1$, i.e. a control cylinder 10 times smaller than the main cylinder.

\begin{figure} 
\centerline{
   \hspace{0.3cm}
   \begin{overpic}[width=6.2cm, trim=10mm 40mm 10mm 170mm, clip=true]{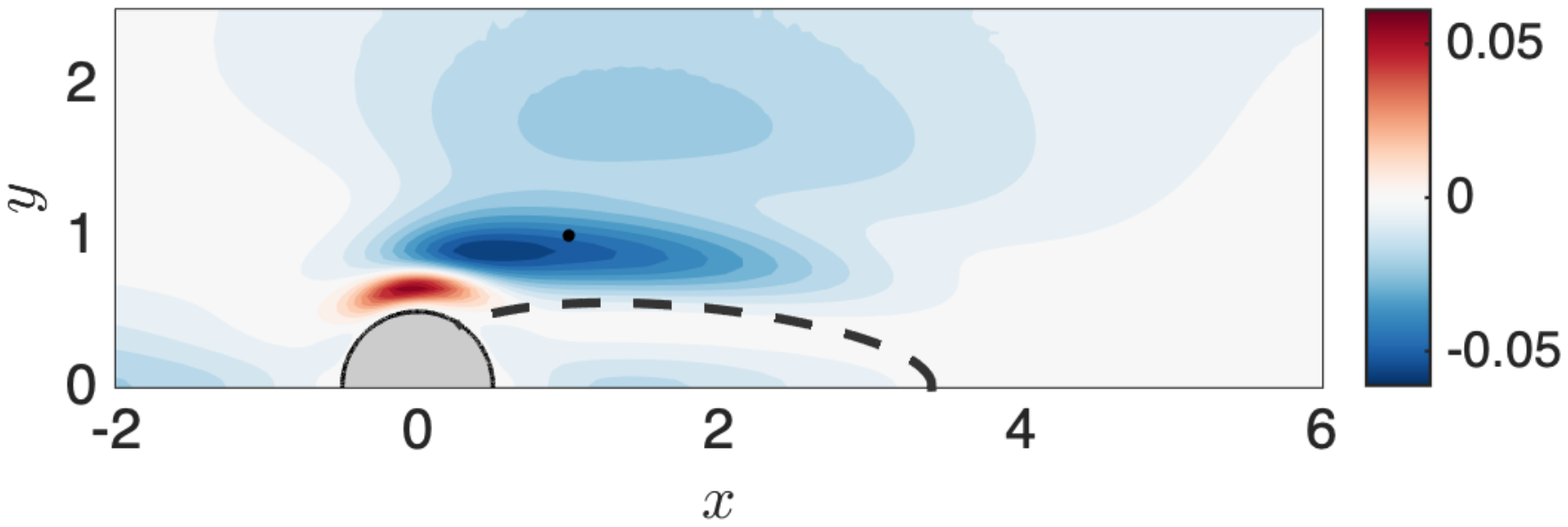}
      \put(-2,30){$(a)$}
   \end{overpic}  
   \begin{overpic}[width=6.2cm, trim=10mm 40mm 10mm 170mm, clip=true]{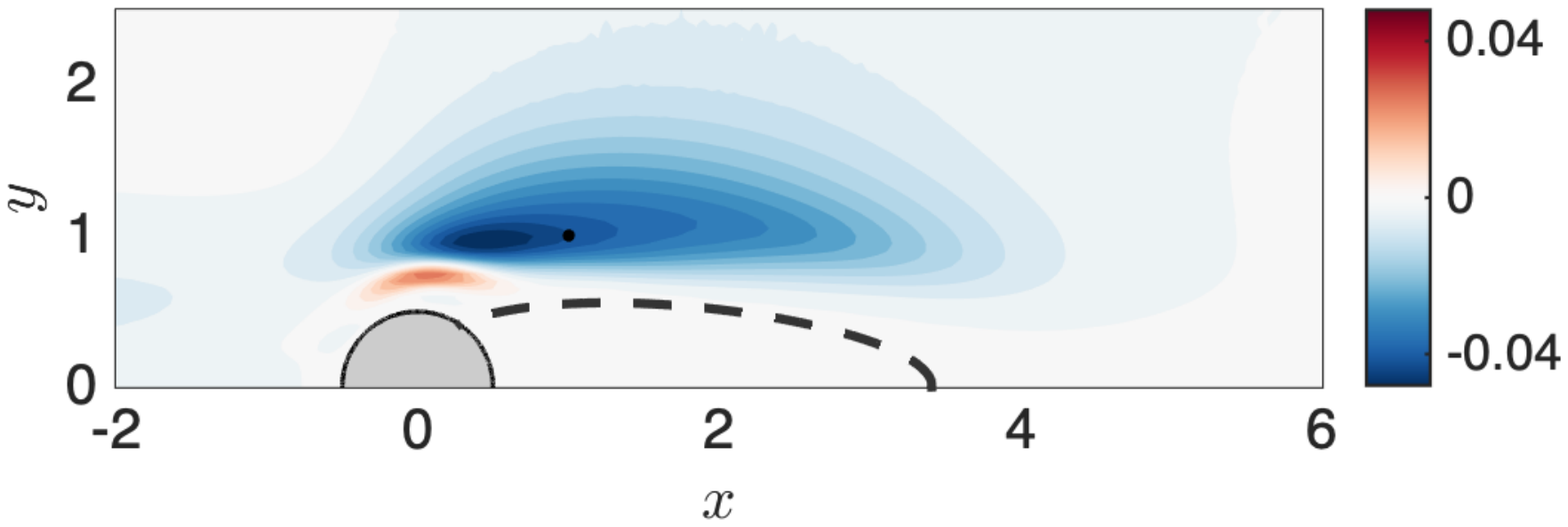}
      \put(-2,30){$(b)$}
   \end{overpic}  
}
\centerline{
   \hspace{0.3cm}
   \begin{overpic}[width=6.2cm, trim=10mm 40mm 10mm 170mm, clip=true]{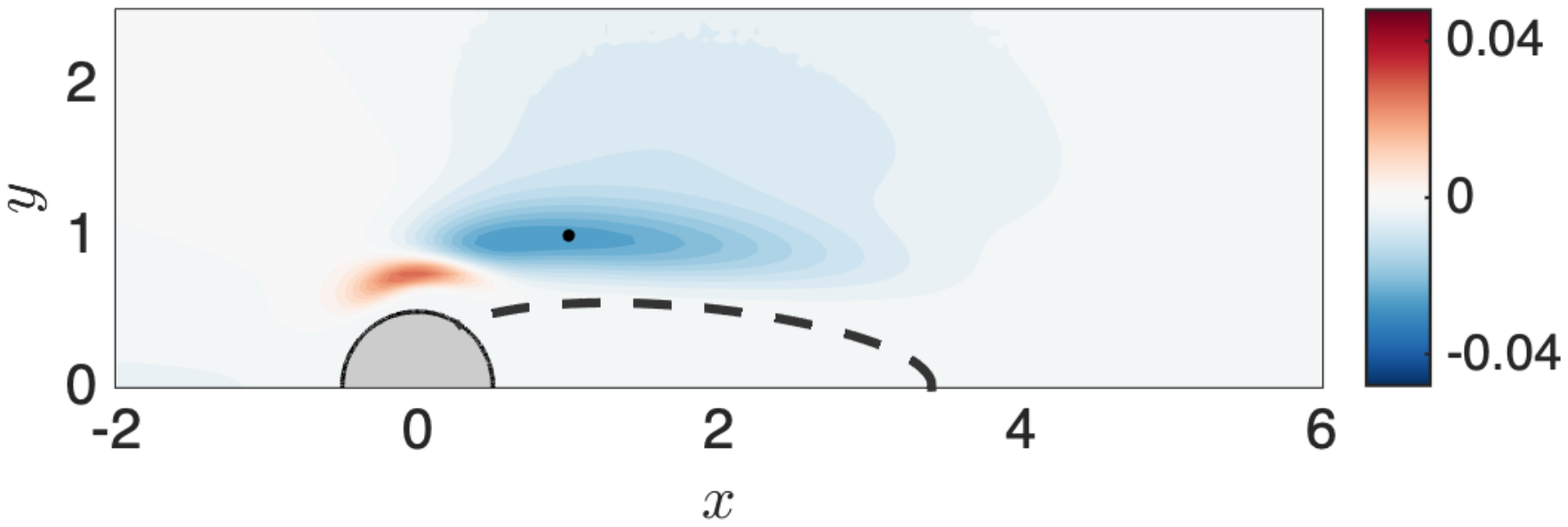}
      \put(-2,30){$(c)$}
   \end{overpic}  
   \begin{overpic}[width=6.2cm, trim=10mm 40mm 10mm 170mm, clip=true]{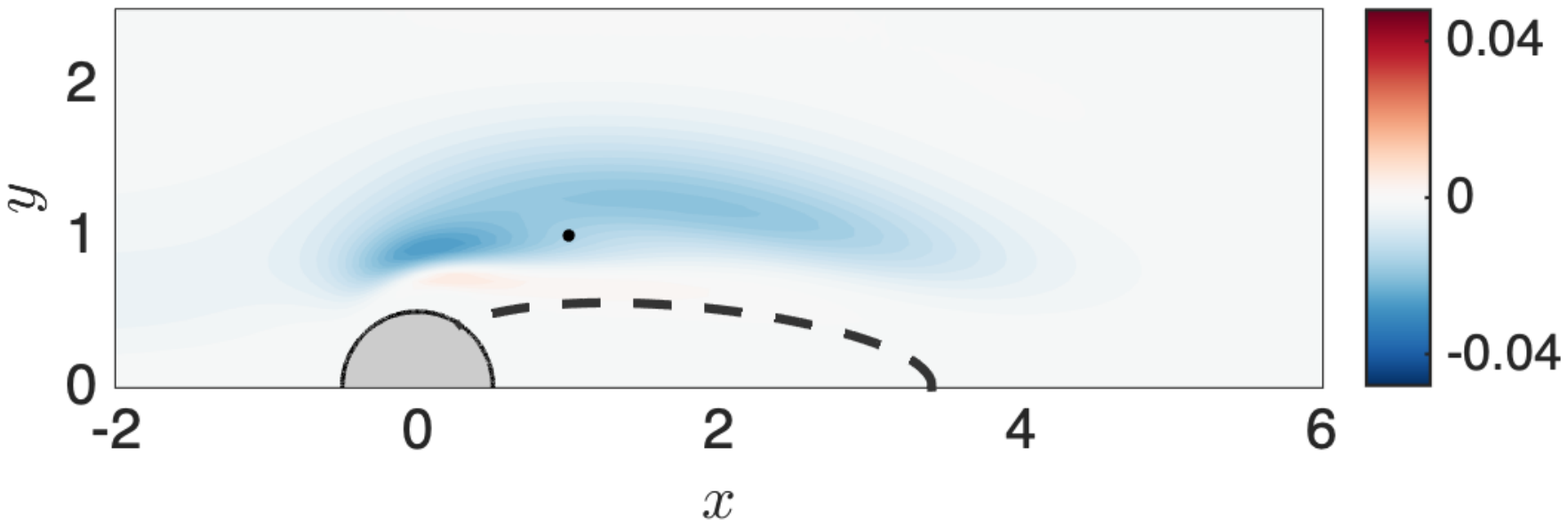}
      \put(-2,30){$(d)$}
   \end{overpic}  
}
\centerline{
   \hspace{0.3cm}
   \begin{overpic}[width=6.2cm, trim=10mm 40mm 10mm 170mm, clip=true]{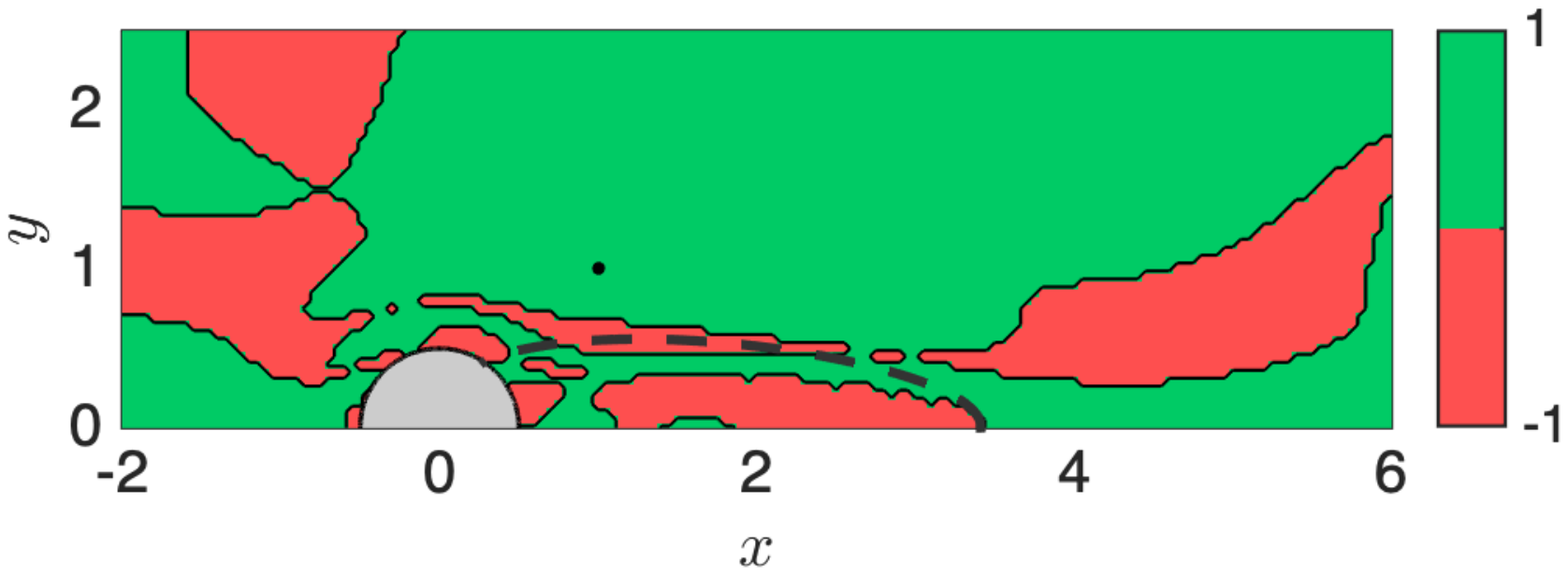}
      \put(-2,30){$(e)$}
   \end{overpic}  
   \hspace{6.2cm}
}
\caption{
Growth rate variation induced by
a small control cylinder 
of diameter $d=0.1$ at $\Rey=50$:
$(a)$~$\epsilon\lambda_{1r}$, and
$(b)$~$\epsilon^2 \lambda_{2r}$.
$(c)$~Term I and $(d)$~term II in the decomposition of $\epsilon^2 \lambda_{2r}$.
$(e)$~Sign of the product $\lambda_{1r} \lambda_{2r}$.
(Figure \ref{fig:F}$(f)$ has no equivalent here because the diameter $d$, and therefore the amplitude $\epsilon$, are fixed.)
The black dot shows the location $\xx_c=(1,1)$ investigated in \S~\ref{sec:analysis}.
}
\label{fig:cyl}
\end{figure}

The first-order growth rate variation induced by the control cylinder is displayed in figure~\ref{fig:cyl}$(a)$.
The map shows a destabilising region on the sides of the main cylinder, stabilising regions on the sides of the recirculation region, and more weakly stabilising regions on the symmetry axis both upstream and downstream of the main cylinder.
This is in agreement with the map obtained by  \cite{Marquet08cyl} (figure~11$(a)$ therein), and is consistent with the map of figure~\ref{fig:F}$(a)$, since the force (\ref{eq:small_cyl_model}) is  oriented mainly along $-x$ outside the recirculation region and mainly along $x$ inside.

Figure~\ref{fig:cyl}$(b)$ shows the second-order growth rate variation. 
The main destabilising and stabilising regions appear rather similar to those of the first-order variation of panel $(a)$. 
This means that, where the signs of those regions do correspond, the second-order variation tends to strengthen the effect of the first-order one. 
A closer look at figure~\ref{fig:cyl}$(e)$ reveals that, where $\lambda_{1r}$ and $\lambda_{2r}$ are not both small, they generally have the same sign. 
Therefore, for a small control cylinder, and considering the variation of $\lambda_r$ up to second order, the situation is one of the following almost everywhere:
(i)~both $\lambda_{1r}$ and $\lambda_{2r}$ are small, so the control cylinder does not modify the growth rate substantially;
(ii)~only $\lambda_{2r}$ is small, so the  effect of the control cylinder is well predicted by $\lambda_{1r}$ alone;
(iii)~$\lambda_{1r}$ and $\lambda_{2r}$ are not small and have the same sign, so the effect of the control cylinder is stronger (more destabilising or more stabilising) than predicted by $\lambda_{1r}$ alone.
One exception is the narrow region where  $\lambda_{2r}>0$ and $\lambda_{1r}$ is small: although first-order sensitivity predicts no effect, the control cylinder is actually destabilising.

The decomposition of $\lambda_{2r}$ into terms I and II in figure~\ref{fig:cyl}$(c)$-$(d)$ shows that the second-order destabilising effect is primarily due to $\UU_2$, while the  second-order stabilising effect is due both to  $\UU_2$ and to the $\UU_1$--$\uu_1$ interaction.

\begin{figure} 
\centerline{
   \hspace{0.6cm}
   \begin{overpic}[width=7cm, trim=10mm 40mm 15mm 170mm, clip=true]{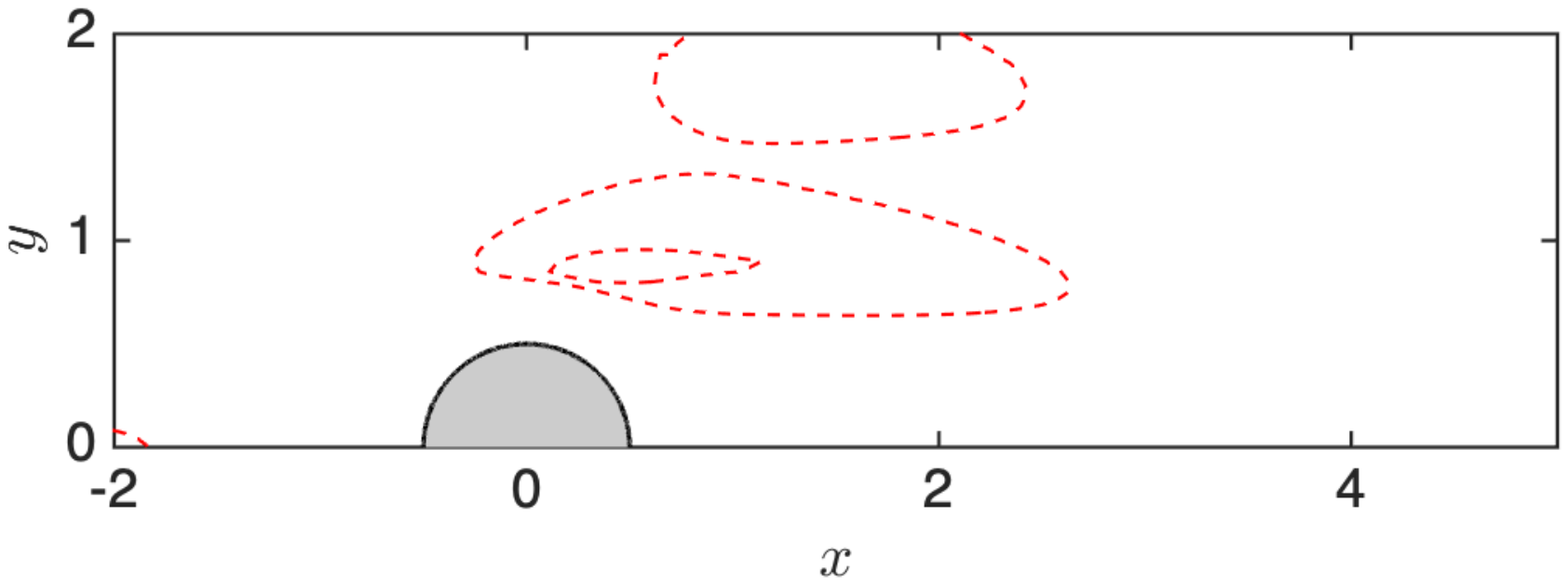}
      \put(-2,31){$(a)$}
      \put(50,23){\tiny \tcr{50}}
      \put(47,18){\tiny \tcr{60}}
   \end{overpic}  
   \hspace{-0.3cm}
   \begin{overpic}[width=7cm, trim=10mm 40mm 15mm 170mm, clip=true]{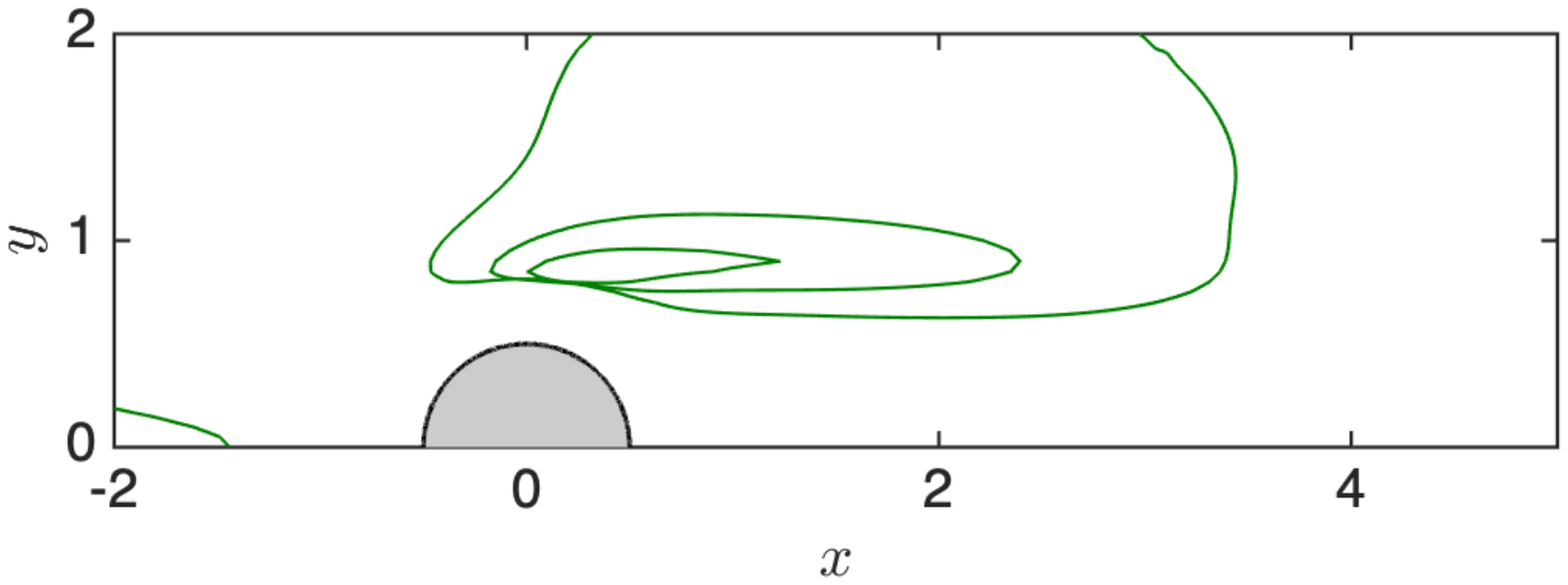}
      \put(-2,31){$(b)$}
      \put(73,21){\tiny \tcgreen{50}}
      \put(60,20){\tiny \tcgreen{60}}
      \put(47,19){\tiny \tcgreen{70}}
   \end{overpic}  
}
\centerline{
   \hspace{0.6cm}
   \begin{overpic}[width=7cm, trim=10mm 40mm 15mm 170mm, clip=true]{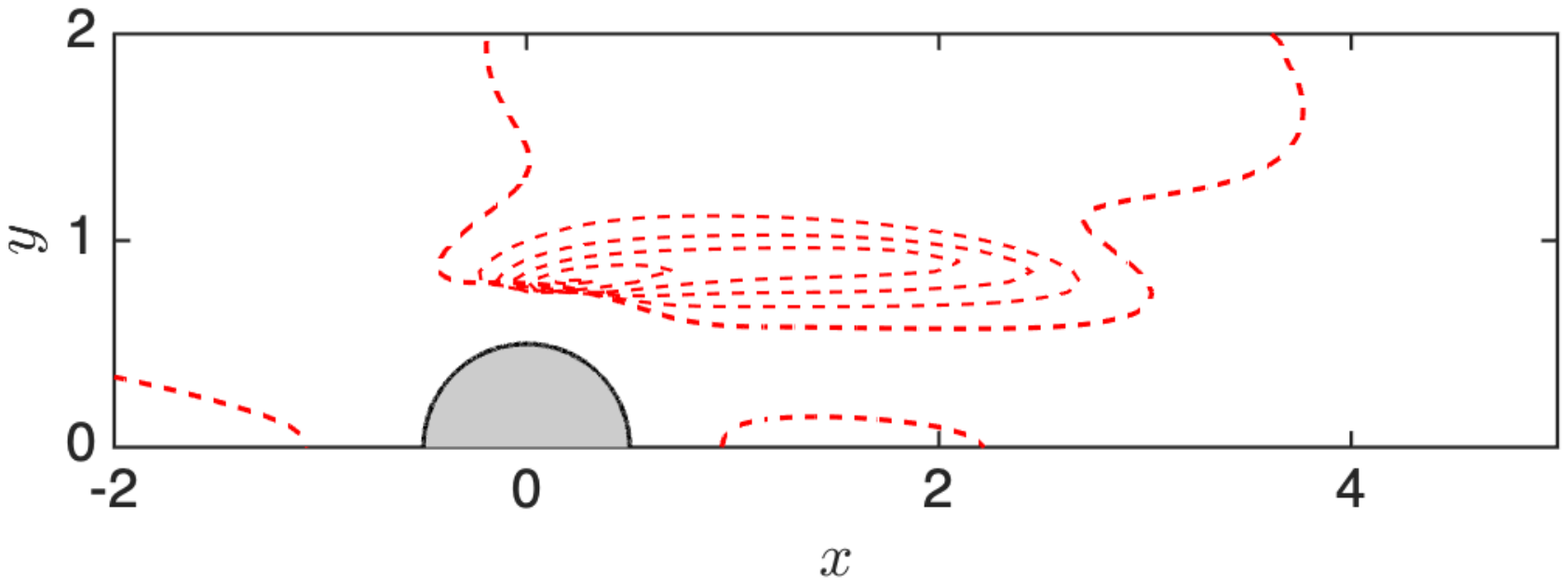}
      \put(-2,31){$(c)$}
      \put(78, 26.0){\tiny \tcr{50}}
      \put(53,   22){\tiny \tcr{60}}
      \put(48, 11.5){\tiny \tcr{50}}
      \put(15,   11.5){\tiny \tcr{50}}
   \end{overpic}  
   \hspace{-0.3cm}
   \begin{overpic}[width=7cm, trim=10mm 40mm 15mm 170mm, clip=true]{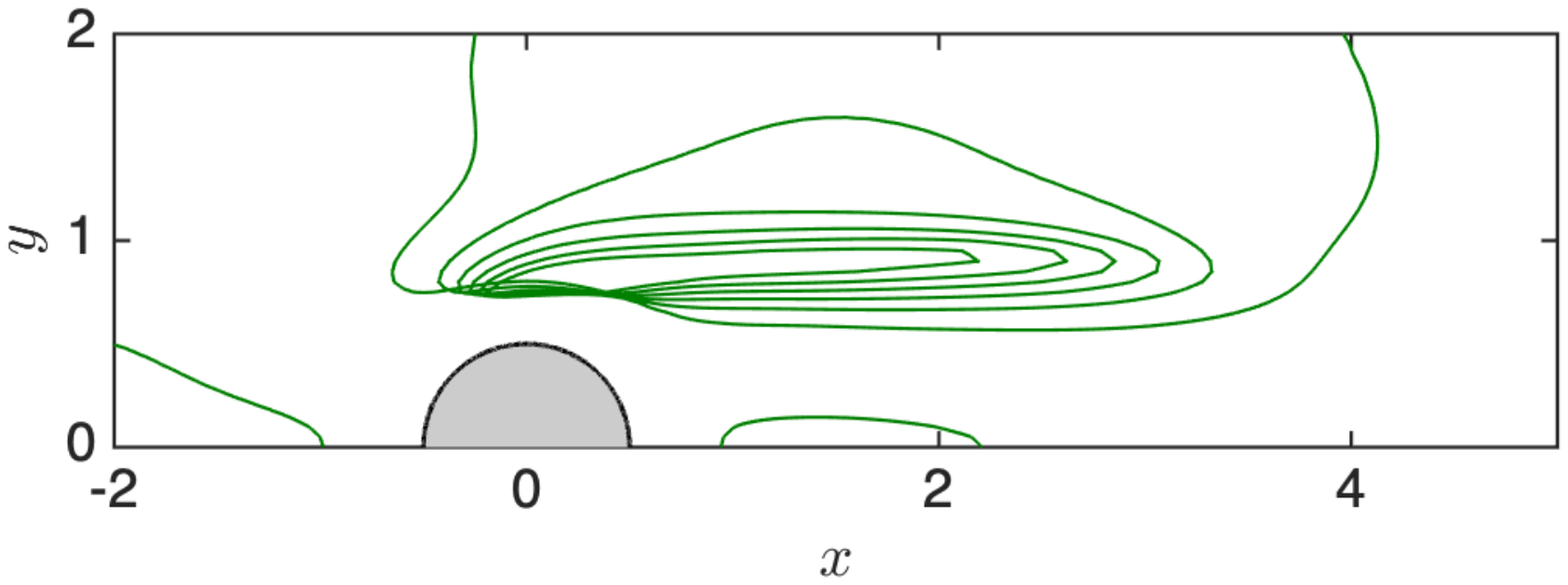}
      \put(-2,31){$(d)$}
      \put(82,   25){\tiny \tcgreen{50}}
      \put(65,   23){\tiny \tcgreen{60}}
      \put(50,   22.5){\tiny \tcgreen{70}}
      \put(48,   11.5){\tiny \tcgreen{50}}
      \put(17,   11.5){\tiny \tcgreen{50}}
   \end{overpic}  
}
\caption{
Passive control with $(a,b)$ one  cylinder or $(c,d)$ two symmetric cylinders  modelled by the force~(\ref{eq:small_cyl_model}) for a diameter $d=0.1$.
On the contours, sensitivity analysis predicts the leading mode to be stabilised, i.e. become exactly neutrally stable. 
$(a,c)$ First-order prediction, $\lambda_{0r} + \epsilon\lambda_{1r} = 0$;
$(b,d)$ second-order prediction, $\lambda_{0r} + \epsilon\lambda_{1r} + \epsilon^2 \lambda_{2r} = 0$.
Reynolds numbers $\Rey=50$, 60, $\ldots$, 100.
Contours are symmetric with respect to $y=0$.
}
\label{fig:stab_contours}
\end{figure}

Figure~\ref{fig:stab_contours}$(a)$-$(b)$ shows the contours where inserting a small control cylinder of diameter $d=0.1$, as described above, is predicted to make the leading mode neutrally stable.
Several Reynolds numbers $\Rey \geq 50$ are considered. 
Inside the regions delimited by these contours, the mode is stable and vortex shedding is expected to be suppressed.
In figure~\ref{fig:stab_contours}$(a)$, only the first-order sensitivity prediction is considered,  $\lambda_{0r} + \epsilon\lambda_{1r}=0$, while in figure~\ref{fig:stab_contours}$(b)$, the second-order correction is included too,  $\lambda_{0r} + \epsilon\lambda_{1r} + \epsilon^2\lambda_{2r}=0$. 
The results compare qualitatively well with the experimental observations of \cite{Strykowski1990} (figure 20 therein): stabilisation is achieved on the side of the recirculation region, in an area that is rather wide at $\Rey=50$ and that becomes smaller as the Reynolds number increases, until shrinking to a single point and vanishing when restabilisation is not possible any more.
Compared to the first-order sensitivity, however, the second-order sensitivity seems to better capture  the results of \cite{Strykowski1990}: in particular, it predicts a wider stabilising area at $\Rey=60$, and  a larger value of the maximum stabilisable Reynolds number $70<\Rey<80$. 

For completeness, figure~\ref{fig:stab_contours}$(c)$-$(d)$ shows stabilising contours for a pair of control cylinders located symmetrically in $(x_c,y_c)$ and $(x_c,-y_c)$, still with $d=0.1$.
In the sensitivity framework, the two cylinders are assumed not to influence each other, which is not satisfied close to the symmetry axis $y=0$.
Unsurprisingly, the main stabilising region is wider but still located on the side of the recirculation region. 
Although conclusions should be drawn with care at larger Reynolds numbers, as the uncontrolled flow becomes linearly unstable to a second two-dimensional mode at $\Rey\simeq 100$ \citep{Verma2011} and to a three-dimensional mode at $\Rey \simeq 190$ \citep{barkley96}, restabilisation can be achieved 
 up to  $\Rey\simeq 100$  and $\Rey > 100$ according to first- and second-order sensitivity predictions, respectively.

%
\subsection{Analysis of the stabilisation induced by a small control cylinder located optimally}
\label{sec:analysis}

In this section, the effect of a small control cylinder is investigated in more detail for the specific control location $\xx_c=(1,1)$, close to the location of largest first- and second-order stabilising effects identified in
\S~\ref{sec:results_pass_device}.

\begin{figure} 
\centerline{
   \begin{overpic}[height=5cm, trim=10mm 65mm 95mm 78mm, clip=true]{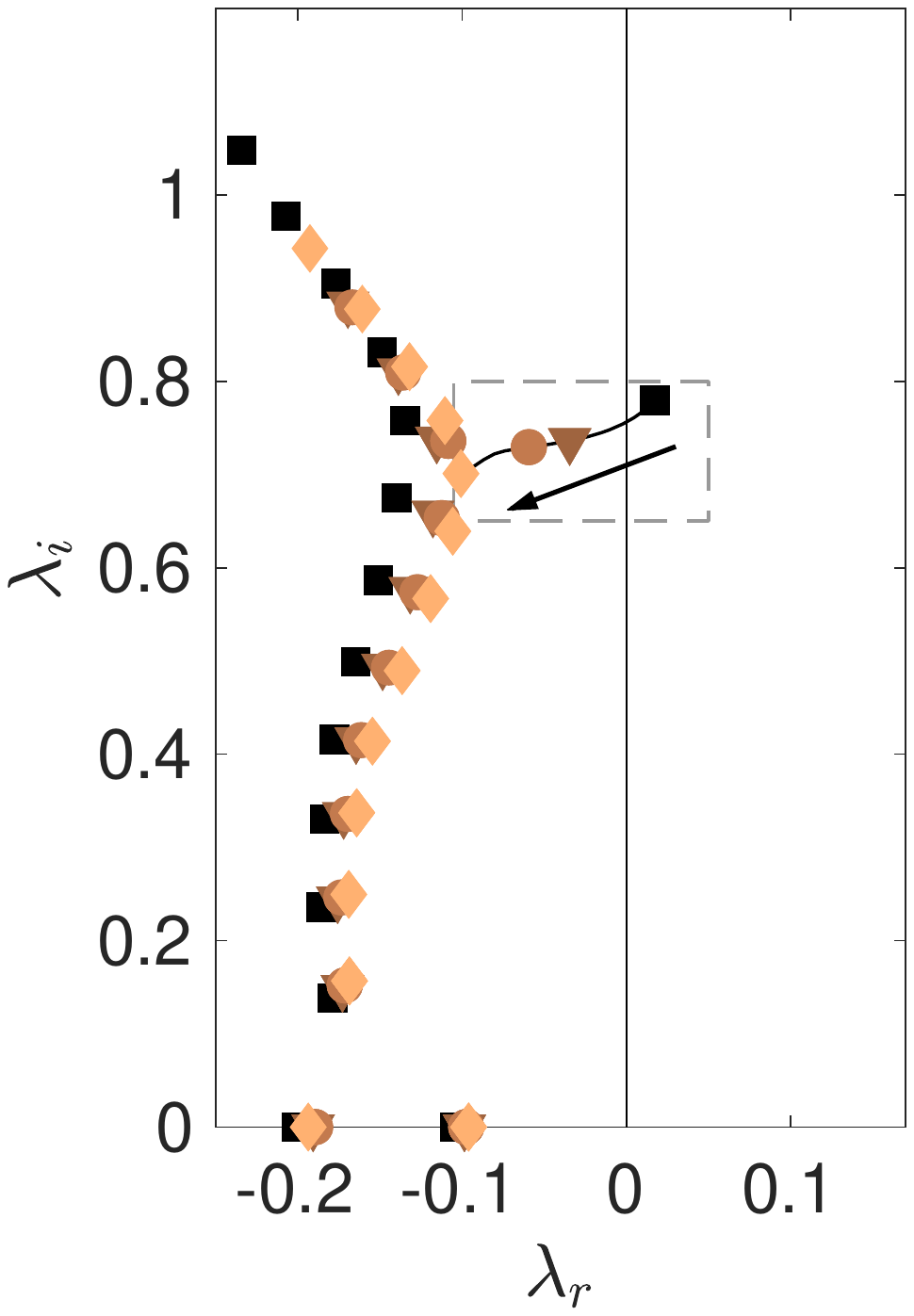}   
      \put(-2,93){$(a)$}  
      \put(44.5,60){\tiny $d$}
   \end{overpic} 
\hspace{0.2cm}
   \begin{overpic}[height=5.1cm, trim=10mm 70mm 15mm 75mm, clip=true]{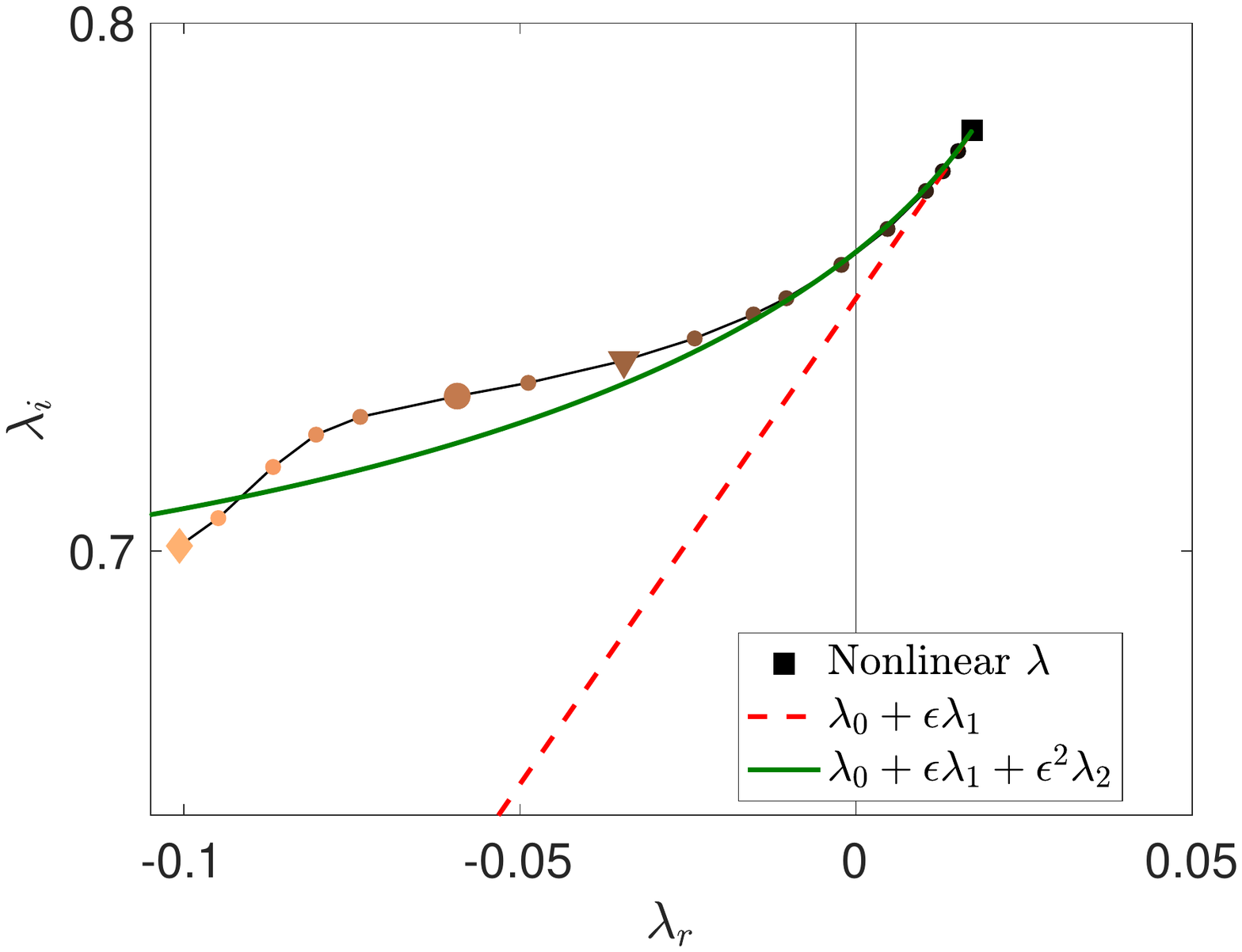}   
      \put(-3,75){$(b)$}        
      \put(73,   70){\tiny $d=0$}
      \put(39.5, 51.5){\tiny $d=0.03$}
      \put(25.0, 48.5){\tiny $d=0.05$}
      \put(13.5, 31.0){\tiny $d=0.1$}
   \end{overpic} 
}
\vspace{-0.2cm}
\caption{
$(a)$~Eigenvalues of the uncontrolled flow (black squares), and of the flow controlled with a small secondary cylinder of diameter $d$
(triangles, $d=0.03$; 
circles, $d=0.05$; 
diamonds, $d=0.1$) located in $\xx_c=(1,1)$. 
$(b)$~Zoomed-in view of the leading eigenvalue (dashed region in panel $a$), with first- and second-order sensitivities (dashed and solid lines, respectively).
$\Rey=50$.
}
\label{fig:spectra_small_cyl}
\end{figure}

Figure~\ref{fig:spectra_small_cyl} shows the eigenspectrum of the flow controlled with a secondary cylinder of increasing diameter $d$.
The leading mode is restabilised for diameters $d \gtrsim 0.004$. 
Other modes remain stable for the whole range of diameters investigated.
As seen in the close-up (figure~\ref{fig:spectra_small_cyl}$b$), the second-order sensitivity (thick solid line) follows closely the actual path of the leading eigenvalue in the complex plane (symbols), accurately capturing both the growth rate and the frequency,  and improving on the first-order prediction (dashed line).

\begin{figure} 
\centerline{
   \begin{overpic}[width=6.3cm, trim=8mm 15mm 25mm 150mm, clip=true]{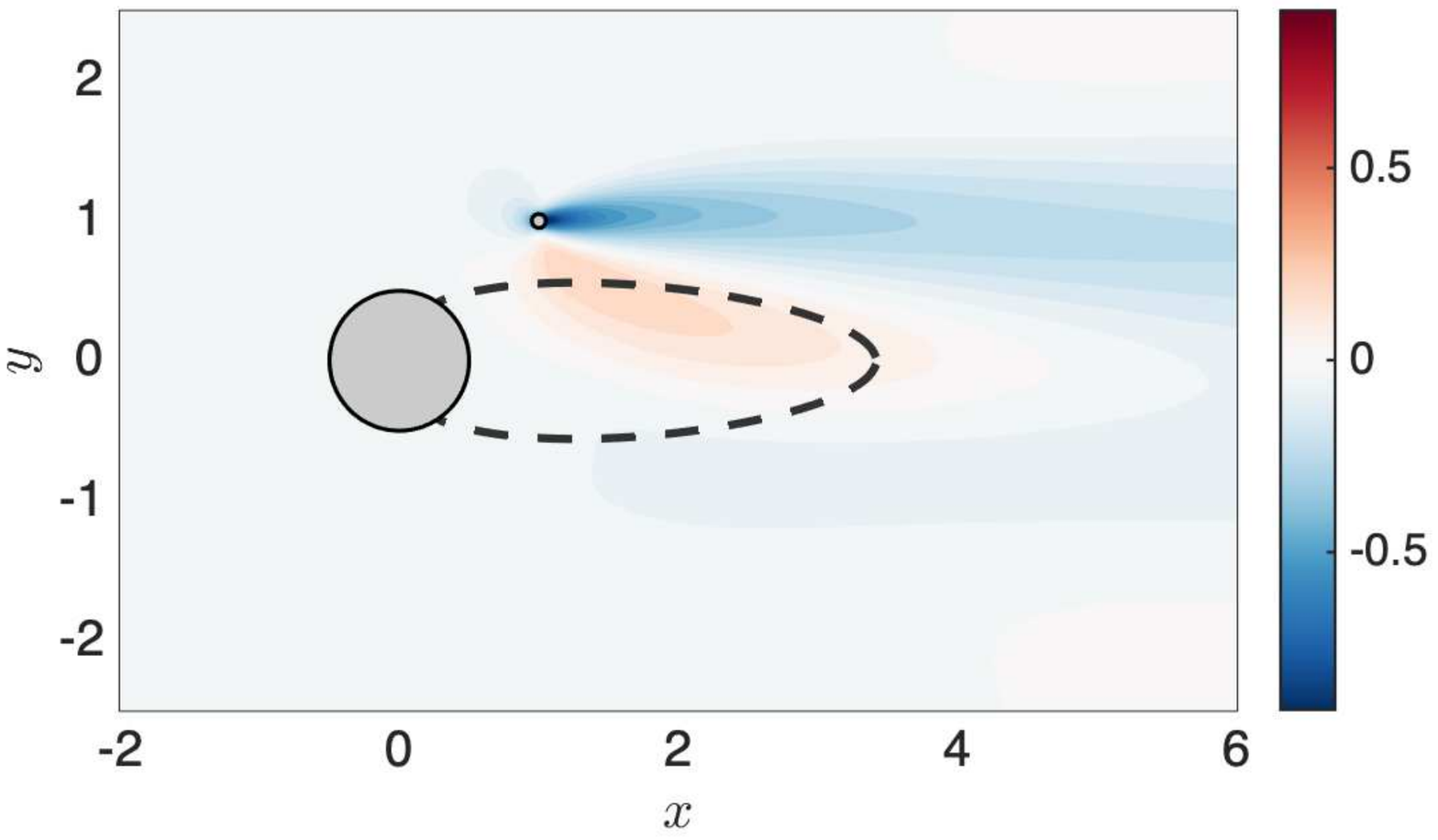}
      \put(-2,57){$(a)$}
   \end{overpic} 
   \hspace{0.1cm}
   \begin{overpic}[width=6.3cm, trim=8mm 15mm 25mm 150mm, clip=true]{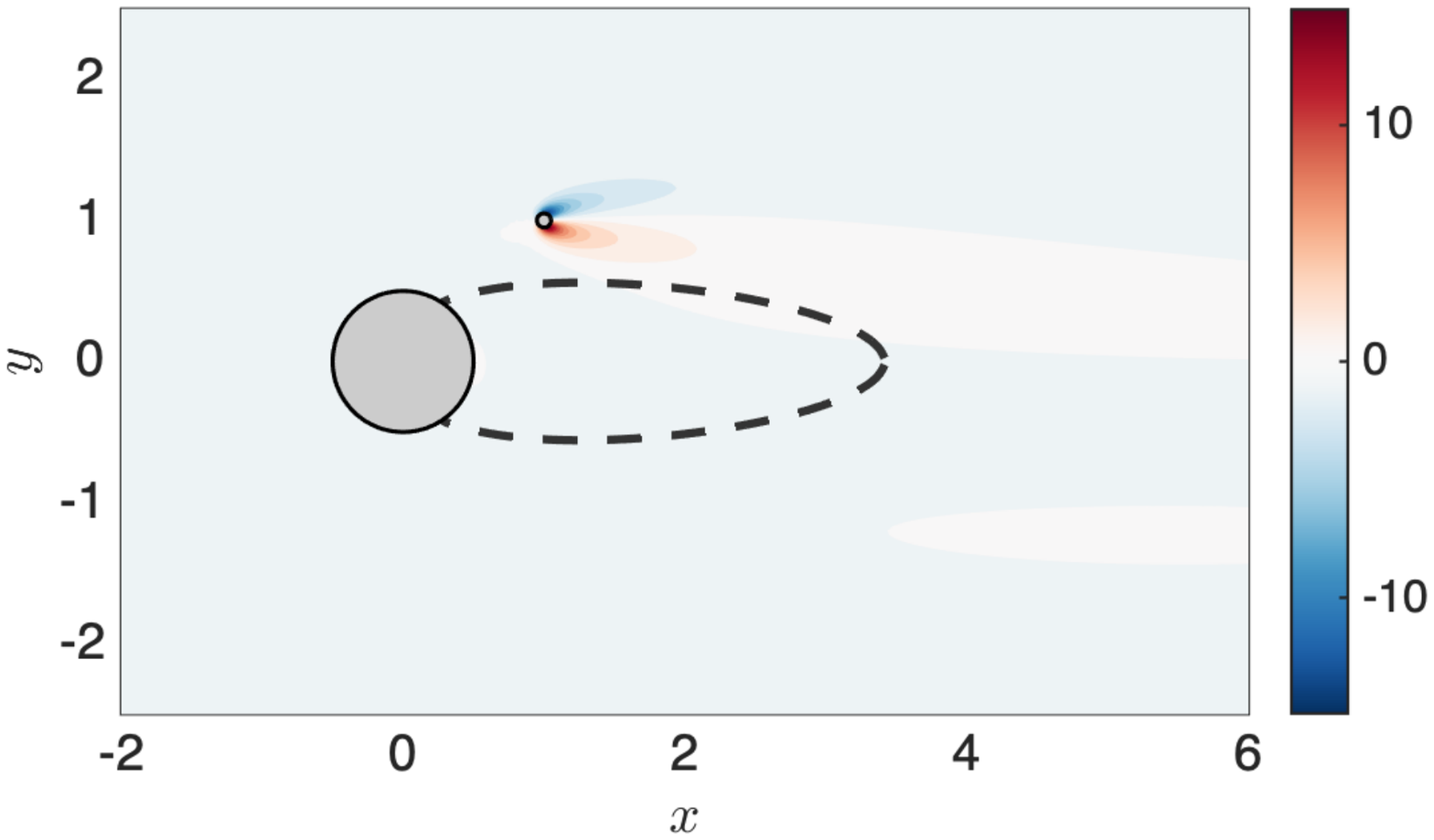}
      \put(-2,57){$(b)$}
   \end{overpic} 
}
\centerline{
   \begin{overpic}[width=6.3cm, trim=8mm 15mm 25mm 150mm, clip=true]{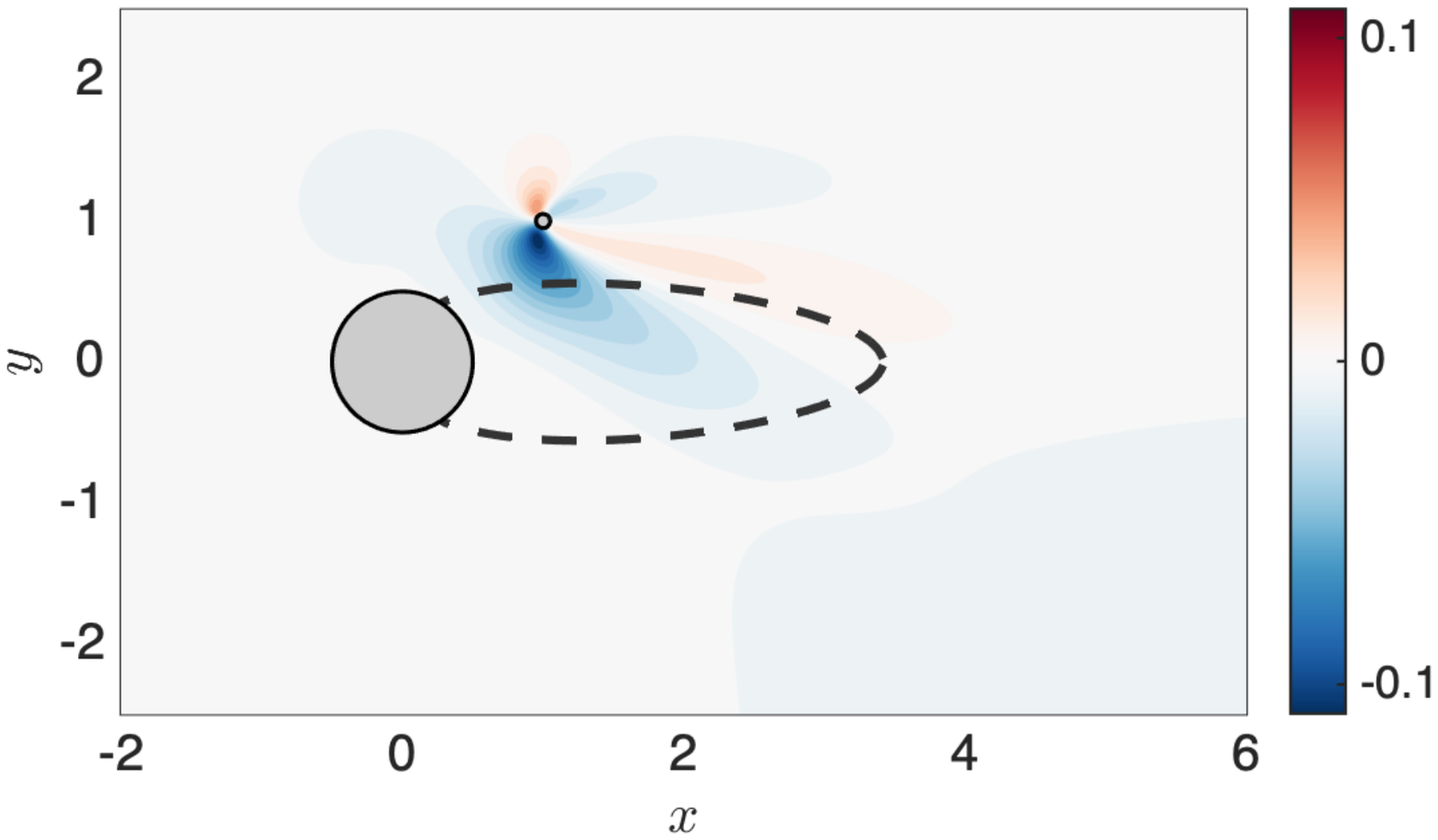}
      \put(-2,57){$(c)$}
   \end{overpic} 
   \hspace{0.1cm}
   \begin{overpic}[width=6.3cm, trim=8mm 15mm 25mm 150mm, clip=true]{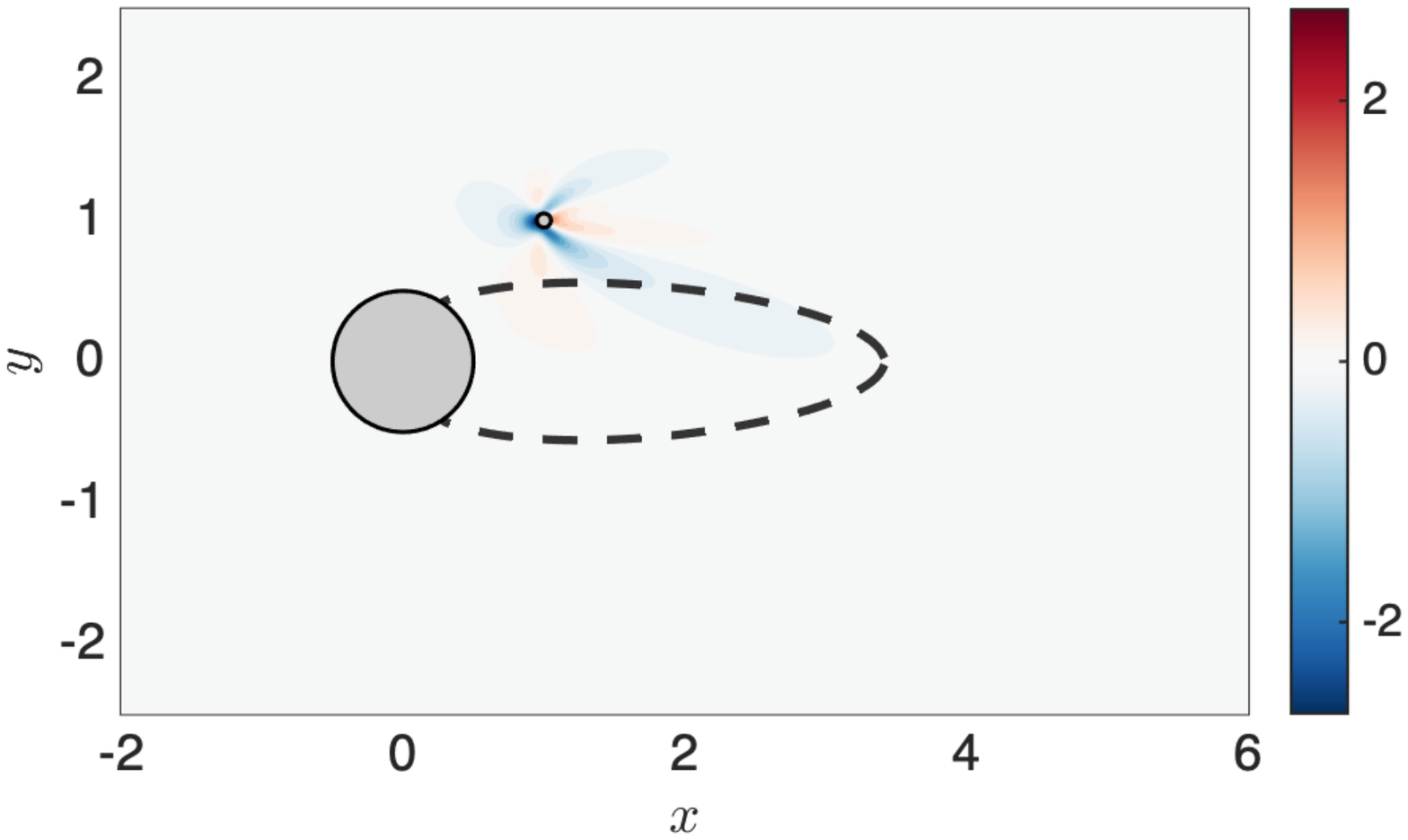}
      \put(-2,57){$(d)$}
   \end{overpic} 
}
\vspace{0.1cm}
\centerline{
   \begin{overpic}[width=10.5cm, trim=15mm 90mm 20mm 73mm, clip=true]{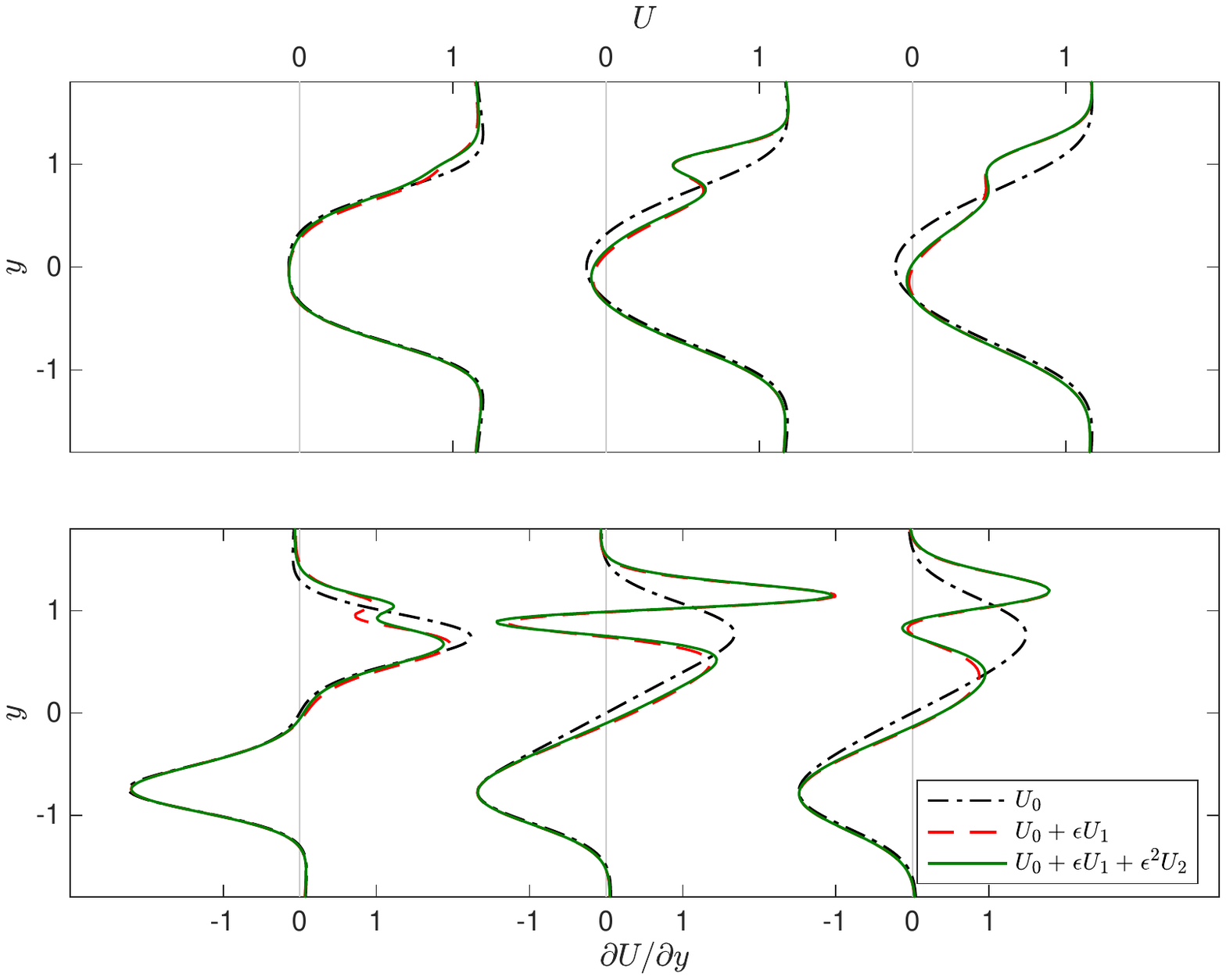}      
      \put(4,68){$(e)$}
      \put(24, 37){\small $x=0.8$}
      \put(47, 37){\small $x=1.5$}
      \put(70.5, 37){\small $x=2$}
   \end{overpic} 
}
\caption{
First- and second-order flow modification.
$(a)$ Streamwise velocity $U_1$,
$(b)$~vorticity $\omega_1$,
$(c)$~streamwise velocity $U_2$, and
$(d)$~vorticity $\omega_2$.
$(e)$~Profiles of streamwise velocity $U$ and horizontal shear $\partial U/\partial y$.
}
\label{fig:U1_U2_Om1_Om2}
\end{figure}

Let us now focus on the diameter $d=0.1$.
As apparent from figure~\ref{fig:cyl},
the predicted first- and second-order growth rate variations are comparable,
\begin{align}
\epsilon\lambda_{1r} = -0.0426, \quad
\epsilon^2\lambda_{2r} = -0.0424, 
\end{align}
and the two second-order contributions are of similar order of magnitude:
$\epsilon^2\lambda_{2r,\text{I}} = -0.0258$,
$\epsilon^2\lambda_{2r,\text{II}} = -0.0167$.
Figure \ref{fig:U1_U2_Om1_Om2} depicts the base flow modification.
At first order, the control cylinder induces a strong velocity deficit $U_1<0$ in its wake, and a slight acceleration $U_1>0$ between the two cylinders (figure \ref{fig:U1_U2_Om1_Om2}$a$).
As a result, two layers of opposite vorticity  emanate from the control cylinder (figure \ref{fig:U1_U2_Om1_Om2}$b$) in a roughly symmetric way.
At second order, velocity and  vorticity are modified more weakly, with a more complicated spatial pattern (figure \ref{fig:U1_U2_Om1_Om2}$c,d$).
The net effect of the control cylinder  is best illustrated by the velocity and shear profiles in figure~\ref{fig:U1_U2_Om1_Om2}$(e)$.
In $x=0.8$, just upstream of the control location $x_c=1$, the flow modification $U_1$ (red dashed line) smooths the velocity profile and  reduces the maximum shear.
Downstream ($x=1.5$ and 2),  the  induced velocity deficit further  reduces  shear in the lower shear layer emanating from the control cylinder, where the positive vorticity $\omega_1$ (figure \ref{fig:U1_U2_Om1_Om2}$b$) counteracts the negative base flow  vorticity $\omega_0$ (figure~\ref{fig:base_flow}).
These observations are consistent with those of \cite{Marquet08cyl}.
In the upper shear layer, however,  the negative  $\omega_1$ adds up to the negative  $\omega_0$,  and shear is strongly increased, well beyond the maximum uncontrolled  shear.
The second-order modification (green solid line) tends to yield an additional reduction in maximum shear, both upstream and downstream of $x_c$, albeit much  smaller.
In light of these  observations, shear alone does not seem to explain entirely  (i)~why $\UU_1$ is stabilising,
and (ii)~why $\UU_2$ brings an additional stabilisation as large as $\UU_1$.

\begin{figure} 
\centerline{
   \hspace{0.2cm}
   \begin{overpic}[width=6.3cm, trim=8mm 15mm 22mm 150mm, clip=true]{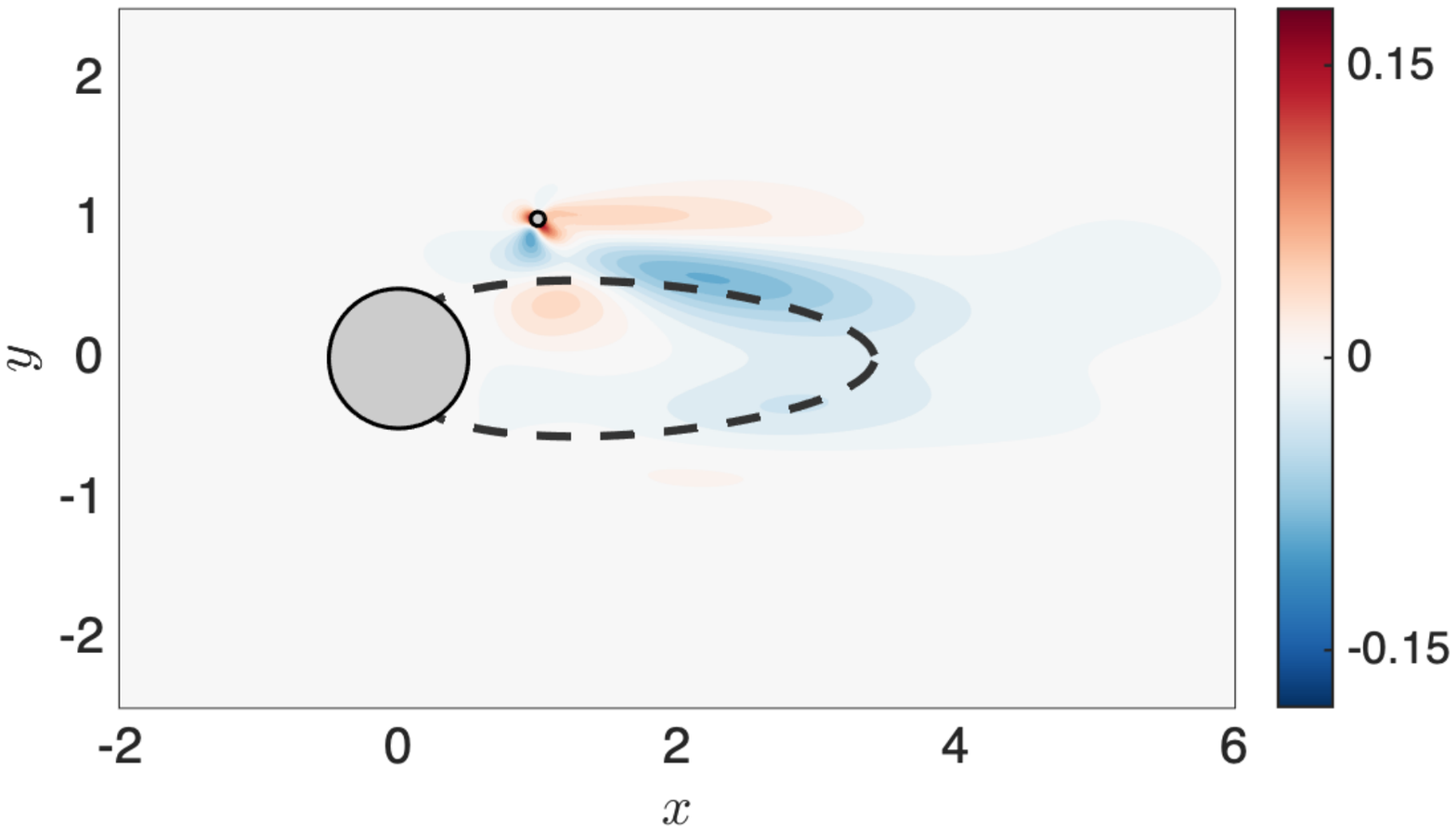}
      \put(-2,57){$(a)$}
   \end{overpic} 
   \hspace{0.2cm}
   \begin{overpic}[width=6.98cm, trim=15mm 125mm 15mm 65mm, clip=true]{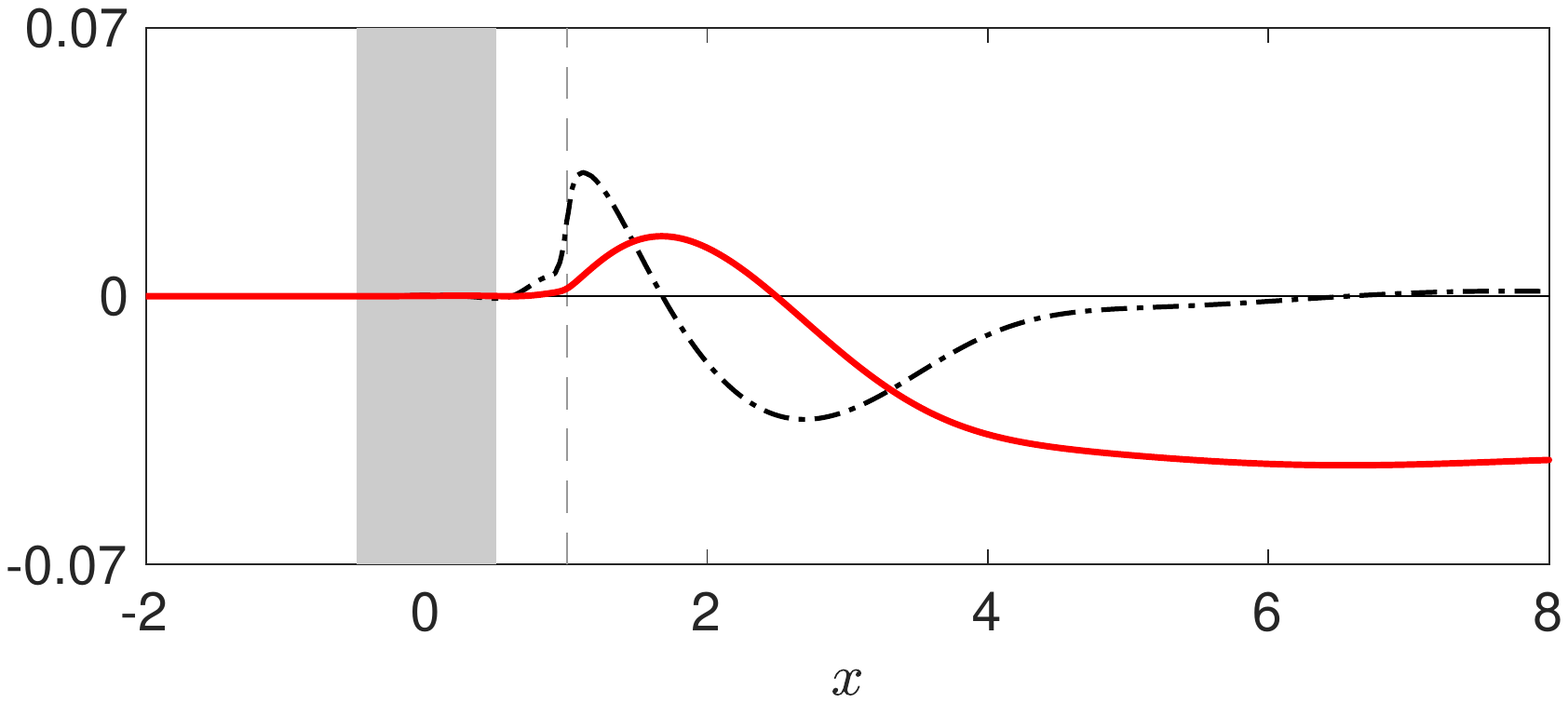}
      \put(-3,51.5){$(b)$}
      \put(76.5,34.){$l_1(x)$}
      \put(65.3,24.5){\tcr{$\int_{-\infty}^x l_1(x') \,\mathrm{d}x'$}}
      \put(81,16.5){\tcr{$\rightarrow \lambda_{1r}$}}
   \end{overpic} 
}
\centerline{
   \hspace{0.2cm}
   \begin{overpic}[width=6.3cm, trim=8mm 15mm 22mm 150mm, clip=true]{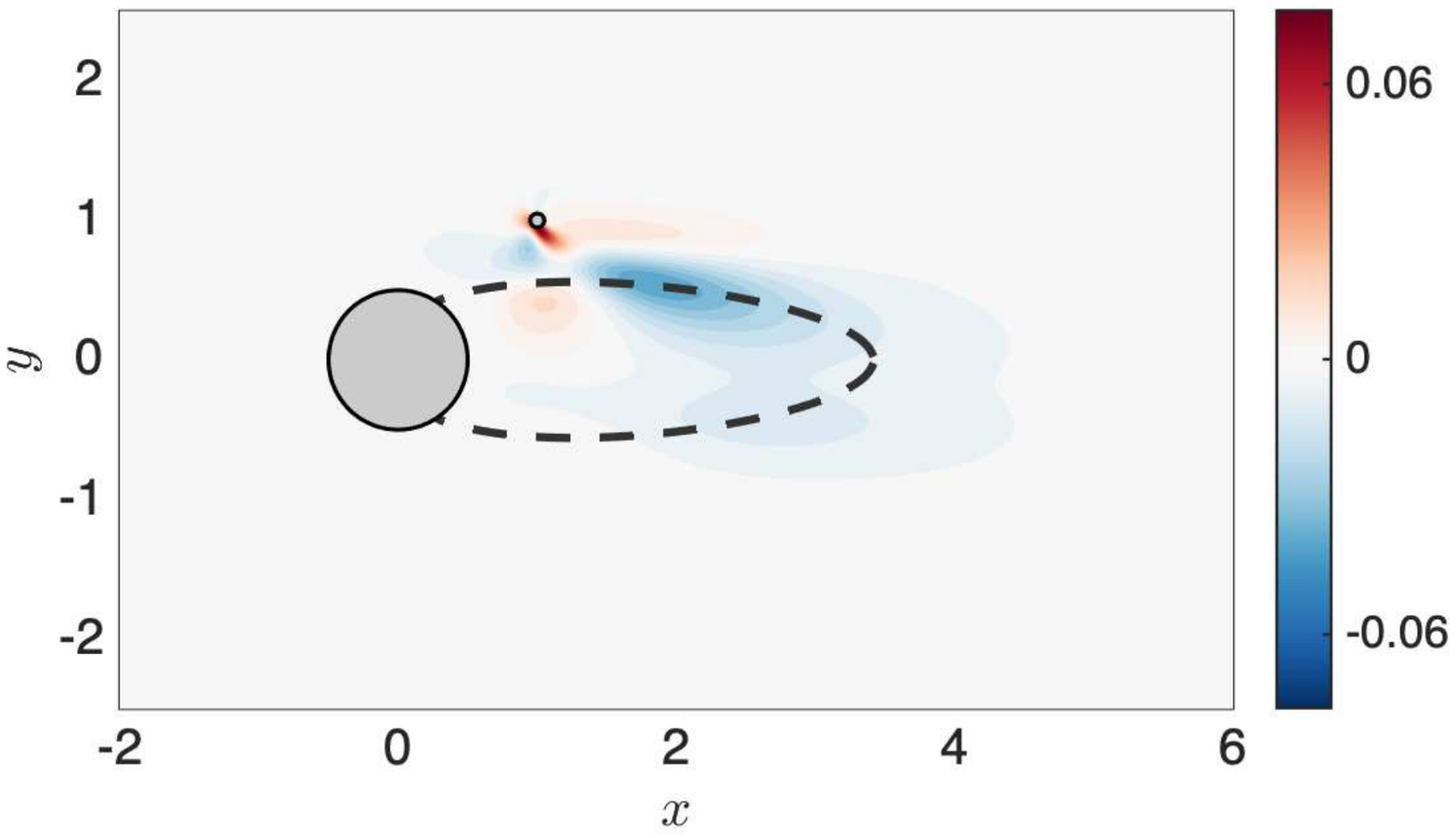}
      \put(-2,57){$(c)$}
   \end{overpic} 
   \hspace{0.2cm}
   \begin{overpic}[width=6.98cm, trim=15mm 125mm 15mm 65mm, clip=true]{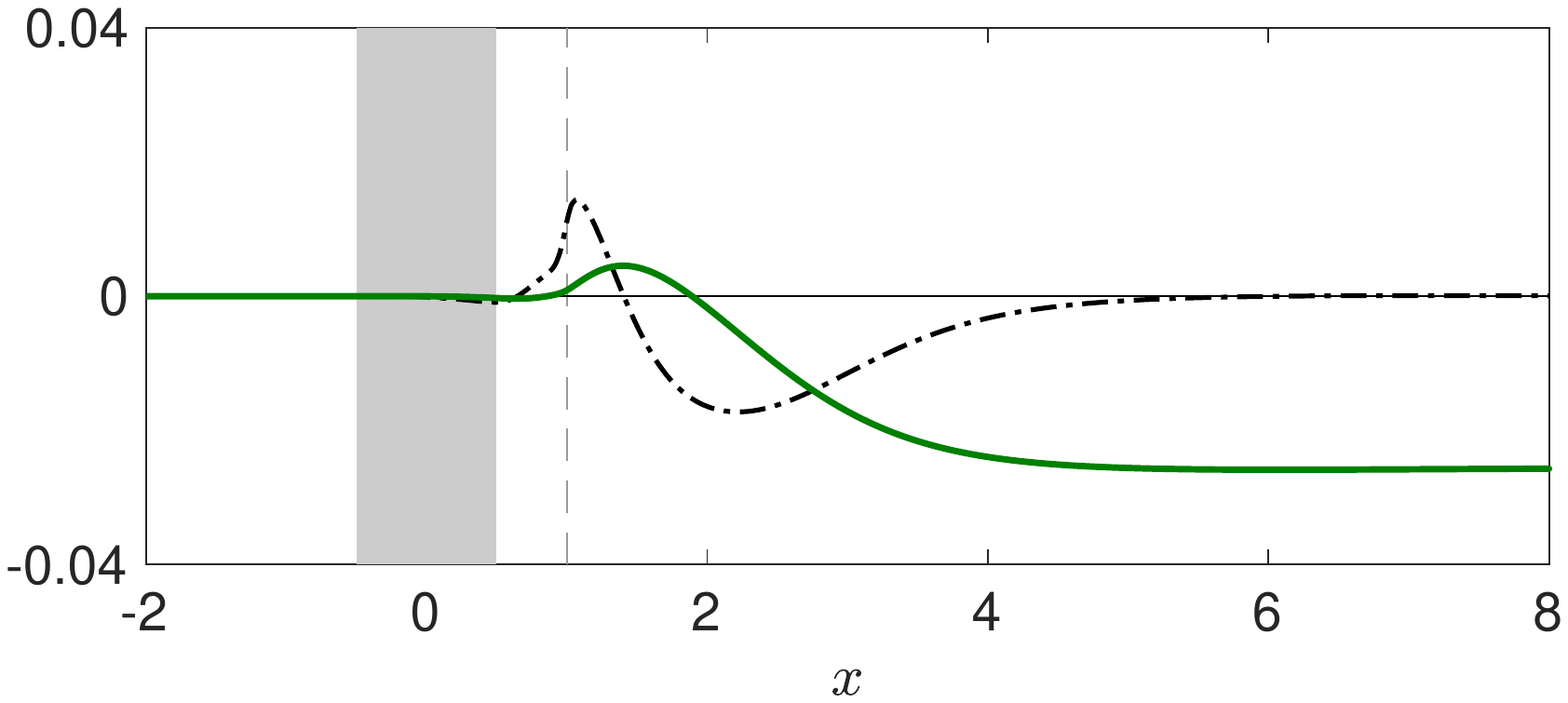}
      \put(-3,51.5){$(d)$}      
      \put(74,33.8){$l_{2,\text{I}}(x)$}
      \put(63,23.7){\tcgreen{$\int_{-\infty}^x l_{2,\text{I}}(x') \,\mathrm{d}x'$}}
      \put(78.5,16.5){\tcgreen{$\rightarrow \lambda_{2r,\text{I}}$}}
   \end{overpic} 
}
\centerline{
   \hspace{0.2cm}
   \begin{overpic}[width=6.3cm, trim=8mm 15mm 22mm 150mm, clip=true]{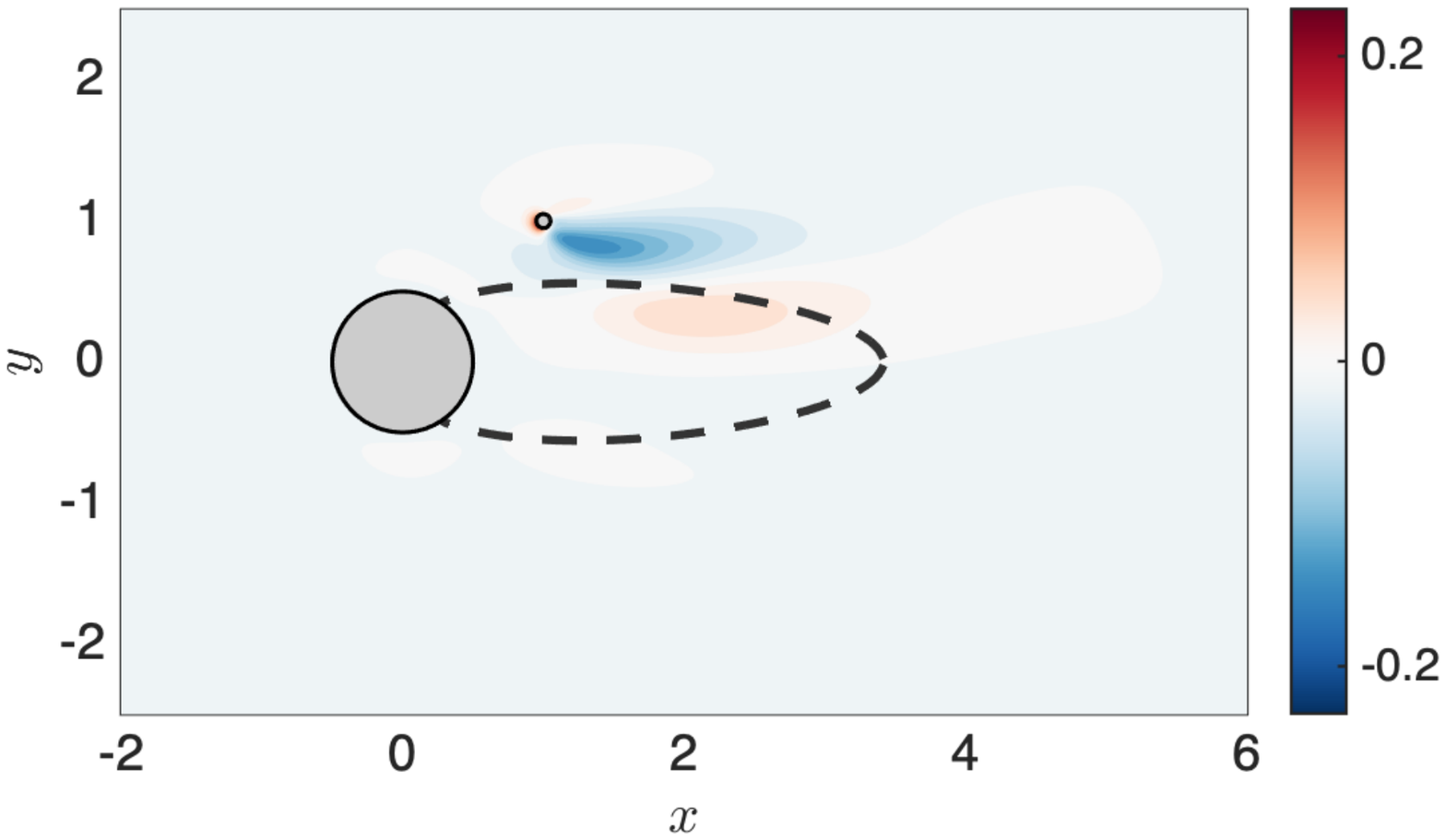}
      \put(-2,57){$(e)$}
   \end{overpic} 
   \hspace{0.2cm}
   \begin{overpic}[width=6.98cm, trim=15mm 125mm 15mm 65mm, clip=true]{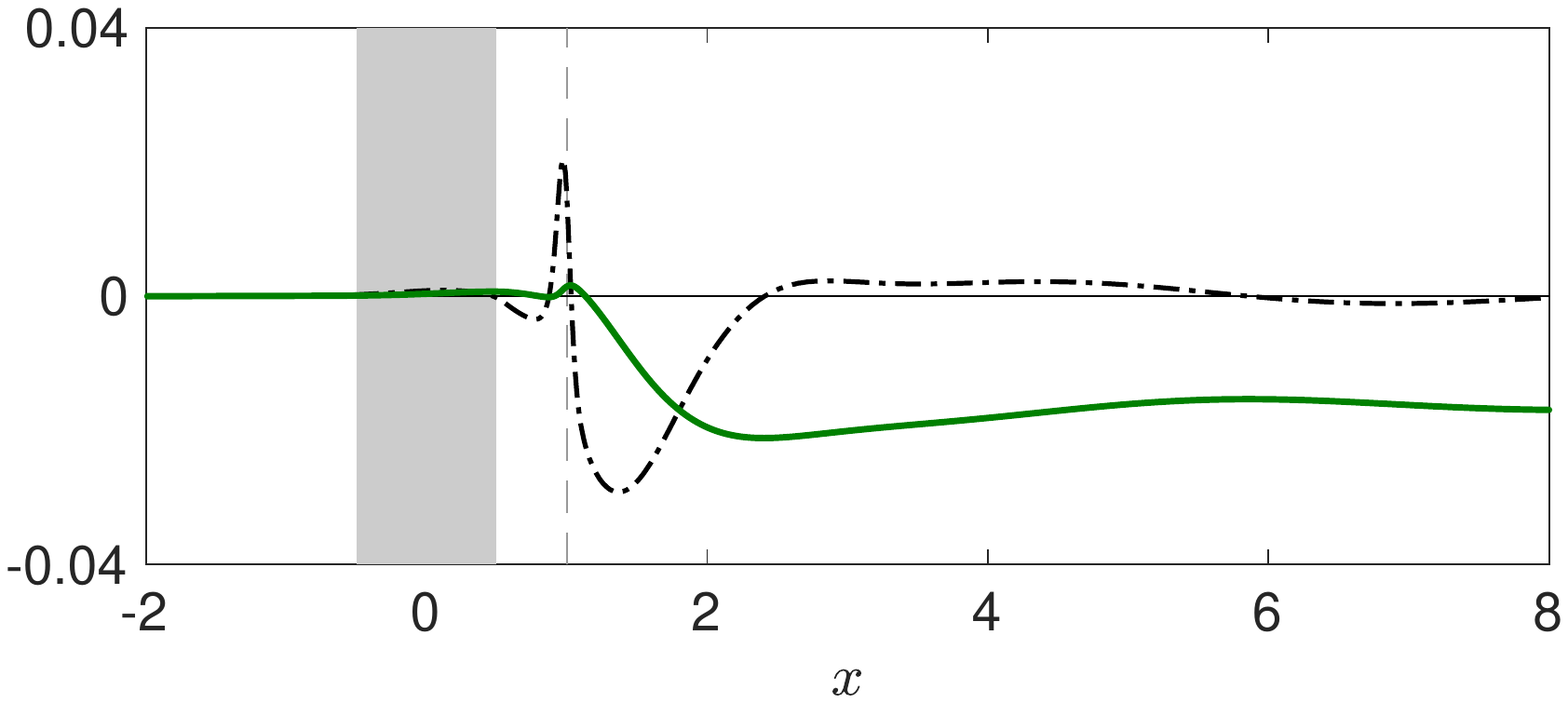}
      \put(-3,51.5){$(f)$}   
      \put(61,34.8){$l_{2,\text{II}}(x)$}
      \put(50,26.4){\tcgreen{$\int_{-\infty}^x l_{2,\text{II}}(x') \,\mathrm{d}x'$}}
      \put(77,20.5){\tcgreen{$\rightarrow \lambda_{2r,\text{II}}$}}
   \end{overpic} 
}
\caption{
Integrands of the first- and second-order growth rate variations $\lambda_{1r}$ and $\lambda_{2r}$ expressed as (\ref{eq:ev1})-(\ref{eq:ev2}), 
for a small control cylinder ($d=0.1$, $\xx_c=(1,1)$) at $\Rey=50$.
$(a)$~Integrand of (\ref{eq:l1});
$(b)$~density $l_1(x)$ (black dash-dotted line) and its cumulative integral (red solid line).
$(c)$~Integrand of (\ref{eq:l2});
$(d)$~density $l_{2,\text{I}}(x)$ (black dash-dotted line) and its cumulative integral (green solid line).
$(e)$~Integrand of (\ref{eq:l3});
$(f)$~density $l_{2,\text{II}}(x)$ (black dash-dotted line) and its cumulative integral (green solid line).
}
\label{fig:integrands}
\end{figure}

Some complementary insight can be gained by looking at regions that contribute to 
the growth rate variation.
Recalling that $\lambda_1$ and $\lambda_2$
are defined by  (\ref{eq:ev1})-(\ref{eq:ev2}) as inner products, it is natural to look at the integrands of $\lambda_1$, $\lambda_{2,\text{I}}$ and $\lambda_{2,\text{II}}$, shown in 
figure~\ref{fig:integrands}$(a,c,e)$.
For a more quantitative picture, it is useful to consider the contribution from each streamwise location $x$:
let us integrate those integrands vertically and define  one-dimensional densities,
\begin{align}
l_1(x) &=  \int_{-\infty}^{\infty} 
\text{Re}\left\{
-\overline \uu_0^\dag \bcdot \left( \AAA_1 \uu_0 \right) \right\} \mathrm{d}y,
\label{eq:l1}
\\
l_{2,\text{I}}(x) &=  \int_{-\infty}^{\infty} 
\text{Re}\left\{
-\overline\uu_0^\dag \bcdot \left(\AAA_2\uu_0 \right)
\right\} \mathrm{d}y,
\label{eq:l2}
\\
l_{2,\text{II}}(x) &= \int_{-\infty}^{\infty} 
\text{Re}\left\{
-\overline \uu_0^\dag \bcdot \left( (\lambda_1\II+\AAA_1)\uu_1 \right)
\right\} \mathrm{d}y.
\label{eq:l3}
\end{align}
By construction, the cumulative integral
$\int_{-\infty}^{x} l_{1}(x') \mathrm{d}x'$ tends to $\lambda_{1r}$ as $x \rightarrow \infty$.
Similarly, the limits of the cumulative integrals of $l_{2,\text{I}}(x)$ and $l_{2,\text{II}}(x)$ are $\lambda_{2r,\text{I}}$ and $\lambda_{2r,\text{II}}$, respectively.
These densities and cumulative integrals are shown in figure~\ref{fig:integrands}$(b,d,f)$ as dash-dotted lines and solid lines, respectively.
All three densities are positive at the control cylinder location;  farther downstream they become negative, in a longer region and with a similar intensity, finally resulting in $\lambda_{1r}<0$ and $\lambda_{2r}<0$.
The two-dimensional integrands are mostly positive in the early wake of the control cylinder, and negative in a wider region running downstream along the separatrix. 
This region can therefore be identified as the main stabilising one when $\lambda_{1r}$ and $\lambda_{2r}$ are understood as inner products 
expressed in terms of
modifications $\AAA_1$, $\AAA_2$ of the linearised NS operator, and  eigenmode modification $\uu_1$.

\begin{figure} 
\centerline{
   \hspace{0.2cm}
   \begin{overpic}[width=6.3cm, trim=8mm 15mm 22mm 150mm, clip=true]{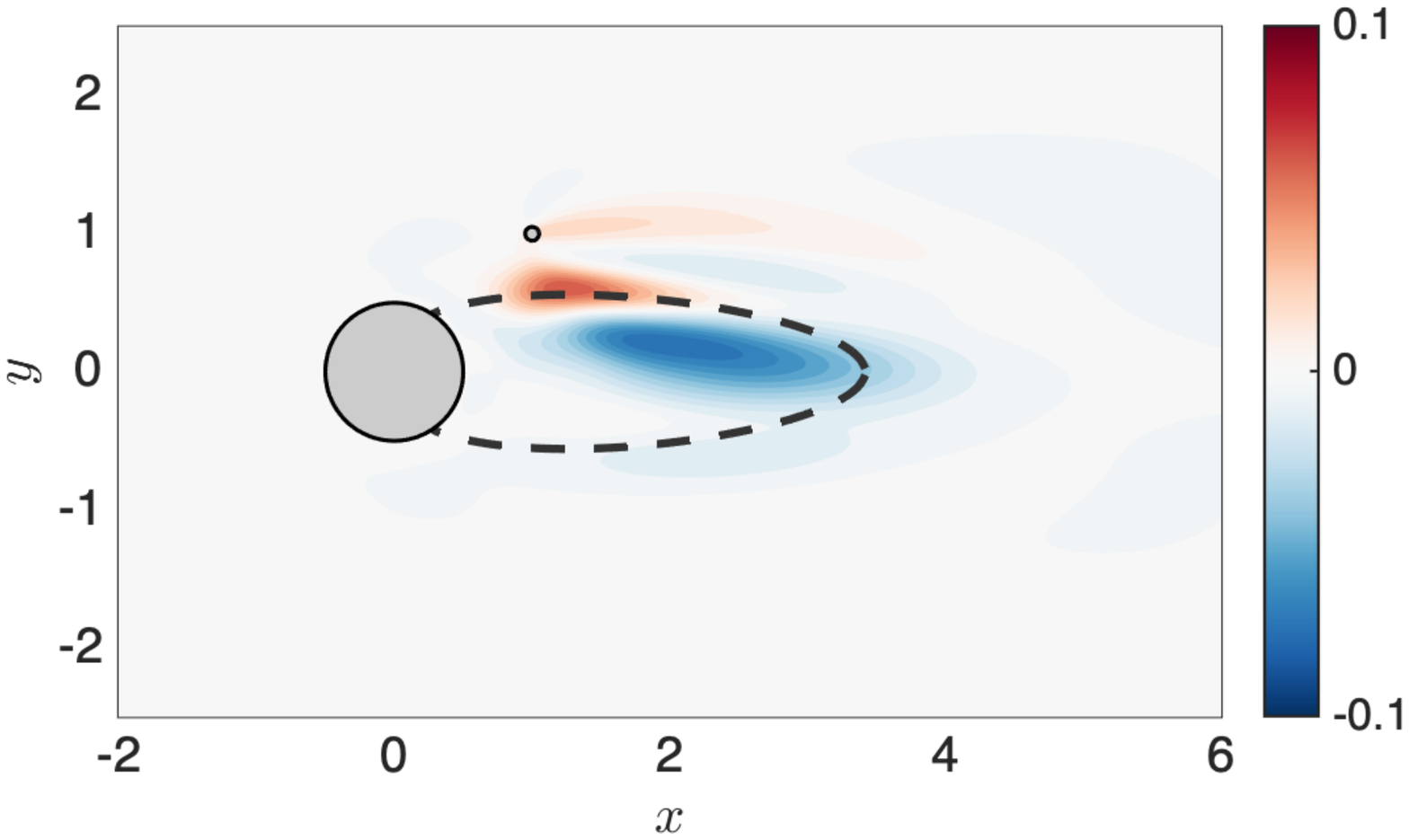}
      \put(-2,57){$(a)$}
   \end{overpic} 
   \hspace{0.2cm}
   \begin{overpic}[width=6.98cm, trim=15mm 125mm 15mm 65mm, clip=true]{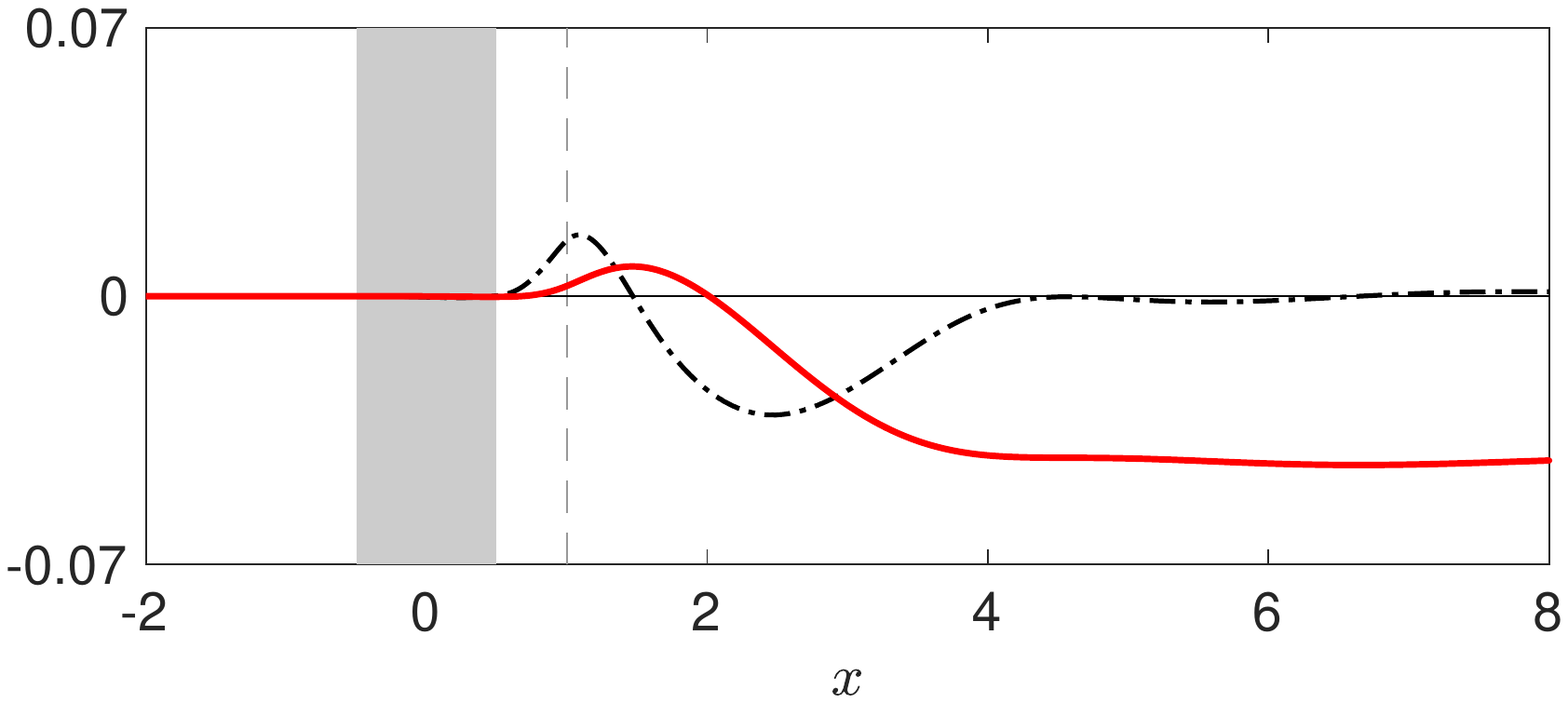}
      \put(-3,51){$(b)$}
      \put(73,34.5){$l'_1(x)$}
      \put(62.3,25){\tcr{$\int_{-\infty}^x l'_1(x') \,\mathrm{d}x'$}}
      \put(81,17.2){\tcr{$\rightarrow \lambda_{1r}$}}
   \end{overpic} 
}
\centerline{
   \hspace{0.2cm}
   \begin{overpic}[width=6.3cm, trim=8mm 15mm 22mm 150mm, clip=true]{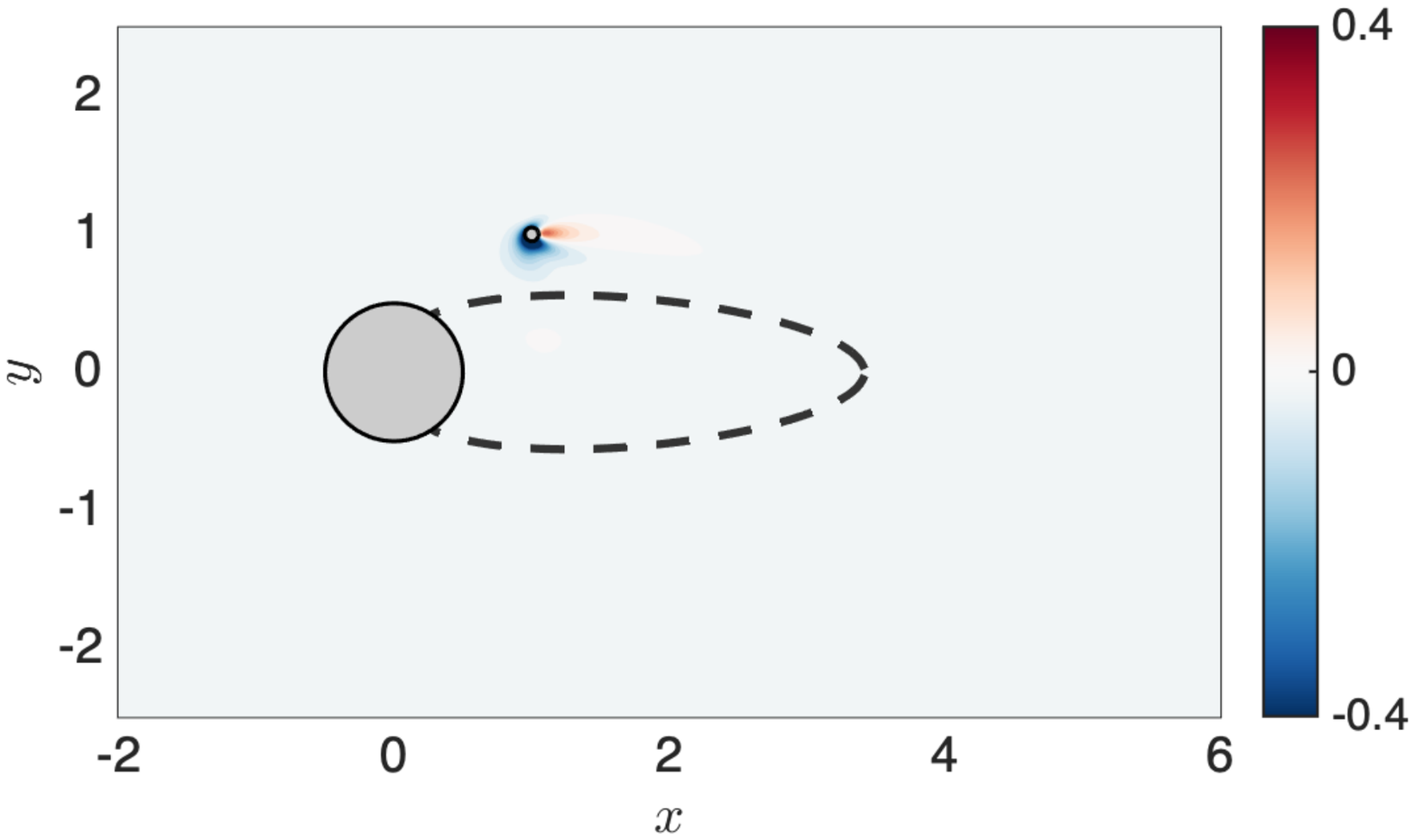}
      \put(-2,57){$(c)$}
   \end{overpic} 
   \hspace{0.2cm}
   \begin{overpic}[width=6.98cm, trim=15mm 125mm 15mm 65mm, clip=true]{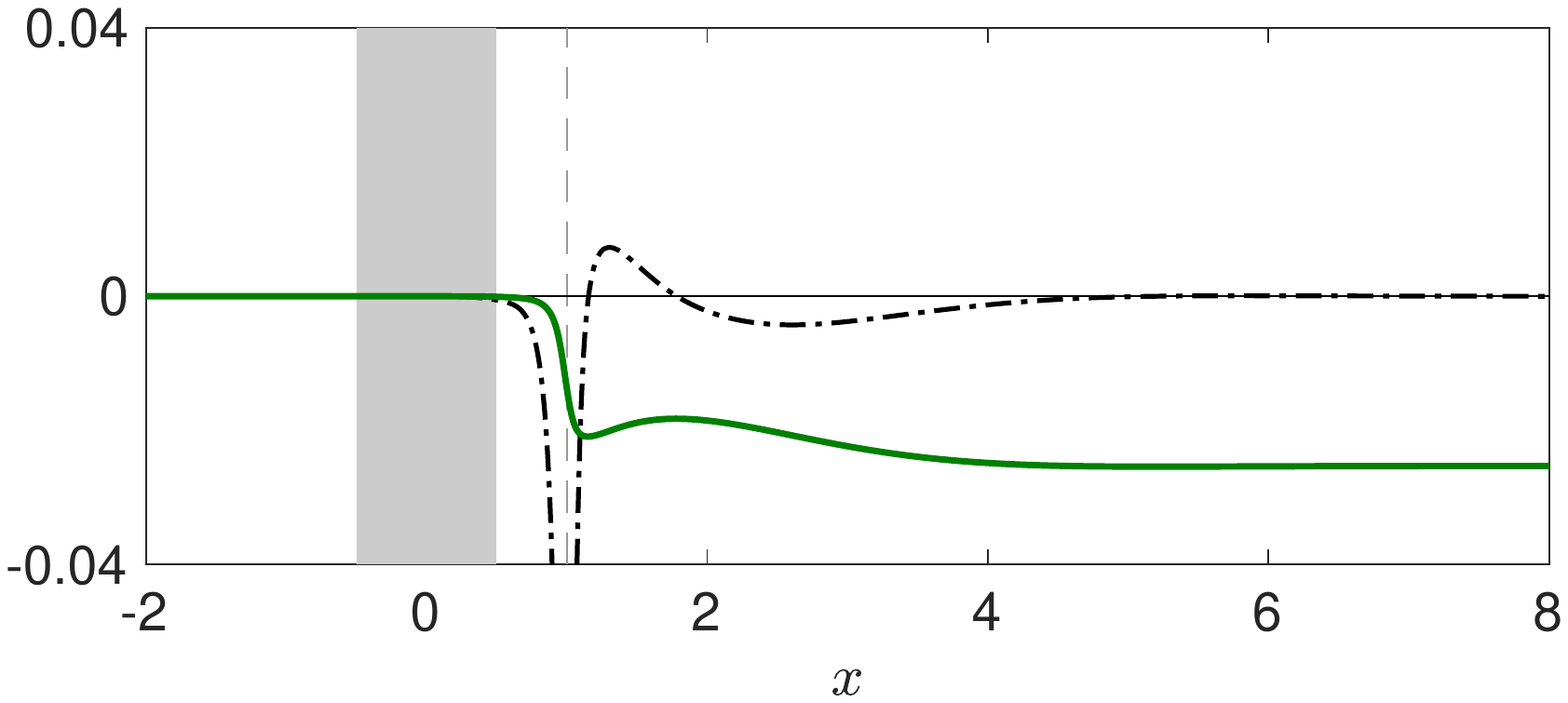}
      \put(-3,51){$(d)$}      
      \put(73.5,35){$l'_{2,\text{I}}(x)$}
      \put(63,24.5){\tcgreen{$\int_{-\infty}^x l'_{2,\text{I}}(x') \,\mathrm{d}x'$}}
      \put(78.5,17.2){\tcgreen{$\rightarrow \lambda_{2r,\text{I}}$}}
   \end{overpic} 
}
\centerline{
   \hspace{0.2cm}
   \begin{overpic}[width=6.3cm, trim=8mm 15mm 22mm 150mm, clip=true]{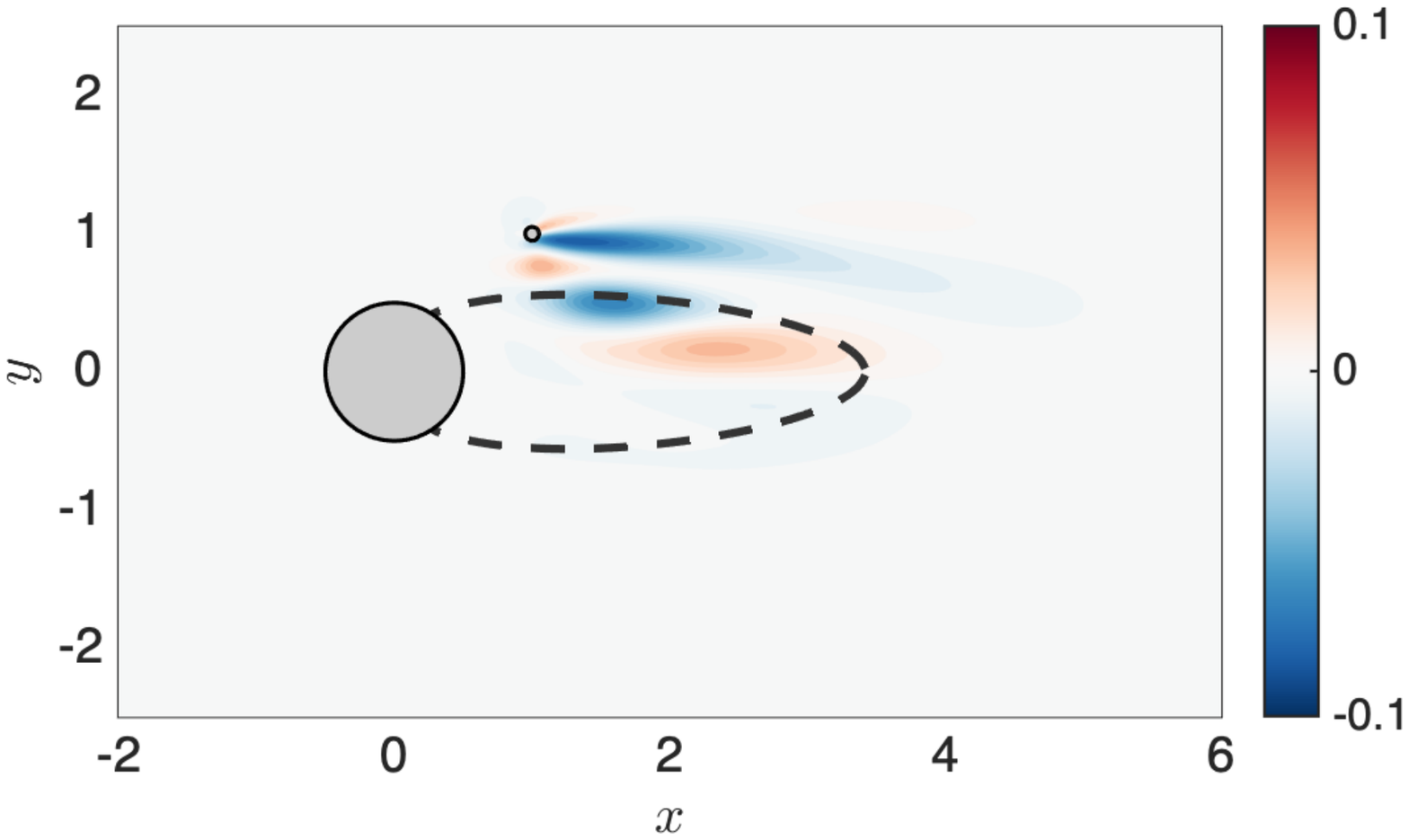}
      \put(-2,57){$(e)$}
   \end{overpic} 
   \hspace{0.2cm}
   \begin{overpic}[width=6.98cm, trim=15mm 125mm 15mm 65mm, clip=true]{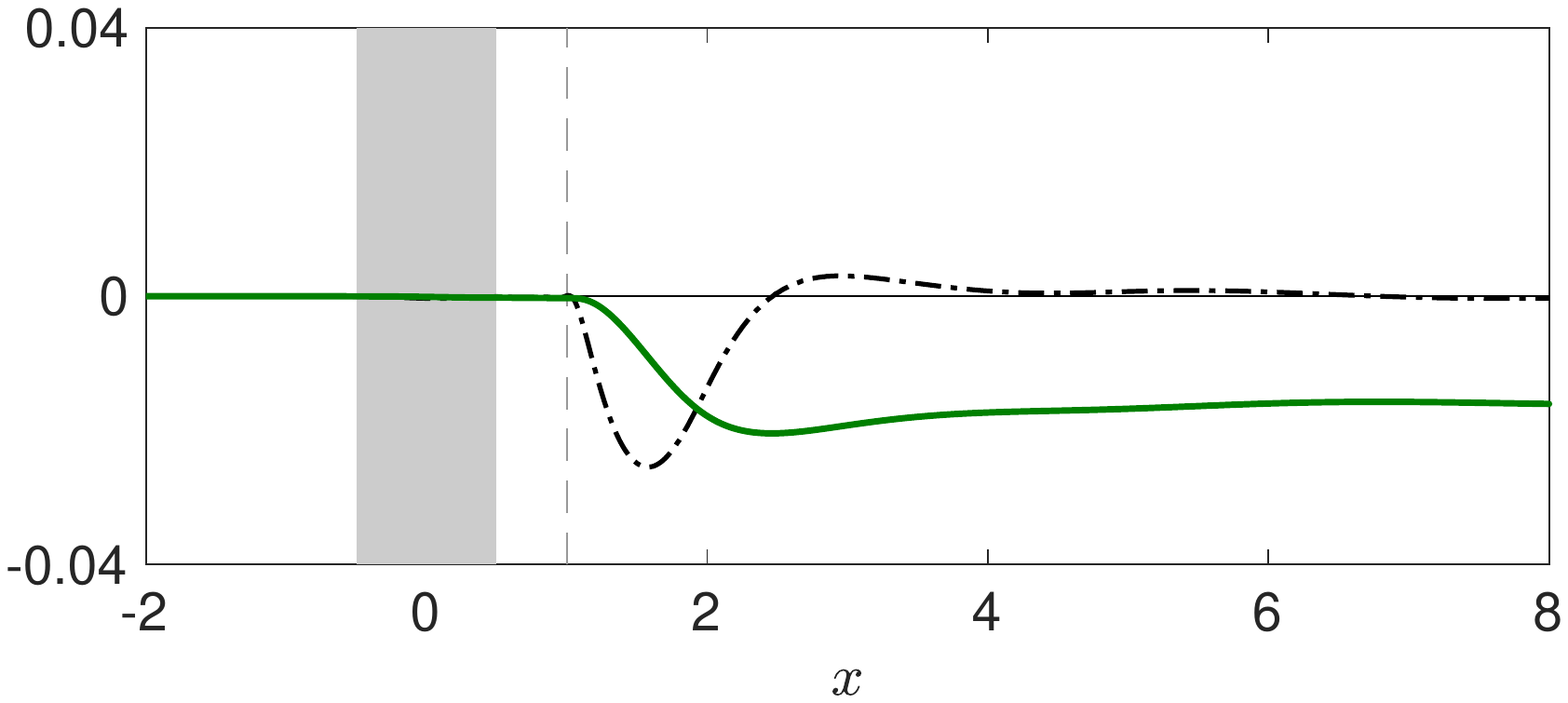}
      \put(-3,51){$(f)$}   
      \put(70.5,35.5){$l'_{2,\text{II}}(x)$}
      \put(60,27.2){\tcgreen{$\int_{-\infty}^x l'_{2,\text{II}}(x') \,\mathrm{d}x'$}}
      \put(77,21.){\tcgreen{$\rightarrow \lambda_{2r,\text{II}}$}}
   \end{overpic} 
}
\caption{
Same as figure \ref{fig:integrands}, for $\lambda_{1r}$ and $\lambda_{2r}$ expressed with sensitivities to first-order base flow modification $\UU_1$.
$(a)$~Integrand of (\ref{eq:l1_2});
$(b)$~density $l'_1(x)$ (black dash-dotted line) and its cumulative integral (red solid line).
$(c)$~Integrand of (\ref{eq:l2_2});
$(d)$~density $l'_{2,\text{I}}(x)$ (black dash-dotted line) and its cumulative integral (green solid line).
$(e)$~Integrand of (\ref{eq:l3_2});
$(f)$~density $l'_{2,\text{II}}(x)$ (black dash-dotted line) and its cumulative integral (green solid line).
}
\label{fig:integrands_sensit}
\end{figure}

It is also possible, and perhaps more informative, to consider  alternative expressions for $\lambda_{1r}$ and $\lambda_{2r}$ where the base flow modification $\UU_1$ appears explicitly. The interested reader is referred to (\ref{eq:first_order_sensit_BFmod}) and (\ref{eq:second_order_sensit_BFmod}) in Appendix~\ref{app:operators}, where the sensitivity operators are derived.
The corresponding integrands are shown in figure~\ref{fig:integrands_sensit}$(a,c,e)$, and the densities
\begin{align}
l'_1(x) &=  \int_{-\infty}^{\infty}  
\text{Re}\left\{
\left( -\overline\LL^\dag \overline\uu_0^\dag \right)
 \bcdot \UU_1  \right\} \mathrm{d}y,
\label{eq:l1_2}
\\
l'_{2,\text{I}}(x) &= \int_{-\infty}^{\infty} 
\text{Re}\left\{ 
\UU_1 \bcdot \left( \KK \UU_1 \right)
\right\} \mathrm{d}y,
\label{eq:l2_2}
\\
l'_{2,\text{II}}(x) &=  \int_{-\infty}^{\infty} 
\text{Re}\left\{
 \UU_1 \bcdot \left( \MM^\dag (\lambda_0\II+\AAA_0)^{-1} \TT \UU_1 \right)
\right\} \mathrm{d}y,
\label{eq:l3_2}
\end{align}
in figure~\ref{fig:integrands_sensit}$(b,d,f)$.
The density $l'_1(x)$ is qualitatively similar to  $l_1(x)$: positive around $x_c$ and negative in a longer region  downstream. 
The two-dimensional integrand, however, exhibits a more complicated structure with alternating positive regions (separatrix and control cylinder wake) and negative regions (especially the recirculation region). This reveals that the main first-order stabilising contribution in terms of flow modification $\UU_1$ comes from the inside the recirculation region, not directly from the control cylinder wake. Again, this is consistent with the observations of \cite{Marquet08cyl}.
Turning now to second order, it appears that $l'_{2,\text{I}}$ and $l'_{2,\text{II}}$ are mostly negative or zero, and only marginally positive.
The integrand of $\lambda_{2r,\text{I}}$ is strongly negative immediately upstream of the control cylinder, while the integrand of $\lambda_{2r,\text{II}}$ is negative in the control cylinder wake and along the separatrix.
Therefore, the main second-order stabilising contribution from the flow modification (quadratic effect of $\UU_1$) comes directly from the control cylinder and its wake.

\section{Optimal control}
\label{sec:opt}

Previous sections have investigated the second-order sensitivity to given, localised controls.
One may wonder about how to design an optimal distributed control $\FF^{opt}$ so as to maximise the growth rate reduction.
This section first recalls how to compute optimal controls  targeting separately the first- and second-order variations, $\lambda_{1r}$ and $\lambda_{2r}$, 
and then presents a method for computing the optimal control targeting at once the total second-order  variation, $\epsilon \lambda_{1r} + \epsilon^2 \lambda_{2r}$. 
This method, borrowed from general linear algebra and applied mathematics, seems rather new in the field of hydrodynamic stability.

\subsection{Optimising first- and second-order variations separately}
\label{sec:opt_1_or_2}

When only the first-order variation $\lambda_{1r}$ is considered, the optimal unit control is proportional to the sensitivity itself \citep{Bottaro03, Boujo15secondorderJFM}:
\begin{eqnarray}
\FF_1^{opt}
= \argminA_{||\FF||=1} \left\{ \lambda_{1r} \right\}
= \argminA_{||\FF||=1} \ps{\SS_{1r}}{\FF}
= -\frac{\SS_{1r}}{||\SS_{1r}||}.
\label{eq:Fopt_1}
\end{eqnarray}
This classical result can be obtained with a Lagrangian method, or as a direct consequence of the Cauchy-Schwarz inequality becoming an equality for two linearly dependent vectors.

When  only the second-order variation $\lambda_{2r}$ is considered (which is relevant when $\lambda_{1r}=0$, for example for the spanwise-periodic control of spanwise-invariant flows), 
the largest growth rate reduction is
\begin{eqnarray}
\min_{||\FF||=1} \left\{ \lambda_{2r} \right\}
= \min_{||\FF||=1} \ps{ \FF}{\SS_{2r} \FF}
=  \min_{||\FF||}   \frac{ \ps{ \FF}{ \frac{1}{2} \left(\SS_{2r}+\SS_{2r}^T\right) \FF} }{\ps{\FF}{\FF}},
\label{eq:Fopt_2}
\end{eqnarray}
i.e. the optimal unit control $\FF_2^{opt}$ is the eigenvector associated with the smallest eigenvalue $\mu$ of the following symmetric eigenvalue problem 
 \citep{Boujo15secondorderJFM, Boujo19sensit}:
\begin{eqnarray} 
\frac{1}{2} \left(\SS_{2r}+\SS_{2r}^T\right) \FF = \mu \FF.
\label{eq:Fopt_3}
\end{eqnarray}

\subsection{Optimising the total second-order variation}
\label{sec:opt_1_and_2}

If now the total second-order variation is to be minimised,
\begin{eqnarray}
\min_{||\FF||=1}  \left\{\epsilon \lambda_{1r} + \epsilon^2 \lambda_{2r} \right\}
=
\min_{||\FF||=1}  \left\{
\epsilon \ps{\SS_{1r}}{\FF} + \epsilon^2 \ps{\FF}{\SS_{2r}\FF}  \right\},
\label{eq:constr_min}
\end{eqnarray}
one can introduce the Lagrangian
\begin{eqnarray}
\mathcal{L}(\FF,\beta) 
&= \epsilon \ps{\SS_{1r}}{\FF} + \epsilon^2 \ps{\FF}{\SS_{2r}\FF} - \beta \left[  \ps{\FF}{\FF} -1 \right], 
\end{eqnarray}
where $\beta$ is  an as yet unknown Lagrange multiplier enforcing the normalisation $||\FF||=1$. 
From the stationarity condition 
$\partial\mathcal{L}/\partial \FF=\00,$ one obtains the following equation for the optimal unit control $\FF_{1+2}^{opt}$:
\begin{eqnarray}
 \epsilon^2 \left(\SS_{2r}+\SS_{2r}^T\right) \FF - 2\beta \FF  = -\epsilon\SS_{1r}.
\label{eq:pb_F}
\end{eqnarray}
One can verify that: 
(i)~in the limit of small control amplitudes, $\epsilon \ll 1$,
the optimal control reduces to $\FF_{1}^{opt}$  proportional to $\SS_{1r}$, as given by (\ref{eq:Fopt_1});
(ii)~in the limit of vanishing first-order sensitivity, $\SS_{1r}=0$,
the optimal control reduces to the
$\FF_{2}^{opt}$ solution of an eigenvalue problem equivalent to (\ref{eq:Fopt_3}). 
In both cases, $\FF_1^{opt}$ and $\FF_2^{opt}$ are independent of the control amplitude $\epsilon$.

In general, however, equation (\ref{eq:pb_F}) for $\FF_{1+2}^{opt}$ is neither a linear system nor an eigenvalue problem,
and $\FF_{1+2}^{opt}$
depends on the amplitude $\epsilon$ considered.
Together with the associated constrained minimisation problem (\ref{eq:constr_min}),
it appears in some least-squares problems, constrained eigenvalue problems and trust-region problems.
It has been studied extensively in the literature, and  several solution techniques are available.
For instance, \cite{GANDER1989} give an iterative method based on solving a so-called explicit secular equation for $\beta$,
but it involves  a full diagonalisation of the operator $\left(\SS_{2r}+\SS_{2r}^T\right)$, which is not tractable in the present study.
Another method consists in finding the smallest $\beta$ solution of the 
implicit secular equation
\begin{align}
\epsilon^2 \SS_{1r}^T \left[ \epsilon^2 
\left( \SS_{2r}+\SS_{2r}^T \right)
-2\beta \II \right]^{-2} \SS_{1r} - 1
= 0.
\end{align}
In either case, the optimal control $\FF_{1+2}^{opt}$ is obtained by substituting the obtained value of $\beta$ in (\ref{eq:pb_F}).

Here, yet another approach from \cite{GANDER1989} is used. 
First, $\FF$ is expressed from (\ref{eq:pb_F}) as 
$  \FF  = - \left[ \epsilon^2 \left(\SS_{2r}+\SS_{2r}^T\right)  - 2\beta \II \right]^{-1} \epsilon\SS_{1r}$,
and the (unit) norm of $\FF$ becomes
\begin{align}
\FF^T \FF 
= 1
=
\epsilon \SS_{1r}^T \left[ \epsilon^2 \left(\SS_{2r}+\SS_{2r}^T\right)  - 2\beta \II \right]^{-2} \epsilon \SS_{1r}
\end{align}
because the operator in square brackets is symmetric.
Second, defining the vector 
\begin{align}
\boldsymbol{\gamma} 
=  \left[\epsilon^2 \left(\SS_{2r}+\SS_{2r}^T\right)  - 2\beta \II \right]^{-1}  \FF
= - \left[\epsilon^2 \left(\SS_{2r}+\SS_{2r}^T\right)  - 2\beta \II \right]^{-2} \epsilon \SS_{1r},
\label{eq:gamma_def}
\end{align}
one can write 
$\FF^T \FF = 1 = - \epsilon \SS_{1r}^T \boldsymbol{\gamma}$
and
\begin{align}
 \left[\epsilon^2 \left(\SS_{2r}+\SS_{2r}^T\right)  - 2\beta \II \right]^{2}\boldsymbol{\gamma} 
&= - \epsilon \SS_{1r}
= - \epsilon \SS_{1r} \FF^T \FF,
\end{align}
finally obtaining 
the quadratic eigenvalue problem
\begin{align}
 \left[\epsilon^2 \left(\SS_{2r}+\SS_{2r}^T\right)  - 2\beta \II \right]^{2}\boldsymbol{\gamma} 
= \epsilon^2 \SS_{1r} \SS_{1r}^T \boldsymbol{\gamma},
\end{align}
to be solved for the smallest eigenvalue $\beta$.
The associated eigenvector $\boldsymbol{\gamma}$
yields the optimal control $\FF_{1+2}^{opt}$  via (\ref{eq:gamma_def}).
In practice, the quadratic eigenvalue problem is transformed into an equivalent linear one,
\begin{align}
\left[ 
\begin{array}{cc}
\epsilon^2 \left(\SS_{2r}+\SS_{2r}^T\right) & -\II\\
-  \epsilon^2 \SS_{1r} \SS_{1r}^T & \epsilon^2 \left(\SS_{2r}+\SS_{2r}^T\right)
\end{array}
\right]
\left( \begin{array}{c}
\boldsymbol{\gamma} \\ \FF
\end{array}\right)
= 
2\beta
\left( \begin{array}{c}
\boldsymbol{\gamma} \\ \FF
\end{array}\right),
\label{eq:quadratic_evp}
\end{align}
which has twice the dimension but can be solved with standard methods.

As mentioned earlier, 
 the optimal control $\FF_{1+2}^{opt}$ is a function of the amplitude considered 
 because  $\epsilon$ is a parameter of the optimisation problem (\ref{eq:quadratic_evp}), which one is free to choose.  
 In the following, let us denote by $\epsilon^*$ the optimisation amplitude.
 A given optimisation amplitude $\epsilon^*$ yields an  optimal unit control $\FF_{1+2}^{opt}$, and  the associated values $\lambda_{1r}$ and $\lambda_{2r}$.
 The control $\epsilon^*\FF_{1+2}^{opt}$ is therefore optimal for this amplitude.
 When implementing this optimal unit control with another amplitude $\epsilon \neq \epsilon^*$, the second-order effect of $\epsilon \FF_{1+2}^{opt}$ will be 
 $\lambda_r = \lambda_{0r} + \epsilon \lambda_{1r} + \epsilon^2 \lambda_{2r}$.
 By construction, this effect will be optimal only for $\epsilon = \epsilon^*$.

\begin{figure} 
\vspace{0.5cm}
\centerline{
   \hspace{-0.85cm}
   \begin{overpic}[height=5cm, trim=5mm 60mm 20mm 85mm, clip=true]{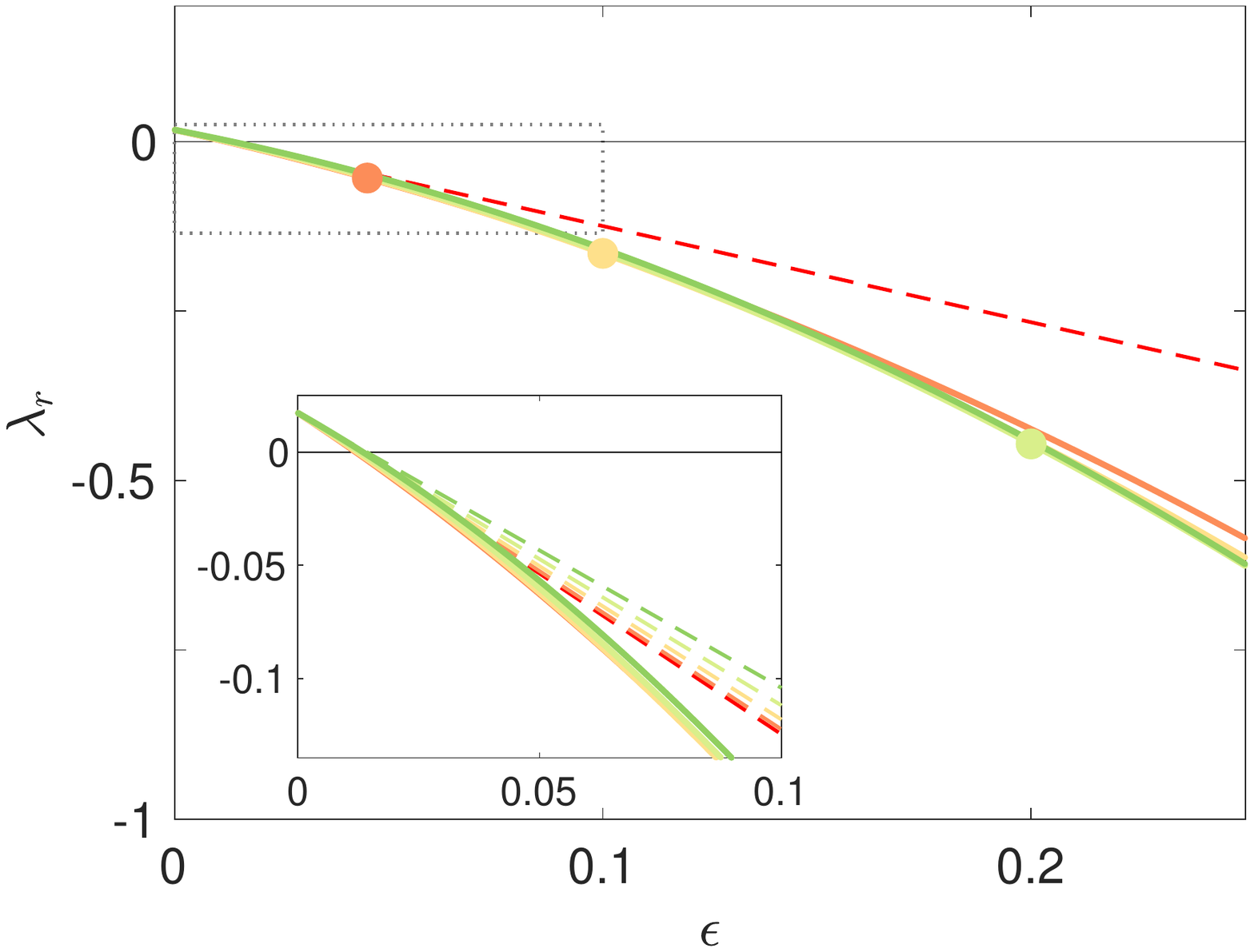}   
      \put(0,81){$(a)$}  
   \end{overpic} 
   \hspace{0.32cm}
   \begin{overpic}[height=5cm, trim=5mm 60mm 20mm 85mm, clip=true]{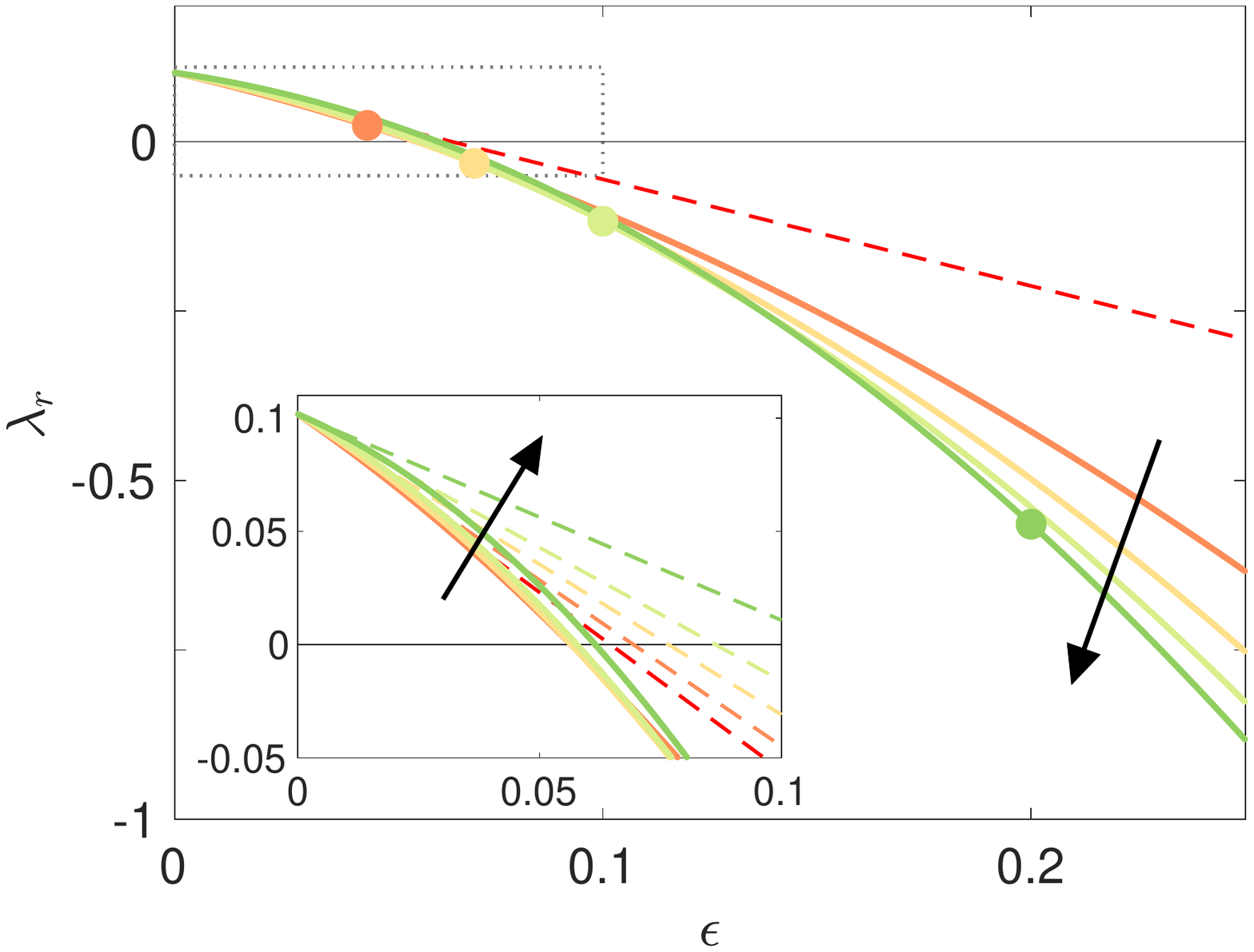} 
      \put(0,81){$(b)$}  
      \put(44.5,47.){\footnotesize $\epsilon^*$}
      \put(78,  29){\footnotesize $\epsilon^*$}
   \end{overpic}  
}
\vspace{-0.4cm}
\centerline{
   \begin{overpic}[width=6.2cm, trim=10mm 25mm 25mm 140mm, clip=true]{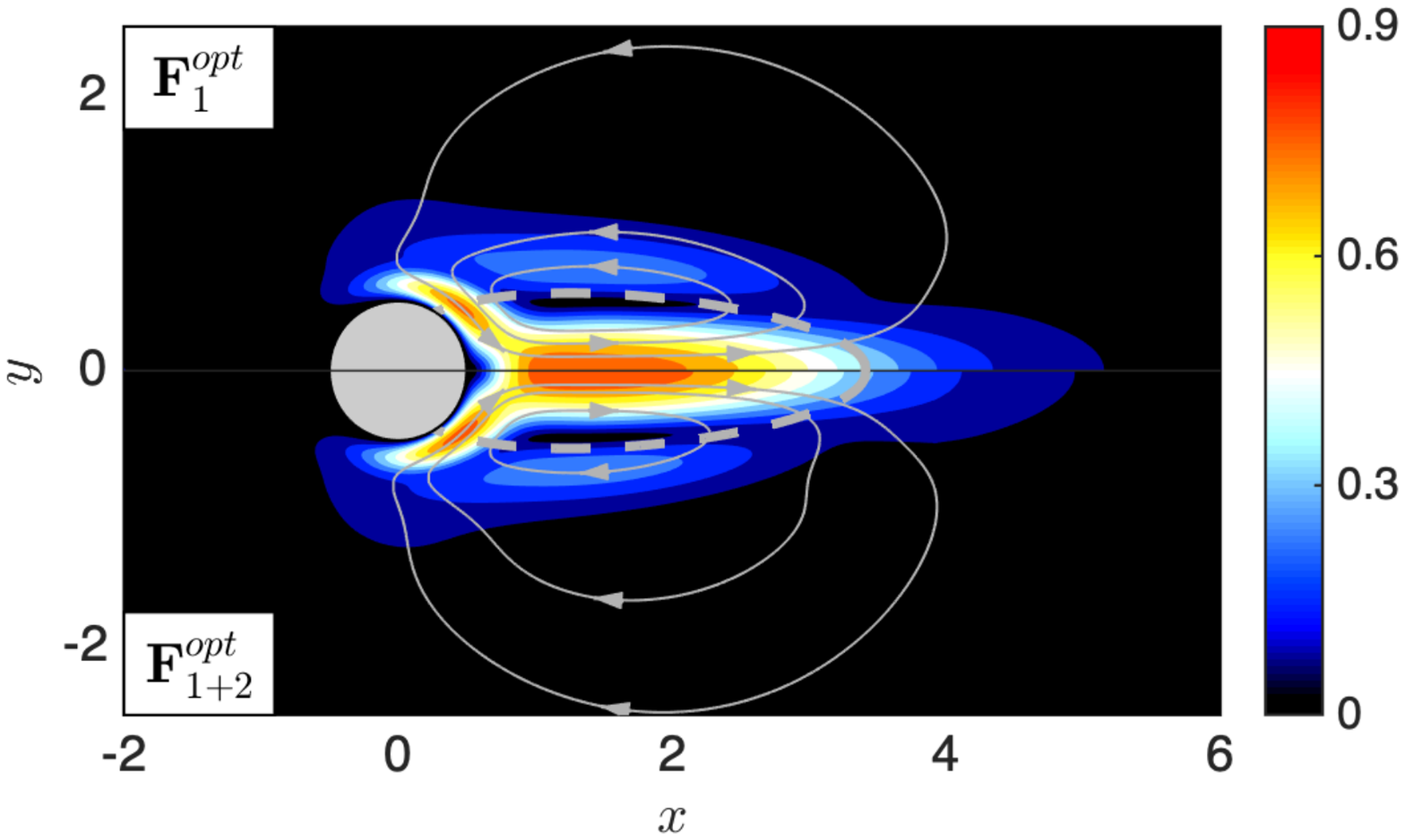}   
      \put(-5,58){$(c)$}  
   \end{overpic} 
   \hspace{0.2cm}
   \begin{overpic}[width=6.2cm, trim=10mm 25mm 25mm 140mm, clip=true]{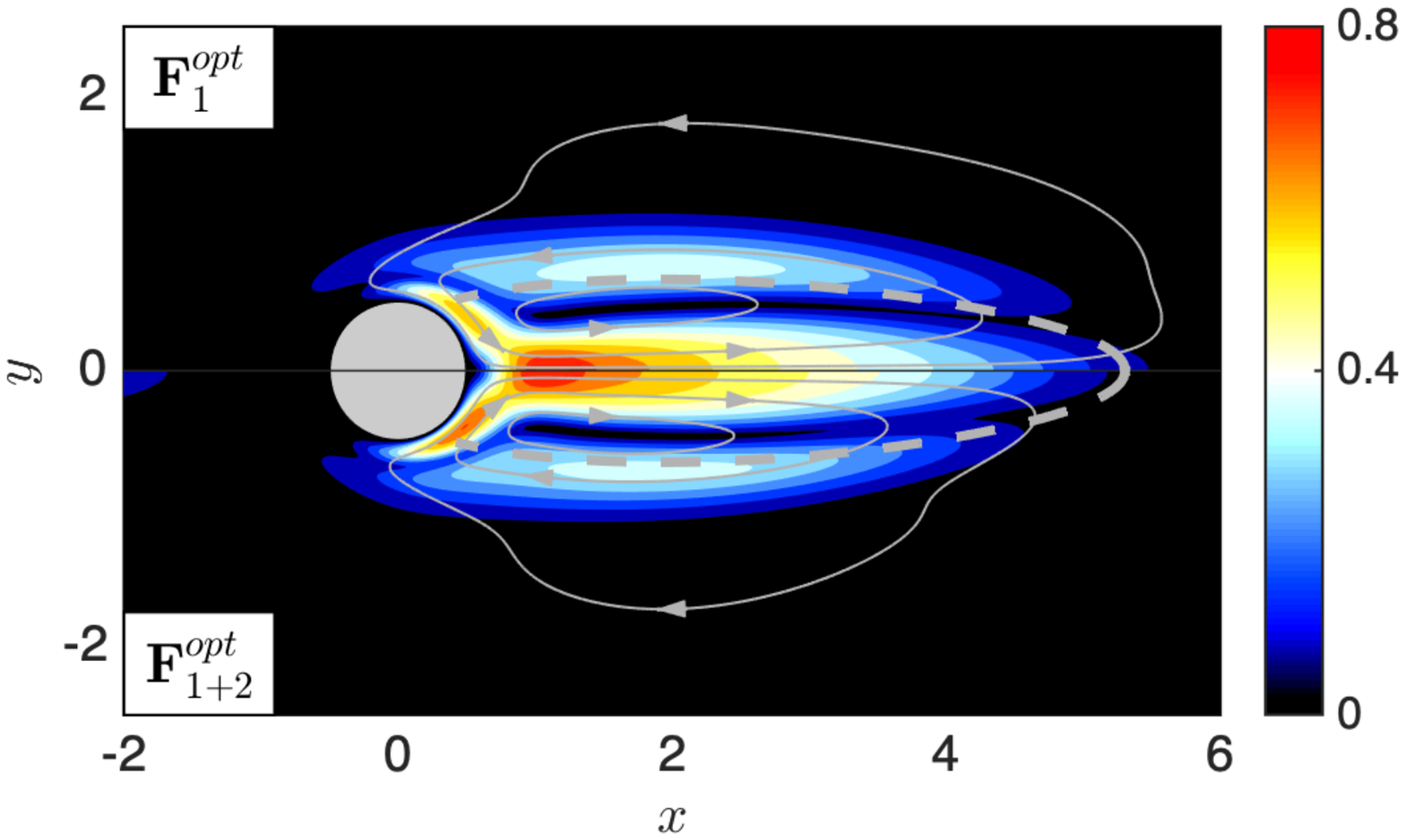}   
      \put(-5,58){$(d)$}  
   \end{overpic}
}
\centerline{   
\begin{overpic}[width=6.2cm, trim=10mm 25mm 25mm 140mm, clip=true]{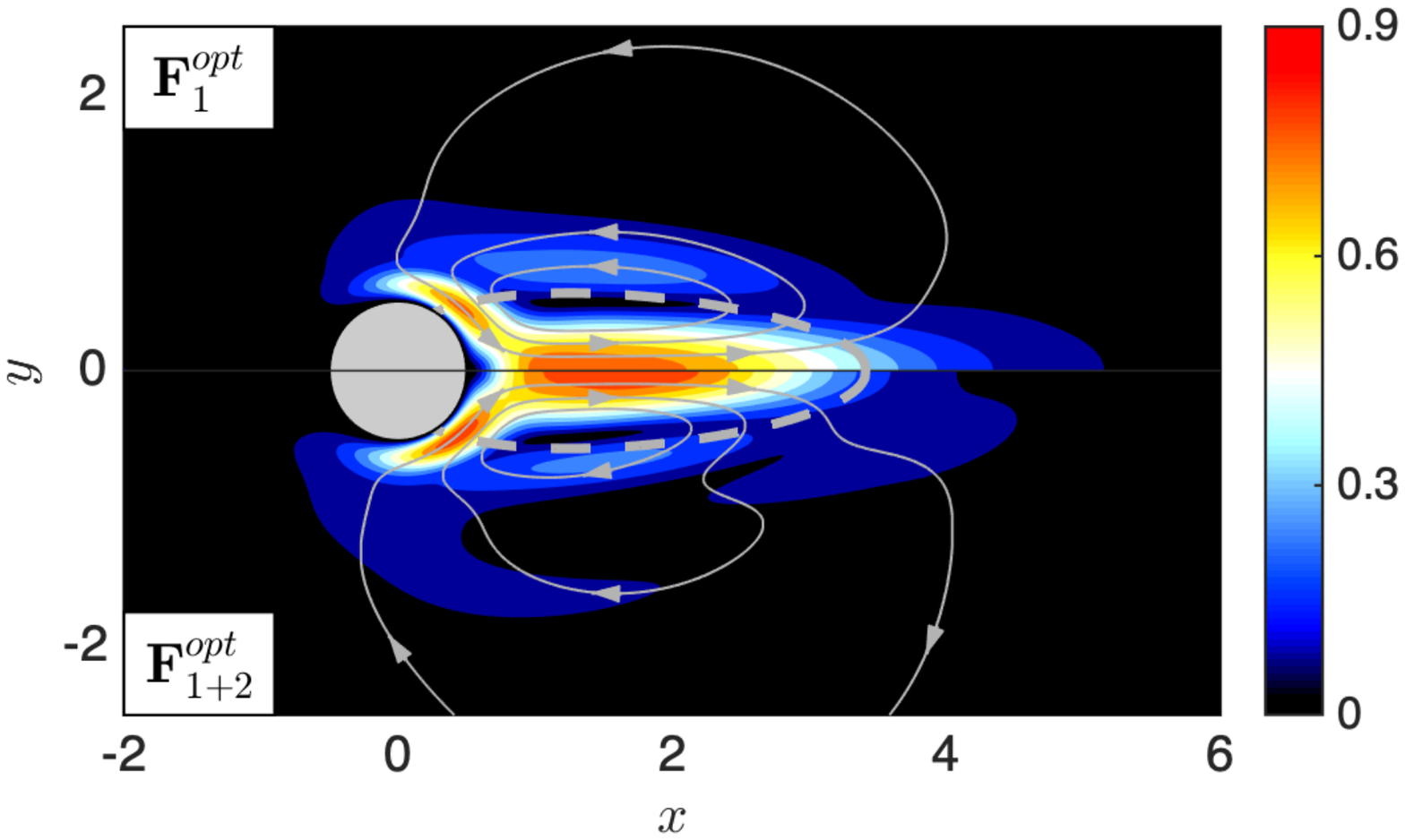}   
      \put(-5,58){$(e)$}   
   \end{overpic} 
   \hspace{0.2cm}
   \begin{overpic}[width=6.2cm, trim=10mm 25mm 25mm 140mm, clip=true]{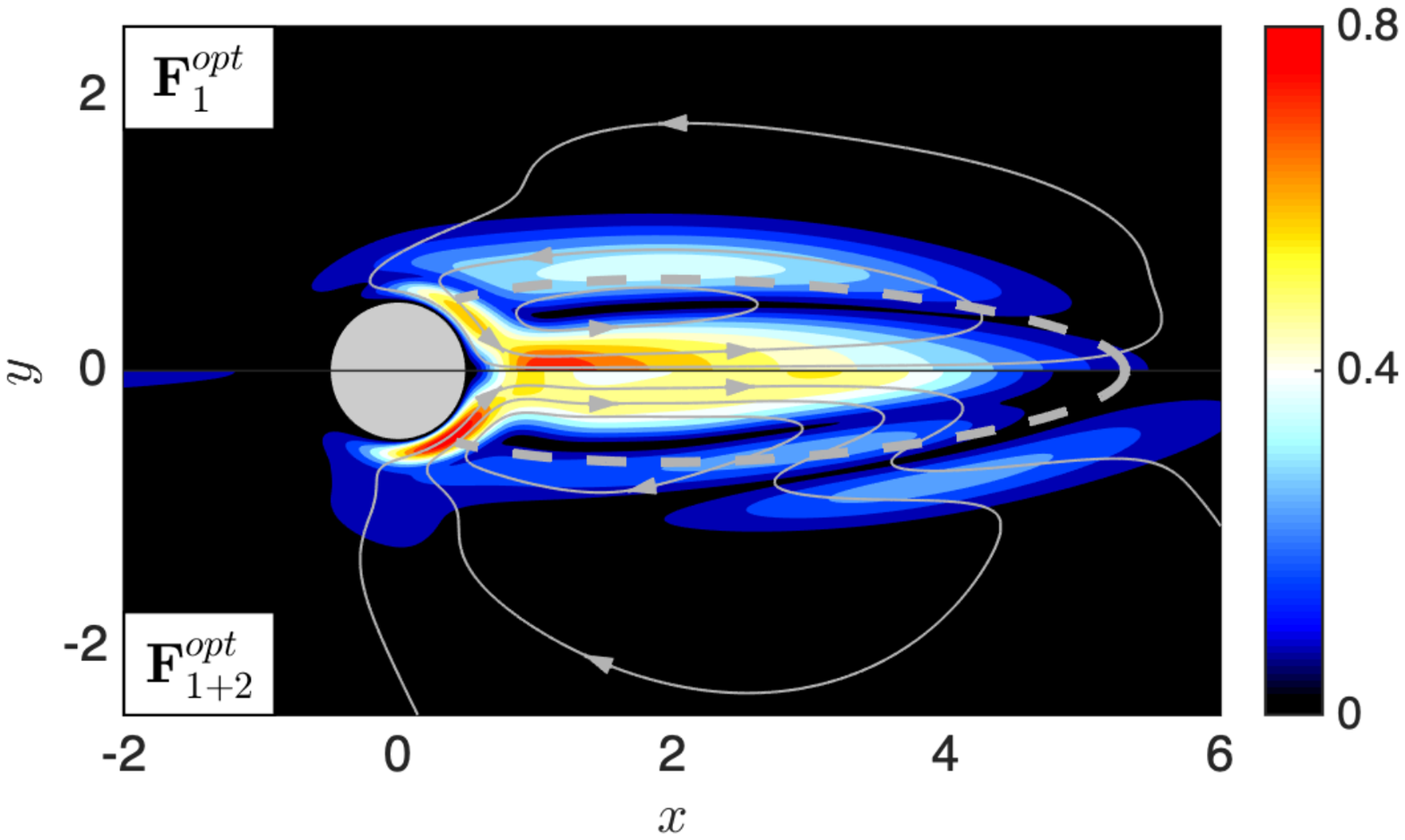}   
      \put(-5,58){$(f)$}  
   \end{overpic} 
}
\centerline{
   \begin{overpic}[width=6.2cm, trim=10mm 25mm 25mm 140mm, clip=true]{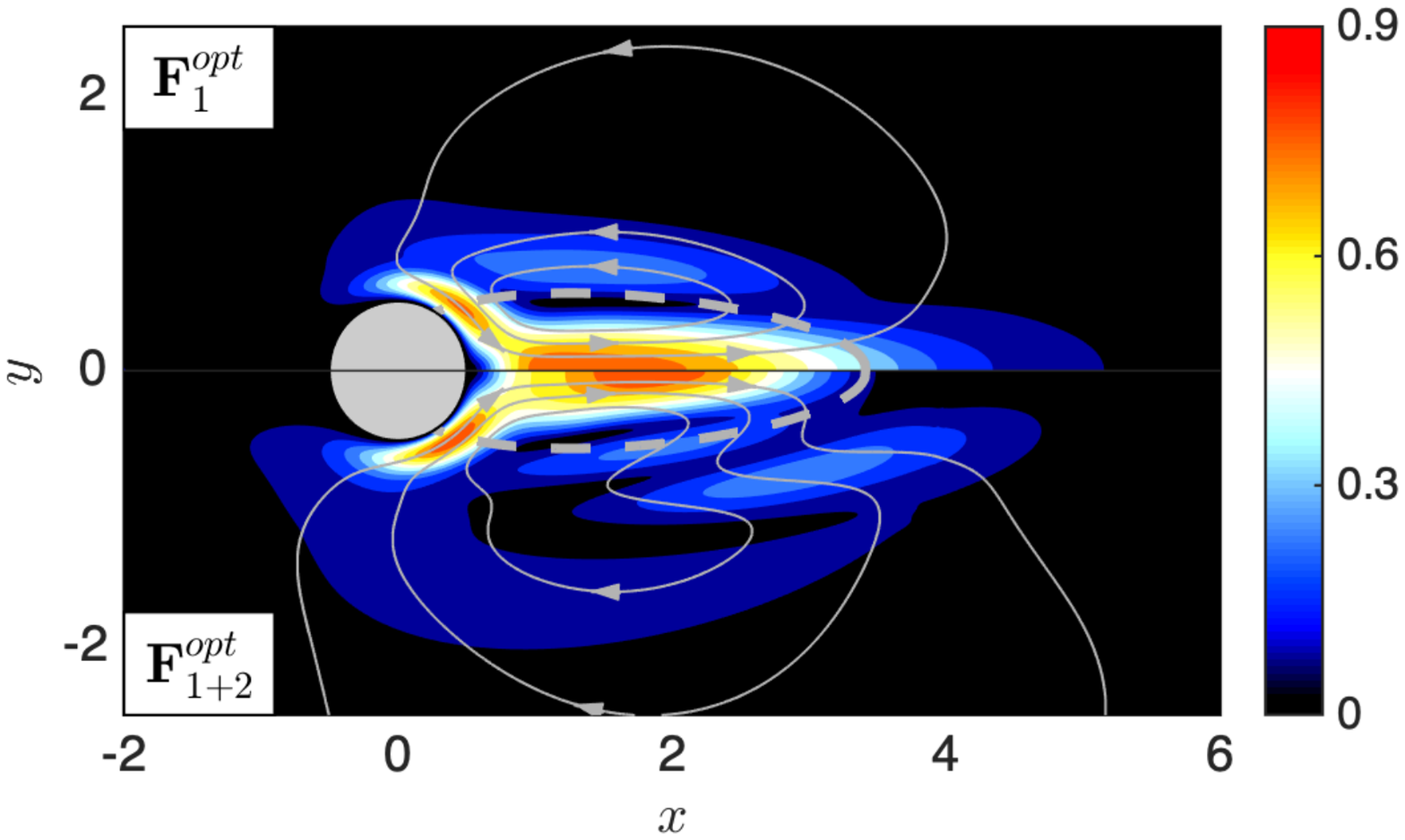}   
      \put(-5,58){$(g)$} 
   \end{overpic} 
   \hspace{0.2cm}
   \begin{overpic}[width=6.2cm, trim=10mm 25mm 25mm 140mm, clip=true]{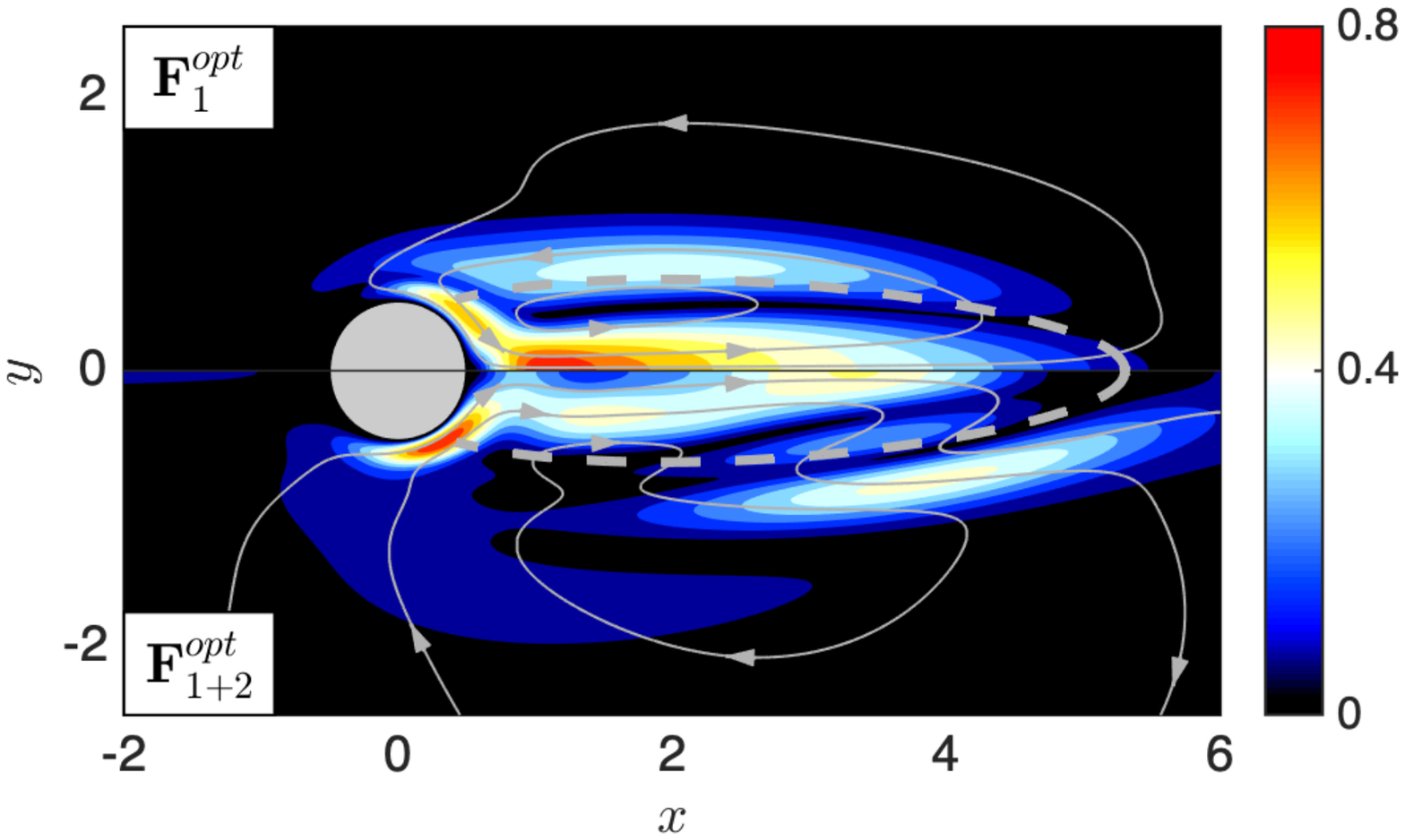}   
      \put(-5,58){$(h)$}  
   \end{overpic} 
}
\caption{
Optimisation of the total second-order variation.
$(a,b)$~Quadratic variation  of the leading growth rate
$\lambda_{0r}+\epsilon\lambda_{1r}+\epsilon^2\lambda_{2r}$ induced by the optimal control 
$\epsilon \FF_{1+2}^{opt}$.
Each solid line corresponds to a  different optimisation amplitude $\epsilon^*$ (symbols).
Dashed line:  linear variation for the first-order optimal $\epsilon \FF_{1}^{opt}$.
Inset: 
close-up of the small-amplitude region, also showing the linear variations (slopes in $\epsilon=0$).
$(c)$-$(h)$~Optimal unit control for first-order growth rate variation only ($\epsilon^*=0$, upper half) and for total first- and second-order growth rate variation ($\epsilon^*>0$, lower half).
Colour, magnitude; streamlines, local orientation.
Optimisation amplitude: 
$(c,d)$~$\epsilon^*=0.02$,
$(e,f)$~$\epsilon^*=0.1$, and
$(g,h)$~$\epsilon^*=0.5$.
Reynolds number: $(a,c,e,g)$ $\Rey=50$, and
$(b,d,f,h)$ $\Rey=80$.
}
\label{fig:Fopt_1_2}
\end{figure}

Figure \ref{fig:Fopt_1_2}$(a,b)$ compares the linear variation of the leading growth obtained with the first-order optimal control $\FF_1^{opt}$ (dashed line), and the quadratic variation obtained with the total second-order optimal control $\FF_{1+2}^{opt}$  (solid lines)
for several optimisation amplitudes  $\epsilon^*$ (symbols).
In all cases, the second-order effect is stabilising ($\lambda_{2r}<0$).
For $\Rey=50$ (figure \ref{fig:Fopt_1_2}$a$), 
changing $\epsilon^*$ makes little difference.
For $\Rey=80$ (figure \ref{fig:Fopt_1_2}$b$), however, 
the impact of $\epsilon^*$ is clearly visible:
controls optimised for larger amplitudes $\epsilon^*$ perform better at large $\epsilon$, but worse at small $\epsilon$ (see inset).
This highlights the flexibility of the method, which allows one to select a  control amplitude and optimise for that specific amplitude.

Figure \ref{fig:Fopt_1_2}$(c)$-$(h)$ shows the unit optimal control for several values of $\epsilon^*$, at $\Rey=50$ (left column) and $\Rey=80$ (right column).
Each panel compares  the first-order  optimal control $\FF_1^{opt}$  (upper half), and
the total second-order optimal control
$\FF_{1+2}^{opt}$  (lower half).
Contours show the magnitude of the vector $\FF$,  streamlines show its orientation.
For small  amplitudes, $\FF_{1+2}^{opt}$ is very similar to $\FF_{1}^{opt}$, as  seen in figure \ref{fig:Fopt_1_2}$(c,d)$ for $\epsilon^*=0.02$.
As the optimisation amplitude increases ($\epsilon^*=0.1$ in figure \ref{fig:Fopt_1_2}$e$,$f$
and
$\epsilon^*=0.5$ in figure \ref{fig:Fopt_1_2}$g$,$h$), 
the optimal control becomes 
weaker in the recirculation region and immediately outside, and stronger in a new area outside the recirculation region, while its overall orientation is preserved.
For finite control amplitudes, making small changes to a control may thus be important, as this can improve its second-order effect.

\section{Conclusion}

A second-order sensitivity operator has been derived and used to predict quadratic eigenvalue variations induced by flow control. 
Introducing suitable adjoint operators, this second-order sensitivity  is made independent of the control.
First- and second-order sensitivity maps have been obtained for the control of the cylinder wake with a steady body force and a model of a small control cylinder, at a much lower computational cost than by recomputing nonlinear controlled flows and eigenmodes.
Considering finite-amplitude control, the range of validity of the first-order sensitivity is  characterised  with a map of `threshold amplitude'.
Regions where the first-order sensitivity underestimates or overestimates the eigenvalue variation up to second order are also conveniently visualised with another dedicated map.
The effect of a small control cylinder tends to be underestimated, such that regions where the flow is fully restabilised become larger when including second-order effects at all the Reynolds numbers investigated.
Decomposing the second-order variation   into two contributions (second-order base flow modification, and interaction between first-order base flow and eigenmode modifications, respectively) reveals that both contributions are equally important in the most sensitive regions.
Analysing the  effect of a small control cylinder located nearly optimally shows that stabilising effects arise from flow modifications in different regions:  inside the recirculation region for first-order stabilisation, immediately upstream of the control cylinder and in its wake for second-order stabilisation.

Finally, with the second-order sensitivity operator available, the optimal control (distributed body force) for stabilisation up to second order is computed.
While the  first-order optimal control is directly proportional to the first-order sensitivity (and independent of the control amplitude), the total second-order optimal control is obtained via a quadratic eigenvalue problem and depends on the amplitude.
As the  amplitude increases, this control becomes  stronger on the sides of the cylinder and the recirculation region, and weaker inside the recirculation region.
Therefore, given a desired amplitude, it is possible to fine-tune the control.

While first-order sensitivity perfectly captures the effect of infinitesimal control on linear stability properties,
this study shows that adjoint-based second-order sensitivity provides a range of useful information for finite-amplitude control, at little extra computational cost.
At some locations in the cylinder flow, e.g. in the shear layers, it seems that higher-order terms would improve the sensitivity prediction. 
This has not been investigated systematically in the present study, so several questions remain open, including the following ones:
In which regions are higher-order effects $\lambda_n$ stronger? 
How does the radius of convergence $r$ of the power expansion vary in space?
Is it possible to relate spatial distributions of $\lambda_n$ and $r$ to any physical mechanism, in this and other flows? 

The present approach can easily be extended to other types of control such as wall blowing/suction and shape deformation.
It is expected to be useful for the passive control of other globally unstable flows, and may  be applied to stable flows too since the resolvent gain (amplification of time-harmonic perturbations) can be expressed as an eigenvalue problem and treated in a similar framework.
It could also be used to speed up the convergence of  gradient-based optimisation when iteratively designing  practical controls aiming for flow stabilisation.

%
\section*{Declaration of Interests}

The author reports no conflict of interest.

\appendix
\section{Second-order sensitivity of the frequency}
\label{app:freq}

Section~\ref{sec:results} focused on the sensitivity of the leading growth rate $\lambda_r$.
For completeness, the sensitivity of the leading mode's frequency $\lambda_i$ is given here.

\subsection{Sensitivity to a steady body force}

\begin{figure} 
\centerline{
   \hspace{0.3cm}
   \begin{overpic}[width=6.2cm, trim=10mm 40mm 10mm 170mm, clip=true]{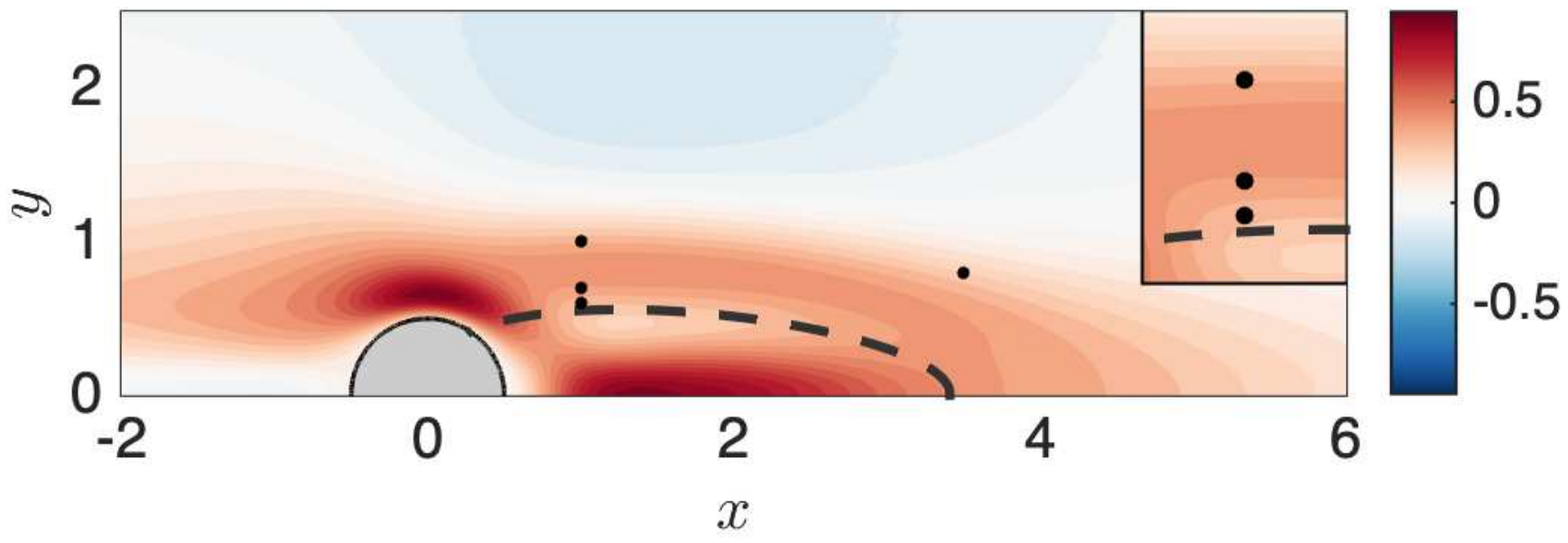}
      \put(-2,30){$(a)$}
   \end{overpic}  
   \begin{overpic}[width=6.2cm, trim=10mm 40mm 10mm 170mm, clip=true]{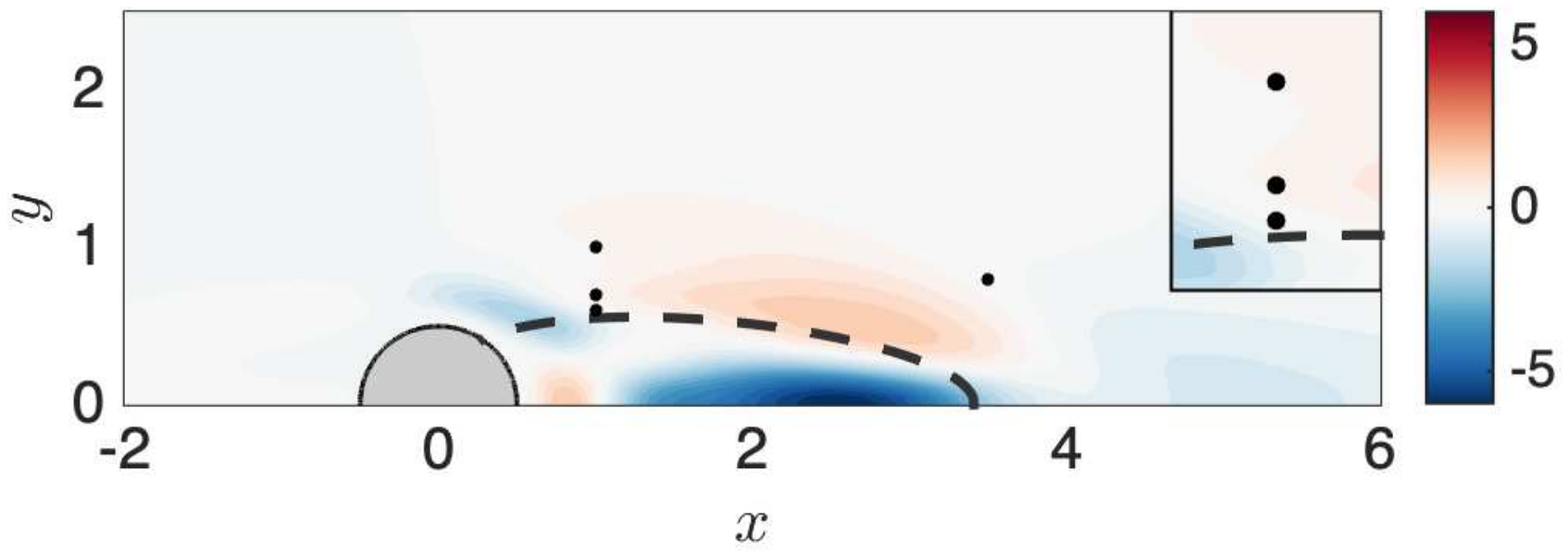}
      \put(-2,30){$(b)$}
   \end{overpic} 
}
\centerline{
   \hspace{0.3cm}
   \begin{overpic}[width=6.2cm, trim=10mm 40mm 10mm 170mm, clip=true]{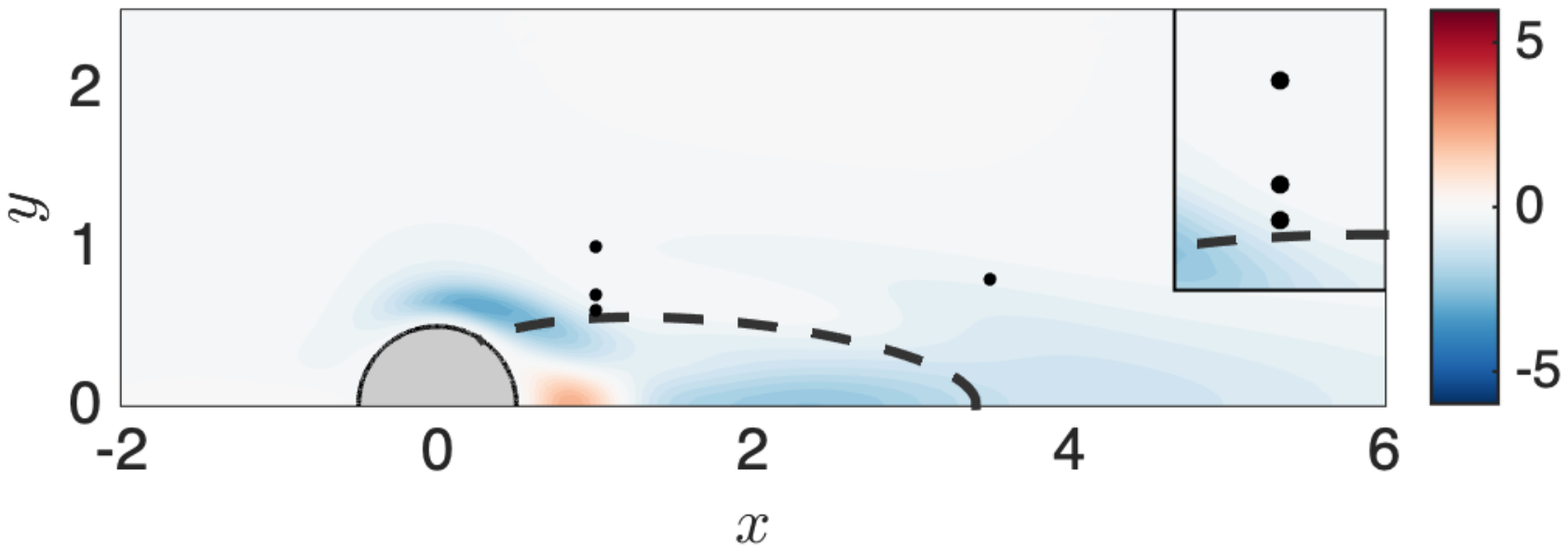}
      \put(-2,30){$(c)$}
   \end{overpic}  
   \begin{overpic}[width=6.2cm, trim=10mm 40mm 10mm 170mm, clip=true]{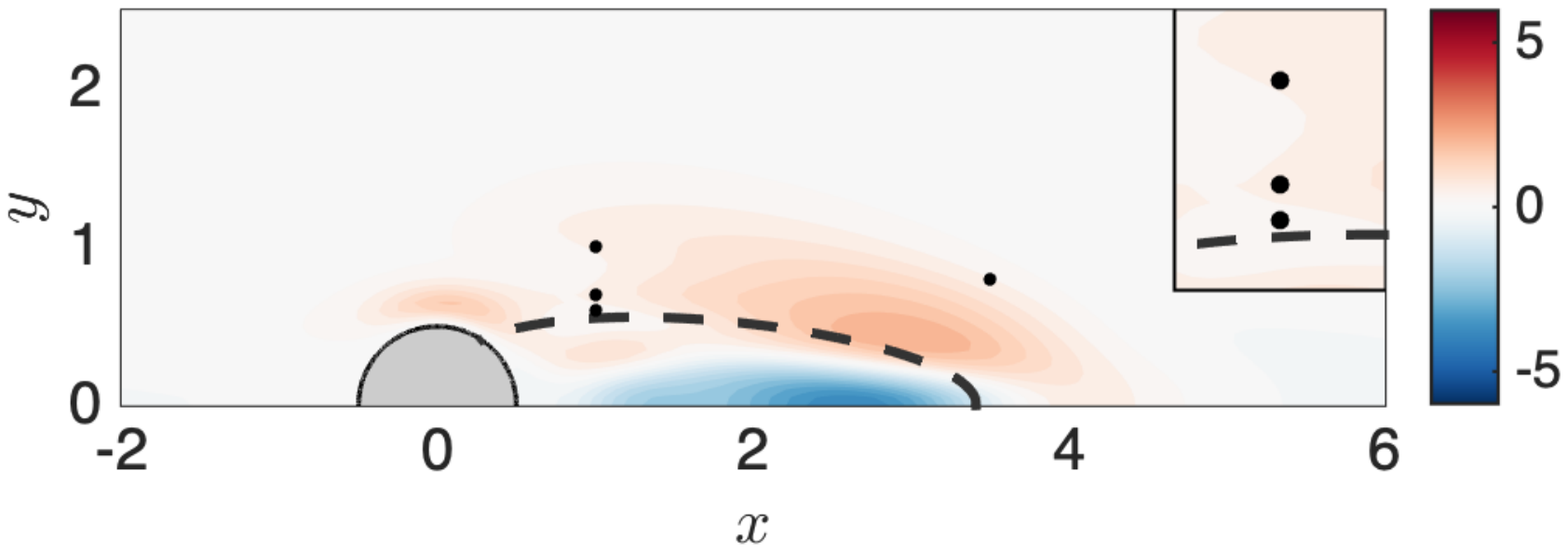}
      \put(-2,30){$(d)$}
   \end{overpic}  
}
\centerline{
   \hspace{0.3cm}
   \begin{overpic}[width=6.2cm, trim=10mm 40mm 10mm 170mm, clip=true]{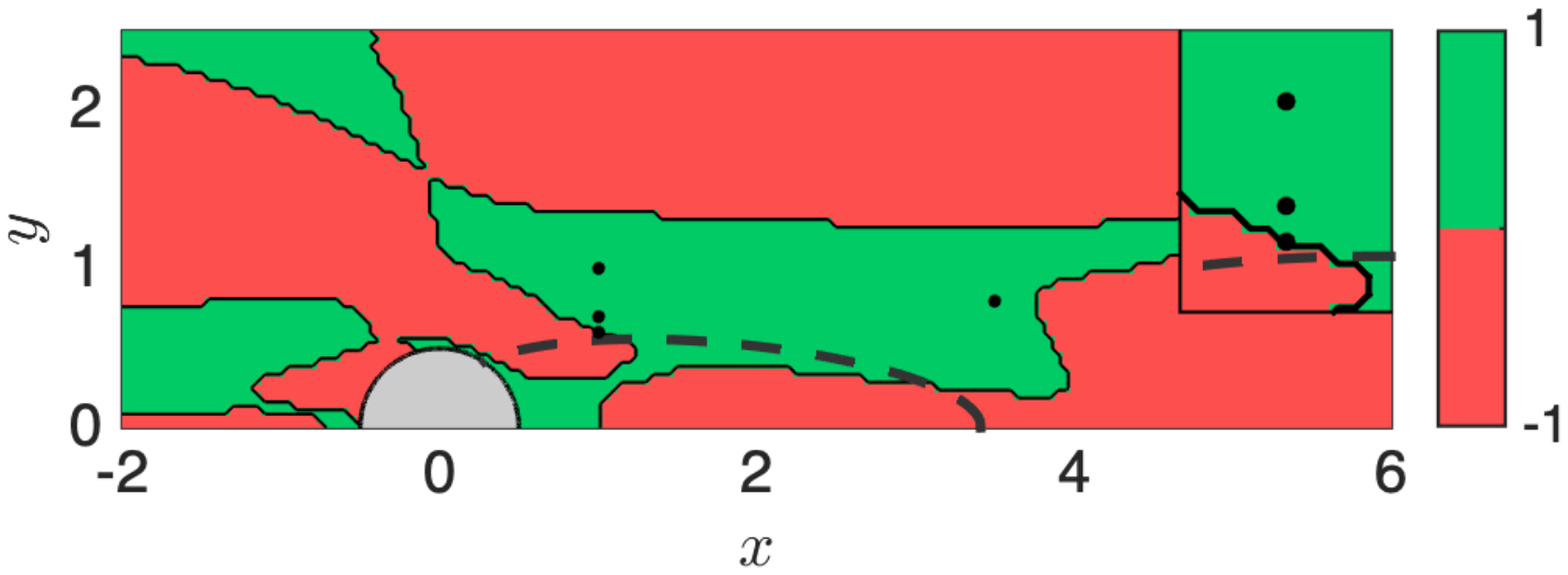}
      \put(-2,30){$(e)$}
   \end{overpic}  
   \begin{overpic}[width=6.2cm, trim=10mm 40mm 10mm 170mm, clip=true]{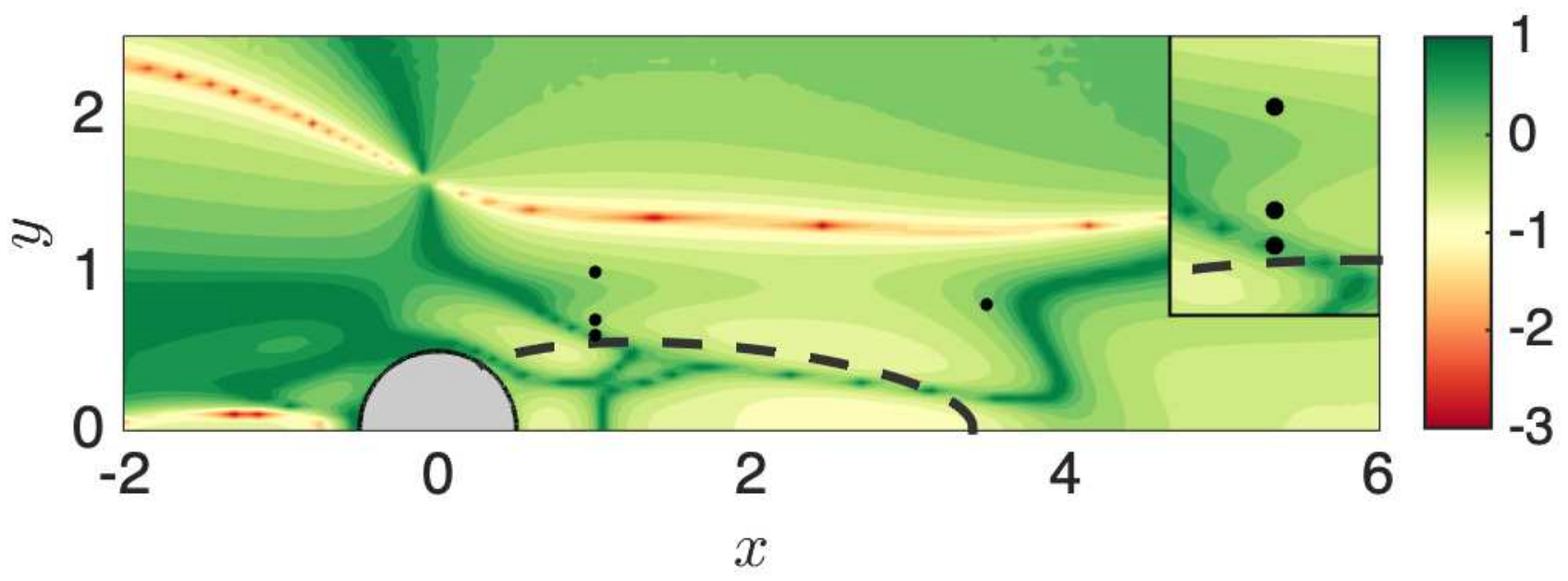}
      \put(-2,30){$(f)$}
   \end{overpic}  
}
\caption{
Same as figure~\ref{fig:F} for the sensitivity of the leading mode's frequency $\lambda_i$ to a localised steady force oriented along the $x$ direction, at
$\Rey=50.$
}
\label{fig:F_freq}
\end{figure}

The sensitivity of the leading mode's  frequency to a steady force $\FF=(\delta(\xx-\xx_c),0)^T$ is shown in figure~\ref{fig:F_freq}.
The following few comments can be made.
\begin{enumerate}
\item
While the first-order sensitivity  is positive almost everywhere (negative if changing the sign of $F_x$), the second-order sensitivity is positive in two distinct regions and negative in two others.

\item
Term I is dominant on the sides of the cylinder and immediately downstream ($x \leq 1$), 
while term II is dominant farther downstream $(x \geq 1$).

\item
The threshold amplitude $\epsilon_t$ is rather large almost everywhere, generally larger than for the leading growth rate (figure~\ref{fig:F}), indicating that second-order effects are less important for the frequency than for the growth rate.
\end{enumerate}

%
\subsection{Sensitivity to a small control cylinder}

\begin{figure} 
\centerline{
   \hspace{0.3cm}
   \begin{overpic}[width=6.2cm, trim=10mm 40mm 10mm 170mm, clip=true]{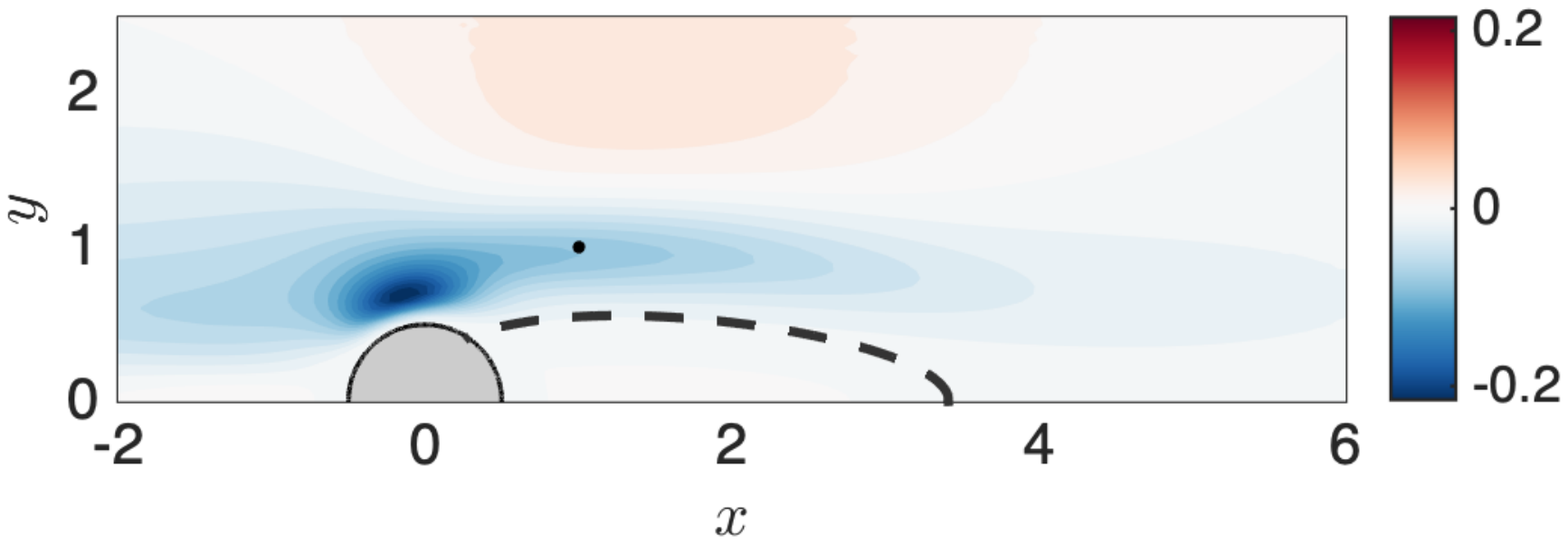}
      \put(-2,30){$(a)$}
   \end{overpic}  
   \begin{overpic}[width=6.2cm, trim=10mm 40mm 10mm 170mm, clip=true]{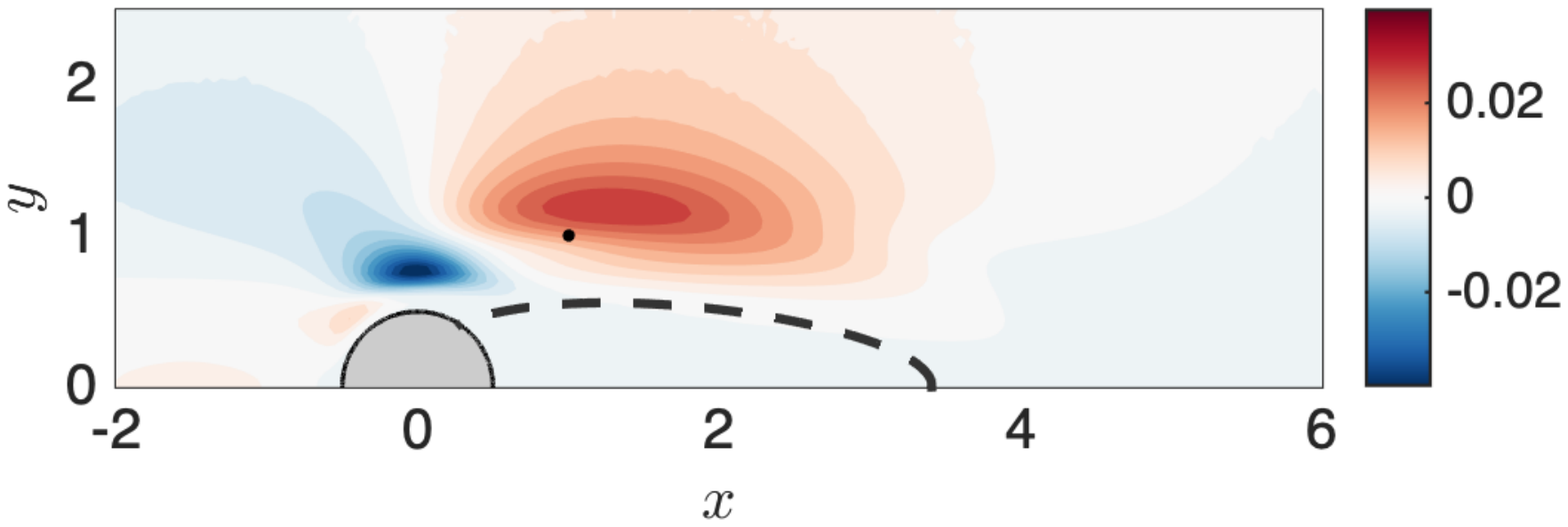}
      \put(-2,30){$(b)$}
   \end{overpic}  
}
\centerline{
   \hspace{0.3cm}
   \begin{overpic}[width=6.2cm, trim=10mm 40mm 10mm 170mm, clip=true]{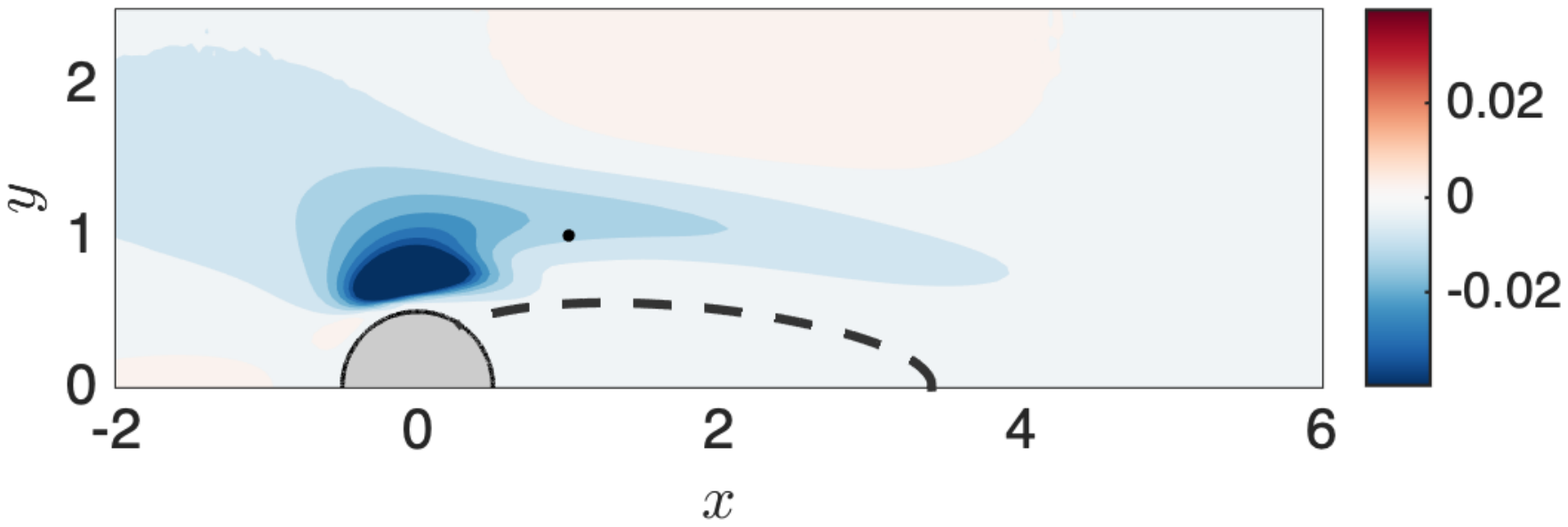}
      \put(-2,30){$(c)$}
   \end{overpic}  
   \begin{overpic}[width=6.2cm, trim=10mm 40mm 10mm 170mm, clip=true]{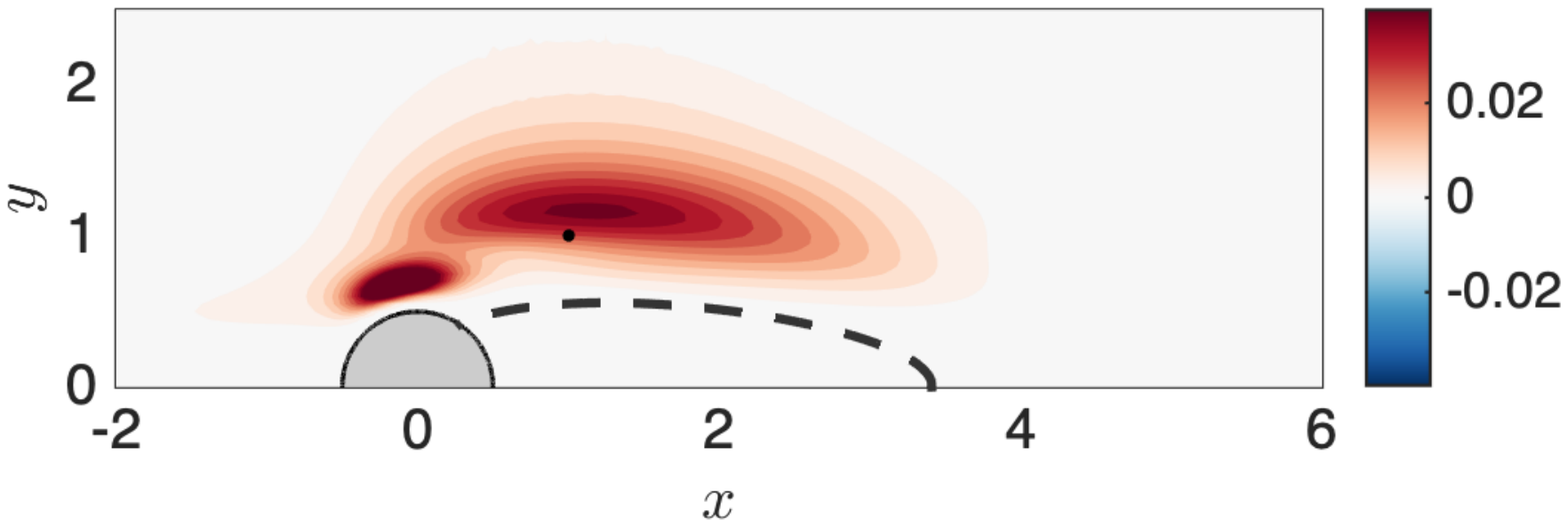}
      \put(-2,30){$(d)$}
   \end{overpic}  
}
\centerline{
   \hspace{0.3cm}
   \begin{overpic}[width=6.2cm, trim=10mm 40mm 10mm 170mm, clip=true]{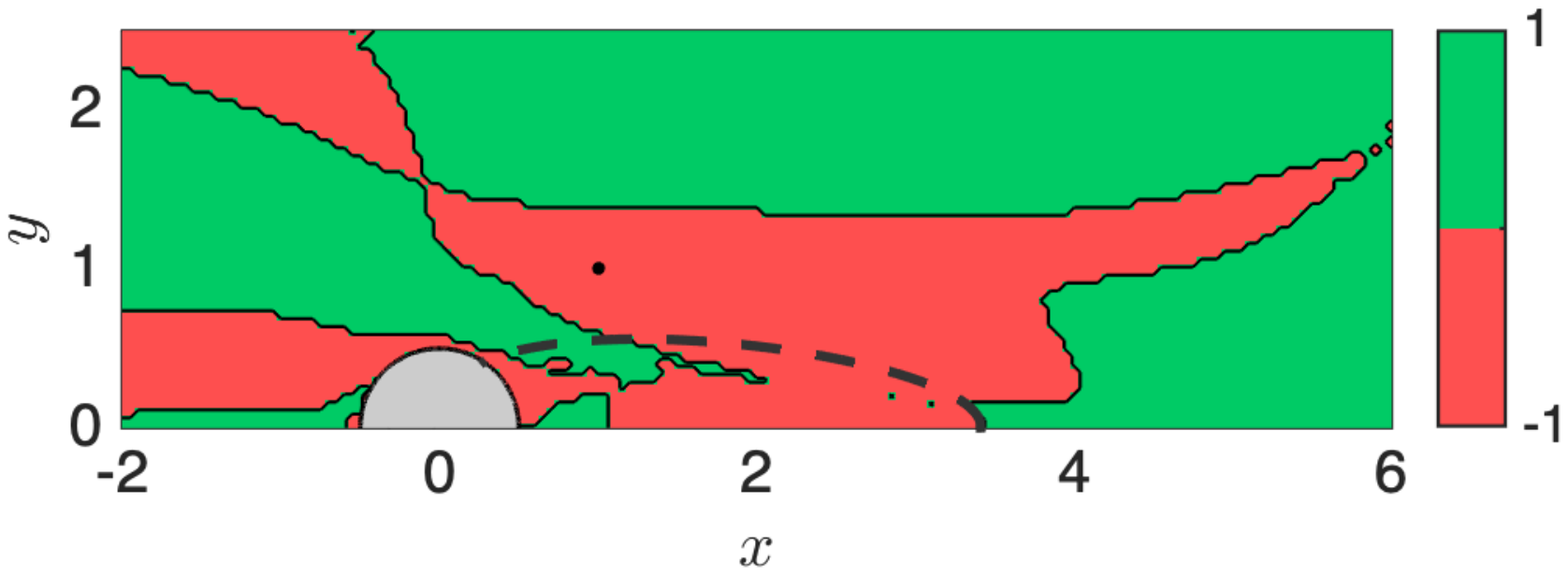}
      \put(-2,30){$(e)$}
   \end{overpic} 
   \hspace{6.2cm} 
}
\caption{
Frequency variation induced by
 a small control cylinder of diameter $d=0.1$, at
$\Rey=50.$
$(a)$~$\epsilon\lambda_{1i}$;
$(b)$~$\epsilon^2 \lambda_{2i}$.
$(c)$~Term I and $(d)$~term II in the decomposition of $\epsilon^2 \lambda_{2i}$.
$(e)$~Sign of the product $\lambda_{1i} \lambda_{2i}$.
(Figure \ref{fig:F_freq}$(f)$ has no equivalent here because the diameter $d$, and therefore the amplitude $\epsilon$, are fixed.)
The black dot shows the location $\xx_c=(1,1)$ investigated in \S~\ref{sec:analysis}.
}
\label{fig:cyl_freq}
\end{figure}

The sensitivity of the leading mode's frequency to a small control cylinder is shown in figure~\ref{fig:cyl_freq}.
The following few comments can be made.
\begin{enumerate}
\item
The first-order sensitivity is negative almost everywhere, while the second-order sensitivity is 
negative on the sides of the cylinder and 
positive on the sides of the recirculation region.
Both sensitivities are small inside the recirculation region.

\item
Term I is dominant on the sides of the cylinder, 
whereas term II is dominant on the sides of both the cylinder and the recirculation region.

\item
The second-order effect is approximately one order of magnitude smaller than the first-order effect. 
This contrasts with the growth rate
(first- and second-order effects of the same order of magnitude; see figure~\ref{fig:cyl}).

\end{enumerate}

%

\section{Derivation of the sensitivity operators}
\label{app:operators}

\subsection{First-order sensitivity operator}

Recall the first-order eigenvalue variation (\ref{eq:ev1}) induced by a steady force $\FF$:
\begin{align}
\lambda_1  &= -\ps{ \uu_0^\dag } {{\AAA}_1 \uu_0}.
\label{eq:ev1_qpp}
\end{align}
Next, define the linear operator $\LL$, which depends only on $\uu_0$, such that
\begin{align}
\AAA_1 \uu_0 = 
   \UU_1 \bcdot\bnabla \uu_0+ \uu_0 \bcdot \bnabla \UU_1  = \LL \UU_1.
\end{align}
Substituting into (\ref{eq:ev1_qpp}) yields
\begin{align}
\lambda_1  &= -\ps{ \uu_0^\dag } {\LL \UU_1}
= -\ps{ \LL^\dag \uu_0^\dag } {\UU_1},
\end{align}
where the adjoint operator of $\LL$ reads
\begin{align}
\LL^\dag=
(*)\bcdot \bnabla \overline\uu_0^T   - \overline\uu_0 \bcdot \bnabla(*).
\end{align}
The first-order sensitivity to base flow modification is therefore
\begin{align}
-\LL^\dag \uu_0^\dag,
\label{eq:first_order_sensit_BFmod}
\end{align}
and since $\UU_1$ is a solution of (\ref{eq:Q1}),
the first-order sensitivity to a steady force $\FF$
is
\begin{align}
\SS_1 = -{\AAA_0^\dag}^{-1} \LL^\dag \uu_0^\dag.
\end{align}
One recognises the usual sensitivities to flow modification and to a steady force \citep{Marquet08cyl, Meliga10}.

\subsection{Second-order sensitivity operator}

Recall the second-order eigenvalue variation (\ref{eq:ev2}) induced by a steady force $\FF$:
\begin{align}
\lambda_2 &= -\ps{ \uu_0^\dag }{
\AAA_2\uu_0 + (\lambda_1\II+\AAA_1)\uu_1 }.
\label{eq:ev2_qpp}
\end{align}
As explained in \S~\ref{sec:steady}, $\uu_1$ is defined up to any component along $\uu_0$ 
(i.e. $\uu_1 = \widetilde\uu_1 + \alpha\uu_0$ such that $\widetilde\uu_1$ has no component along $\uu_0$).
Injecting and developing yields
\begin{align}
\lambda_2 &= -\ps{ \uu_0^\dag }{
\AAA_2\uu_0}
- \ps{ \uu_0^\dag}{(\lambda_1\II+\AAA_1) \left( \widetilde\uu_1 + \alpha \uu_0\right)}
\nonumber\\
&= -\ps{ \uu_0^\dag }{
\AAA_2\uu_0}
- \ps{ \uu_0^\dag}{(\lambda_1\II+\AAA_1)  \widetilde\uu_1 }
- \alpha \ps{ \uu_0^\dag}{(\lambda_1\II+\AAA_1) \uu_0 }
\nonumber\\
&= -\ps{ \uu_0^\dag }{
\AAA_2\uu_0}
- \lambda_1 \underbrace{ \ps{ \uu_0^\dag}{\widetilde\uu_1 } }_{=0}
- \ps{ \uu_0^\dag}{ \AAA_1  \widetilde\uu_1 }
- \alpha \ps{ \uu_0^\dag}{(\lambda_0\II+\AAA_0) \uu_1 }
\nonumber\\
&= -\ps{ \uu_0^\dag }{
\AAA_2\uu_0}
- \ps{ \uu_0^\dag}{ \AAA_1  \widetilde\uu_1 }
- \alpha \ps{ \underbrace{ (\lambda_0\II+\AAA_0)^\dag \uu_0^\dag}_{=0} }{ \uu_1 },
\end{align}
so the arbitrary component $\alpha\uu_0$ does not modify  $\lambda_2$.
The second term can be rewritten in terms of $\uu_0$ by recalling that $\widetilde\uu_1$ is a solution of (\ref{eq:evp1}):
\begin{align}
\lambda_2 &= -\ps{ \uu_0^\dag }{
\AAA_2\uu_0}
+ \ps{ \uu_0^\dag}{ \AAA_1 (\lambda_0\II+\AAA_0)^{-1}(\lambda_1\II+\AAA_1) \uu_0}.
\end{align}
Next, defining the linear operators $\TT$ and $\MM$, which depend only on $\uu_0$ and $\uu_0^\dag$, respectively, such that
\begin{align}
(\lambda_1 \II + \AAA_1) \uu_0
&= \lambda_1 \uu_0 +  \UU_1 \bcdot\bnabla \uu_0+ \uu_0 \bcdot \bnabla \UU_1 
= \TT \UU_1,
\label{eq:T}
\\
\AAA_1^\dag \uu_0^\dag 
&=
-\UU_1 \bcdot\bnabla \uu_0^\dag + \uu_0^\dag \bcdot \bnabla \UU_1^T 
= \MM \UU_1,
\label{eq:M}
\end{align}
substituting into (\ref{eq:ev2_qpp}) and noting that $\AAA_2\uu_0 = \LL\UU_2$ yields
\begin{align}
\lambda_2 &=  
-\ps{ \uu_0^\dag }{\LL\UU_2}
+
\ps{ \MM \UU_1 }{  (\lambda_0\II+\AAA_0)^{-1} \TT \UU_1}
\nonumber
\\
&=
-\ps{ \LL^\dag \uu_0^\dag }{\UU_2}
+
\ps{ \UU_1 }{ \MM^\dag (\lambda_0\II+\AAA_0)^{-1} \TT \UU_1},
\end{align}
where the adjoint operator of $\MM$ reads
\begin{align}
\MM^\dag=
 -(*) \bcdot \bnabla\overline\uu_0^\dag
 -(*) \bcdot \bnabla{\overline\uu_0^\dag}^T.
\end{align}
Since $\UU_2$ is a solution of (\ref{eq:Q2}),
one can rewrite the first term,
\begin{align}
\lambda_2 &=  
\ps{ \LL^\dag \uu_0^\dag }{ \AAA_0^{-1}(\UU_1 \bcdot \bnabla \UU_1)}
+
\ps{ \UU_1 }{ \MM^\dag (\lambda_0\II+\AAA_0)^{-1} \TT \UU_1}
\nonumber \\
 &=  
\ps{ \UUa }{ (\UU_1 \bcdot \bnabla \UU_1)^T}
+
\ps{ \UU_1 }{ \MM^\dag (\lambda_0\II+\AAA_0)^{-1} \TT \UU_1},
\end{align}
where $\UUa$ is a solution of 
\be
\AAA_0^\dag \UUa = \LL^\dag  \uu_0^\dag. 
\ee 
Finally, introducing the linear operator 
\begin{align}
\qquad \KK
= 
\overline{\UU^\dag} \bcdot \bnabla(*)^T 
\end{align}
allows one to rearrange
the first term:
\begin{align}
\lambda_2 &= 
\ps{ \UU_1 }{ \KK \UU_1 }
+
\ps{ \UU_1 }{ \MM^\dag (\lambda_0\II+\AAA_0)^{-1} \TT \UU_1}.
\end{align}
The second-order sensitivity to base flow modification is therefore
\begin{align}
\KK + \MM^\dag (\lambda_0\II+\AAA_0)^{-1} \TT,
\label{eq:second_order_sensit_BFmod}
\end{align}
and since $\UU_1$ is a solution of (\ref{eq:Q1}),
the second-order sensitivity to a steady force $\FF$ is
\begin{align}
\SS_2 = {\AAA_0^\dag}^{-1}  \left( 
\underbrace{\KK}_{\text{I}} + 
\underbrace{\MM^\dag (\lambda_0\II+\AAA_0)^{-1} \TT}_{\text{II}} 
\right) \AAA_0^\dag.
\label{eq:S2_detailed}
\end{align}
Like in (\ref{eq:ev2}), term I is the effect of $\UU_2$ and term II is the effect of the $\UU_1$--$\uu_1$ interaction,

\section{Application to other sensitivity problems: the example of the resolvent gain}
\label{app:gain}

The method reported in this paper can  easily be adapted to compute second-order sensitivity in other problems if the quantity of interest is defined by an eigenvalue problem.
This is the case of the resolvent gain, a measure of the linear amplification of a time-harmonic perturbation or external forcing. 
The resolvent gain is particularly relevant to linearly stable flows, as it captures  non-normal effects not accessible to modal stability analysis.
The main steps of the method are outlined here.

Consider a harmonic forcing $\ff'(\xx,t) = \ff(\xx) e^{i\omega t} + c.c.$ applied to a linearly stable base flow $\UU(\xx)$. 
In the stationary regime, small-amplitude perturbations are harmonic  at the same frequency, $\uu'(\xx,t) = \uu(\xx) e^{i\omega t} + c.c.$, and their  linear evolution is described by 
\begin{align} 
i\omega\uu +  \UU \bcdot \bnabla\uu  + \uu \bcdot \bnabla\UU    +   \bnabla  p - \nnu \bnabla^2  \uu = \ff.
\end{align}
In other words, the problem is defined as
\begin{align} 
\NN(\UU) &= \00,
\\
(i\omega \II+\AAA) \uu &=\ff.  
\end{align}
The linear gain at the frequency of interest is the ratio of the norm of the response to the norm of the forcing,
$G(\omega) = ||\uu|| / ||\ff||$,
which can be recast as
\begin{align}
G^2(\omega) = \dfrac{||\uu||^2}{||\ff||^2}
= \dfrac{ \ps{\RR^\dag \RR \ff}{\ff} }{ \ps{\ff}{\ff} }
\end{align}
upon defining the resolvent operator $\RR(\omega) = (i\omega \II+\AAA)^{-1}$ such that $\uu=\RR \ff$, and its adjoint operator $\RR^\dag$.
At a given frequency, the gain is maximised by the optimal forcing,
\begin{align}
G_{opt}^2(\omega)  = \max_{\ff} \dfrac{||\uu||^2}{||\ff||^2}
= \dfrac{||\uu_{opt}||^2}{||\ff_{opt}||^2}
= \dfrac{ \ps{\RR^\dag \RR\ff_{opt}}{\ff_{opt}} }{ \ps{\ff_{opt}}{\ff_{opt}} },
\end{align}
which can be solved via the eigenvalue problem
\begin{align}
\RR^\dag \RR \ff = G^2 \ff,
\label{eq:RRf}
\end{align}
i.e. a problem similar to (\ref{eq:evp_compact}), where the operator and the eigenvalue are now $\RR^\dag \RR$  and $-G^2$, respectively.

When a small-amplitude steady control is applied on the base flow, 
\begin{align} 
\NN(\UU) &= \epsilon \FF,
\label{eq:ctrl_BF}
\end{align}
 the base flow, the linear response and the resolvent gain are modified and can  be expressed as power series expansions,
\begin{align} 
\UU&=\UU_0+\epsilon\UU_1+\epsilon^2\UU_2+\ldots,
\\
\uu& = \uu_0+\epsilon\uu_1+\epsilon^2\uu_2+\ldots,
\\
G^2 &= G^2_0 +\epsilon G^2_1 +\epsilon^2 G^2_2 +\ldots,
\end{align} 
and one would like to predict the first ans second-order gain variations 
$G_1$ and $G_2$.
Two cases should be distinguished: either
(i)~the harmonic forcing $\ff$ is prescribed, 
or 
(ii)~the optimal gain is of interest and the optimal forcing $\ff_{opt}$ is itself modified by the control as
\begin{align} 
\ff_{opt} & = \ff_{0} + \epsilon \ff_{1} + \epsilon^2 \ff_{2} + \ldots.
\end{align} 
Let us consider for now the most general case
(ii).
Injecting the above expansions in (\ref{eq:RRf})-(\ref{eq:ctrl_BF})
yields
equations (\ref{eq:Q0})-(\ref{eq:Q2}) for $\UU_0$, $\UU_1$ and $\UU_2$,
and the following equations analogous to (\ref{eq:evp0})-(\ref{eq:evp2}) for the response:
\begin{align} 
\left[ (\RR^\dag \RR)_0-G_0^2 \II \right] \ff_0 &= \00,
\label{eq:RR0}
\\
\left[ (\RR^\dag \RR)_0-G_0^2 \II \right] \ff_1 &= -\left[(\RR^\dag \RR)_1-G_1^2\II \right] \ff_0,
\label{eq:RR1}
\\
 \left[(\RR^\dag \RR)_0-G_0^2\II \right] \ff_2 &=
-\left[(\RR^\dag \RR)_1-G_1^2\II \right] \ff_1 
-\left[(\RR^\dag \RR)_2-G_2^2\II \right] \ff_0.
\label{eq:RR2}
\end{align}
In the derivation of the above equations, 
the expansion $\RR = \RR_0 + \epsilon \RR_1 + \epsilon^2 \RR_2 + \ldots$ has been injected into $\RR^\dag \RR$, giving
\begin{align} 
(\RR^\dag \RR)_0 &= \RR_0^\dag \RR_0,
\\
(\RR^\dag \RR)_1 &= \RR_0^\dag \RR_1 + \RR_1^\dag \RR_0,
\\
(\RR^\dag \RR)_2 &= \RR_0^\dag \RR_2 + \RR_1^\dag \RR_1 + \RR_2^\dag \RR_0,
\end{align}
and the expansion
 $\AAA = \AAA_0 + \epsilon \AAA_1 + \epsilon^2 \AAA_2 + \ldots$ has been injected into $\RR=(i\omega\II+\AAA)^{-1}$, allowing one to identify
\begin{align} 
\RR_0 &=\RR,
\\
\RR_1 &= -\RR_0 \AAA_1 \RR_0,
\\
\RR_2 &= -\RR_0 \AAA_2 \RR_0.
\end{align}
Projecting (\ref{eq:RR1})-(\ref{eq:RR2}) on the adjoint forcing $\ff^\dag=\ff$ (note that $\RR^\dag \RR$ is self-adjoint) 
and choosing the normalisation $\ps{ \ff_0 }{ \ff_0 }=1$ 
yields the expressions of the desired gain variations, similar to (\ref{eq:ev1})-(\ref{eq:ev2}):
\begin{align} 
G_1^2
&=
\ps{ \ff_0 }{ (\RR^\dag \RR)_1 \ff_0 },
\label{eq:G1_expl}
\\
G_2^2 &=
\ps{ \ff_0 }{ (\RR^\dag \RR)_2 \ff_0  +
\left[(\RR^\dag \RR)_1 - G_1^2 \II \right]  \ff_1 }.
\label{eq:G2_expl}
\end{align} 
For a given control $\FF$, one can easily compute the base flow modifications $\UU_1$ and $\UU_2$,
build the operators $\AAA_1$, $\AAA_2$, $\RR_1$ and $\RR_2$,
compute the forcing modification $\ff_1$, and  finally calculate the first- and second-order gain variations $G_1^2$ and $G_2^2$.

More interestingly, it is  possible to recast these variations as
\begin{align}
G_1^2 &= \ps{\SS_1}{\FF},
\label{eq:sensit1_G}
\\
G_2^2 &= \ps{\FF}{\SS_2\FF},
\label{eq:sensit2_G}
\end{align}
where the sensitivity operators $\SS_1$ and $\SS_2$ depend only on the uncontrolled base flow $\UU_0$ and the forcing $\ff_0$.
The derivation involves introducing suitable adjoint operators, along the same lines as
 the derivation of the sensitivity operators for $\lambda_1$ and $\lambda_2$.
The final result reads
\begin{align}
\SS_1 = -2 G_0^2 
\text{Re}\left\{ 
{\AAA_0^\dag}^{-1} \LL^\dag \ff_0
\right\}
\end{align}
for the first-order sensitivity, where one recognises the usual sensitivity to a steady force \citep{Brandt11},
and 
\begin{align}
\SS_2 = {\AAA_0^\dag}^{-1}  \left( 
\underbrace{ 2 G_0^2 
\text{Re}\left\{ \KK \right\}
+ \LL^\dag \RR_0^\dag \RR_0 \LL}_{\text{I}}
+ \underbrace{ \MM^\dag  \left[ \RR_0^\dag \RR_0 - G_0^2 \II\right]^{-1} \TT }_{\text{II}}
\right) \AAA_0^\dag
\label{eq:S2_detailed_gain}
\end{align}
for the second-order sensitivity, 
where  $\KK$, $\LL$, $\MM$ and $\TT$ are now defined by 
\begin{align}
\KK &= \overline{\UU^\dag} \bcdot \bnabla(*)^T, 
\quad \mbox{where} \quad
\AAA_0^\dag \UUa  = \LL^\dag  \ff_0, \\
\AAA_1      \uu_0 &= \LL \UU_1, \\
-(\RR^\dag \RR)_1 \ff_0 &= \MM \UU_1, \\
\left[ (\RR^\dag \RR)_1 - G_1^2 \II\right] \ff_0 &= \TT \UU_1.
\end{align}

Comparing with the second-order eigenvalue sensitivity (\ref{eq:S2_detailed}), it appears that term II
(from the $\UU_1$--$\uu_1$ interaction) is directly analogous,
while term I (from $\UU_2$) contains  an analogous part depending on $\KK$ but also  an additional part.

Coming back to case (i), where the harmonic forcing $\ff$ is fixed, the second-order gain variation becomes
$G_2^2 = \ps{\ff}{ (\RR^\dag\RR)_2 \ff}$,  term II is null, and the second-order sensitivity operator reduces to
\begin{align}
\SS_2 = {\AAA_0^\dag}^{-1}  \left( 
 2 G_0^2 
\text{Re}\left\{ \KK \right\}
+ \LL^\dag \RR_0^\dag \RR_0 \LL
\right) \AAA_0^\dag.
\end{align}


\bibliographystyle{jfm}
\bibliography{ALL}

\begin{thebibliography}{42}
\expandafter\ifx\csname natexlab\endcsname\relax\def\natexlab#1{#1}\fi
\def\au#1{#1} \def\ed#1{#1} \def\yr#1{#1}\def\at#1{#1}\def\jt#1{\textit{#1}}
  \def\bt#1{#1}\def\bvol#1{\textbf{#1}} \def\vol#1{#1} \def\pg#1{#1}
  \def\publ#1{#1}\def\arxiv#1{#1}\def\org#1{#1}\def\st#1{\textit{#1}}

\bibitem[Barkley \& Henderson(1996)]{barkley96}
{\sc \au{Barkley, D.} \& \au{Henderson, R.~D.}} \yr{1996}
  \at{Three-dimensional {F}loquet stability analysis of the wake of a circular
  cylinder}.  \jt{Journal of Fluid Mechanics}  \bvol{322},  \pg{215--241}.

\bibitem[Bewley {\em et~al.\/}(2001)Bewley, Moin \& Temam]{Bewley01}
{\sc \au{Bewley, T.R.}, \au{Moin, P.} \& \au{Temam, R.}} \yr{2001}
  \at{{DNS}-based predictive control of turbulence: an optimal benchmark for
  feedback algorithms}.  \jt{Journal of Fluid Mechanics}  \bvol{447},
  \pg{179--225}.

\bibitem[Bottaro {\em et~al.\/}(2003)Bottaro, Corbett \& Luchini]{Bottaro03}
{\sc \au{Bottaro, A.}, \au{Corbett, P.} \& \au{Luchini, P.}} \yr{2003}  \at{The
  effect of base flow variation on flow stability}.  \jt{Journal of Fluid
  Mechanics}  \bvol{476},  \pg{293--302}.

\bibitem[Boujo {\em et~al.\/}(2015)Boujo, Fani \&
  Gallaire]{Boujo15secondorderJFM}
{\sc \au{Boujo, E.}, \au{Fani, A.} \& \au{Gallaire, F.}} \yr{2015}
  \at{Second-order sensitivity of parallel shear flows and optimal
  spanwise-periodic flow modifications}.  \jt{Journal of Fluid Mechanics}
  \bvol{782},  \pg{491–514}.

\bibitem[Boujo {\em et~al.\/}(2019)Boujo, Fani \& Gallaire]{Boujo19sensit}
{\sc \au{Boujo, E.}, \au{Fani, A.} \& \au{Gallaire, F.}} \yr{2019}
  \at{Second-order sensitivity in the cylinder wake: Optimal spanwise-periodic
  wall actuation and wall deformation}.  \jt{Physical Review Fluids}  \bvol{4},
   \pg{053901}.

\bibitem[Boujo \& Gallaire(2014)]{Boujo14lengthJFM}
{\sc \au{Boujo, E.} \& \au{Gallaire, F.}} \yr{2014}  \at{Controlled
  reattachment in separated flows: a variational approach to recirculation
  length reduction}.  \jt{Journal of Fluid Mechanics}  \bvol{742},
  \pg{618--635}.

\bibitem[Boujo \& Sellier(2019)]{BoujoSellier2019}
{\sc \au{Boujo, E.} \& \au{Sellier, M.}} \yr{2019}  \at{Pancake making and
  surface coating: Optimal control of a gravity-driven liquid film}.
  \jt{Physical Review Fluids}  \bvol{4},  \pg{064802}.

\bibitem[Brandt {\em et~al.\/}(2011)Brandt, Sipp, Pralits \& Marquet]{Brandt11}
{\sc \au{Brandt, L.}, \au{Sipp, D.}, \au{Pralits, J.O.} \& \au{Marquet, O.}}
  \yr{2011}  \at{Effect of base-flow variation in noise amplifiers: the
  flat-plate boundary layer}.  \jt{Journal of Fluid Mechanics}  \bvol{687},
  \pg{503--528}.

\bibitem[Chomaz({2005})]{Chomaz05}
{\sc \au{Chomaz, J.M.}} \yr{{2005}}  \at{{Global instabilities in spatially
  developing flows: Non-normality and nonlinearity}}.  \jt{{Annual Review of
  Fluid Mechanics}}  \bvol{{37}},  \pg{{357--392}}.

\bibitem[{Cossu}(2014)]{Cossu14secondorder}
{\sc \au{{Cossu}, C.}} \yr{2014}  \at{{On the stabilizing mechanism of {2D}
  absolute and global instabilities by {3D} streaks}}.  \jt{ArXiv e-prints} ,
  \arxiv{arXiv: 1404.3191}.

\bibitem[Del~Guercio {\em et~al.\/}(2014{\natexlab{{\em a\/}}})Del~Guercio,
  Cossu \& Pujals]{DelGuercio2014globalwake}
{\sc \au{Del~Guercio, G.}, \au{Cossu, C.} \& \au{Pujals, G.}}
  \yr{2014{\natexlab{{\em a\/}}}}  \at{Optimal perturbations of non-parallel
  wakes and their stabilizing effect on the global instability}.  \jt{Physics
  of Fluids}  \bvol{26}~(2),  \pg{024110}.

\bibitem[Del~Guercio {\em et~al.\/}(2014{\natexlab{{\em b\/}}})Del~Guercio,
  Cossu \& Pujals]{DelGuercio2014globalcyl}
{\sc \au{Del~Guercio, G.}, \au{Cossu, C.} \& \au{Pujals, G.}}
  \yr{2014{\natexlab{{\em b\/}}}}  \at{Optimal streaks in the circular cylinder
  wake and suppression of the global instability}.  \jt{Journal of Fluid
  Mechanics}  \bvol{752},  \pg{572--588}.

\bibitem[Del~Guercio {\em et~al.\/}(2014{\natexlab{{\em c\/}}})Del~Guercio,
  Cossu \& Pujals]{DelGuercio2014parallel}
{\sc \au{Del~Guercio, G.}, \au{Cossu, C.} \& \au{Pujals, G.}}
  \yr{2014{\natexlab{{\em c\/}}}}  \at{Stabilizing effect of optimally
  amplified streaks in parallel wakes}.  \jt{Journal of Fluid Mechanics}
  \bvol{739},  \pg{37--56}.

\bibitem[Fani {\em et~al.\/}(2013)Fani, Camarri \& Salvetti]{Fani2013}
{\sc \au{Fani, A.}, \au{Camarri, S.} \& \au{Salvetti, M.~V.}} \yr{2013}
  \at{Investigation of the steady engulfment regime in a three-dimensional
  {T}-mixer}.  \jt{Physics of Fluids}  \bvol{25}~(6),  \pg{064102}.

\bibitem[Finn(1953)]{Finn1953}
{\sc \au{Finn, R.~K.}} \yr{1953}  \at{Determination of the drag on a cylinder
  at low {R}eynolds numbers}.  \jt{Journal of Applied Physics}  \bvol{24}~(6),
  \pg{771--773}.

\bibitem[Foures {\em et~al.\/}(2014)Foures, Caulfield \& Schmid]{Foures14}
{\sc \au{Foures, D.P.G.}, \au{Caulfield, C.P.} \& \au{Schmid, P.J.}} \yr{2014}
  \at{Optimal mixing in two-dimensional plane {P}oiseuille flow at finite
  {P}éclet number}.  \jt{Journal of Fluid Mechanics}  \bvol{748},
  \pg{241--277}.

\bibitem[Gander {\em et~al.\/}(1989)Gander, Golub \& von Matt]{GANDER1989}
{\sc \au{Gander, W.}, \au{Golub, G.~H.} \& \au{von Matt, U.}} \yr{1989}  \at{A
  constrained eigenvalue problem}.  \jt{Linear Algebra and its Applications}
  \bvol{114-115},  \pg{815 -- 839}, special Issue Dedicated to Alan J. Hoffman.

\bibitem[Giannetti \& Luchini(2007)]{Giannetti07}
{\sc \au{Giannetti, F.} \& \au{Luchini, P.}} \yr{2007}  \at{Structural
  sensitivity of the first instability of the cylinder wake}.  \jt{Journal of
  Fluid Mechanics}  \bvol{581},  \pg{167--197}.

\bibitem[Hecht(2012)]{Hecht2012}
{\sc \au{Hecht, F.}} \yr{2012}  \at{New development in {F}ree{F}em++}.
  \jt{Journal of Numerical Mathematics}  \bvol{20}~(3-4),  \pg{251--265}.

\bibitem[Hill(1992)]{Hill92AIAA}
{\sc \au{Hill, D.~C.}} \yr{1992}  \at{A theoretical approach for analyzing the
  restabilization of wakes}.  \jt{AIAA 92-0067} .

\bibitem[Hinch(1991)]{Hinch1991}
{\sc \au{Hinch, E.J.}} \yr{1991} {\em Perturbation Methods\/}.
  \publ{Cambridge: Cambridge University Press}.

\bibitem[Hwang {\em et~al.\/}(2013)Hwang, Kim \& Choi]{Hwang2013}
{\sc \au{Hwang, Y.}, \au{Kim, J.} \& \au{Choi, H.}} \yr{2013}
  \at{Stabilization of absolute instability in spanwise wavy two-dimensional
  wakes}.  \jt{Journal of Fluid Mechanics}  \bvol{727},  \pg{346--378}.

\bibitem[Jameson {\em et~al.\/}(1998)Jameson, Martinelli \& Pierce]{Jameson98}
{\sc \au{Jameson, A.}, \au{Martinelli, L.} \& \au{Pierce, N.A.}} \yr{1998}
  \at{Optimum aerodynamic design using the {N}avier–{S}tokes equations}.
  \jt{Theoretical and Computational Fluid Dynamics}  \bvol{10}~(1-4),
  \pg{213--237}.

\bibitem[Lions(1971)]{Lions71}
{\sc \au{Lions, J.L.}} \yr{1971} {\em Optimal control of systems governed by
  partial differential equations\/}.  \publ{New York: Springer-Verlag}.

\bibitem[Luchini \& Bottaro(2014)]{Luchini14}
{\sc \au{Luchini, P.} \& \au{Bottaro, A.}} \yr{2014}  \at{Adjoint equations in
  stability analysis}.  \jt{Annual Review of Fluid Mechanics}  \bvol{46}~(1),
  \pg{493--517}.

\bibitem[Magri \& Juniper(2013)]{Magri13}
{\sc \au{Magri, L.} \& \au{Juniper, M.~P.}} \yr{2013}  \at{Sensitivity analysis
  of a time-delayed thermo-acoustic system via an adjoint-based approach}.
  \jt{Journal of Fluid Mechanics}  \bvol{719},  \pg{183--202}.

\bibitem[Marquet {\em et~al.\/}(2008)Marquet, Sipp \& Jacquin]{Marquet08cyl}
{\sc \au{Marquet, O.}, \au{Sipp, D.} \& \au{Jacquin, L.}} \yr{2008}
  \at{Sensitivity analysis and passive control of cylinder flow}.  \jt{Journal
  of Fluid Mechanics}  \bvol{615},  \pg{221--252}.

\bibitem[Meliga {\em et~al.\/}(2014)Meliga, Boujo, Pujals \&
  Gallaire]{Meliga14}
{\sc \au{Meliga, P.}, \au{Boujo, E.}, \au{Pujals, G.} \& \au{Gallaire, F.}}
  \yr{2014}  \at{Sensitivity of aerodynamic forces in laminar and turbulent
  flow past a square cylinder}.  \jt{Physics of Fluids}  \bvol{26}~(10),
  \pg{104101}.

\bibitem[Meliga {\em et~al.\/}(2010)Meliga, Sipp \& Chomaz]{Meliga10}
{\sc \au{Meliga, P.}, \au{Sipp, D.} \& \au{Chomaz, J.-M.}} \yr{2010}
  \at{Open-loop control of compressible afterbody flows using adjoint methods}.
   \jt{Physics of Fluids}  \bvol{22}~(5),  \pg{054109}.

\bibitem[Mensah {\em et~al.\/}(2020)Mensah, Orchini \& Moeck]{Mensah2020JSV}
{\sc \au{Mensah, G.~A.}, \au{Orchini, A.} \& \au{Moeck, J.~P.}} \yr{2020}
  \at{Perturbation theory of nonlinear, non-self-adjoint eigenvalue problems:
  Simple eigenvalues}.  \jt{Journal of Sound and Vibration}  \bvol{473},
  \pg{115200}.

\bibitem[Mohammadi \& Pironneau(2001)]{Pironneau01}
{\sc \au{Mohammadi, B.} \& \au{Pironneau, O.}} \yr{2001} {\em Applied shape
  optimization for fluids\/}.  \publ{Oxford: Clarendon Press}.

\bibitem[Monokrousos {\em et~al.\/}(2011)Monokrousos, Bottaro, Brandt, Di~Vita
  \& Henningson]{Monokrousos11}
{\sc \au{Monokrousos, A.}, \au{Bottaro, A.}, \au{Brandt, L.}, \au{Di~Vita, A.}
  \& \au{Henningson, D.S.}} \yr{2011}  \at{Nonequilibrium thermodynamics and
  the optimal path to turbulence in shear flows}.  \jt{Physical Review Letters}
   \bvol{106},  \pg{134502}.

\bibitem[Pringle \& Kerswell(2010)]{Pringle2010}
{\sc \au{Pringle, C.C.T.} \& \au{Kerswell, R.R.}} \yr{2010}  \at{Using
  nonlinear transient growth to construct the minimal seed for shear flow
  turbulence}.  \jt{Physical Review Letters}  \bvol{105},  \pg{154502}.

\bibitem[Schmid(2007)]{Schmid2007}
{\sc \au{Schmid, P.~J.}} \yr{2007}  \at{Nonmodal stability theory}.  \jt{Annu.
  Rev. Fluid Mech.}  \bvol{39},  \pg{129--162}.

\bibitem[Sipp \& Lebedev(2007)]{Sipp2007}
{\sc \au{Sipp, D.} \& \au{Lebedev, A.}} \yr{2007}  \at{Global stability of base
  and mean flows: a general approach and its applications to cylinder and open
  cavity flows}.  \jt{Journal of Fluid Mechanics}  \bvol{593},  \pg{333--358}.

\bibitem[Strykowski \& Sreenivasan(1990)]{Strykowski1990}
{\sc \au{Strykowski, P.~J.} \& \au{Sreenivasan, K.~R.}} \yr{1990}  \at{On the
  formation and suppression of vortex `shedding' at low {R}eynolds numbers}.
  \jt{Journal of Fluid Mechanics}  \bvol{218},  \pg{71--107}.

\bibitem[Tammisola(2017)]{Tammisola2017}
{\sc \au{Tammisola, O.}} \yr{2017}  \at{Optimal wavy surface to suppress vortex
  shedding using second-order sensitivity to shape changes}.  \jt{European
  Journal of Mechanics - B/Fluids}  \bvol{62}~(Supplement C),  \pg{139--148}.

\bibitem[Tammisola {\em et~al.\/}(2014)Tammisola, Giannetti, Citro \&
  Juniper]{Tammisola2014}
{\sc \au{Tammisola, O.}, \au{Giannetti, F.}, \au{Citro, V.} \& \au{Juniper,
  M.~P.}} \yr{2014}  \at{Second-order perturbation of global modes and
  implications for spanwise wavy actuation}.  \jt{Journal of Fluid Mechanics}
  \bvol{755},  \pg{314--335}.

\bibitem[Tchoufag {\em et~al.\/}(2013)Tchoufag, Magnaudet \&
  Fabre]{Tchoufag2013}
{\sc \au{Tchoufag, J.}, \au{Magnaudet, J.} \& \au{Fabre, D.}} \yr{2013}
  \at{Linear stability and sensitivity of the flow past a fixed oblate
  spheroidal bubble}.  \jt{Physics of Fluids}  \bvol{25}~(5),  \pg{054108}.

\bibitem[Tritton(1959)]{Tritton1959}
{\sc \au{Tritton, D.J.}} \yr{1959}  \at{Experiments on the flow past a circular
  cylinder at low {R}eynolds numbers}.  \jt{Journal of Fluid Mechanics}
  \bvol{6},  \pg{547--567}.

\bibitem[Tuckerman \& Barkley(2000)]{Tuckerman2000timestepping}
{\sc \au{Tuckerman, L.~S.} \& \au{Barkley, D.}} \yr{2000} Bifurcation analysis
  for timesteppers.  \bt{In {\em Numerical Methods for Bifurcation Problems and
  Large-Scale Dynamical Systems\/} (ed. \ed{E.~Doedel \& L.~S. Tuckerman})},
  \pg{pp. 453--466}.  \publ{New York, NY: Springer New York}.

\bibitem[Verma \& Mittal(2011)]{Verma2011}
{\sc \au{Verma, A.} \& \au{Mittal, S.}} \yr{2011}  \at{A new unstable mode in
  the wake of a circular cylinder}.  \jt{Physics of Fluids}  \bvol{23}~(12),
  \pg{121701}.

\end{thebibliography}

\end{document}